\documentclass[letterpaper,11pt]{article}
\usepackage{jheppub}
\usepackage{xcolor}
\usepackage{braket}
\usepackage{graphicx}
\usepackage{soul}
\usepackage{amssymb,amsmath}

\newcommand{\pd}{\partial}
\newcommand{\nn}{\nonumber\\}
\DeclareMathOperator{\tr}{tr}
\usepackage{verbatim}
\usepackage{anyfontsize}
\usepackage{dsfont}
\usepackage{tikz}
\usepackage{sidecap}
\usepackage{subfig}
\sidecaptionvpos{figure}{t}
\usepackage[english]{babel}
\usepackage{enumitem}  

\usepackage{graphicx, xcolor, varwidth}
\usepackage{setspace}
\usepackage[permil]{overpic}
\usepackage{graphicx}
\usepackage{mathtools}
\usepackage{wasysym}
\usepackage{graphicx}% Include figure files
\usepackage{placeins}
\usepackage{bm}
\newcommand{\be}{\begin{equation}}
\newcommand{\ee}{\end{equation}}

\usepackage{amsmath}
\usepackage{amsfonts}
\usepackage{amssymb}
\usepackage{graphicx}
\usepackage{wrapfig}
\usepackage{braket}
\usepackage{hyperref}
\usepackage{amsthm,verbatim,cancel}
\usepackage{color}
\usepackage{caption}

\newcommand{\bea}{\begin{eqnarray}}
\newcommand{\eea}{\end{eqnarray}}

\usepackage{tikz}
\usetikzlibrary{positioning,decorations.pathmorphing}
\usepackage{hyperref}
\usepackage{amssymb}
\usepackage{amsmath}
\usepackage{color}

\newcommand \mathtikz[1] {\quad \vcenter{\hbox{\tikz{#1}}} \quad}

\newcommand\idC[2] { %Closed identity
\begin{scope}[xshift=#1,yshift=#2]
\filldraw[left color=lightgray, right color=white] (-0.25,0) -- (0.25,0) -- (0.25,-1) to [in=-90,out=-90] (-0.25,-1) -- (-0.25,0);
\filldraw[left color=white,right color=lightgray] (0,0) ellipse (0.25 and 0.1);, 
\draw[dotted] (0.25,-1) arc (0:180:0.25 and 0.1);
\end{scope}
}

\newcommand\idA[2] { %Open identity
\begin{scope}[xshift=#1,yshift=#2]
\filldraw[fill=white,draw=black] (-0.25,0) rectangle (0.25,-1);
\end{scope}
}

\newcommand\muC[2]{ % Closed product
\begin{scope}[xshift=#1,yshift=#2]
\filldraw[left color=lightgray, right color=white] (-0.25,0) to [out=-90,in=180] (0,-0.33) to [in=-90,out=0] (0.25,0) to  (0.75,0) to [in=90,out=-90] (0.25,-1) to [out=-90,in=-90] (-0.25,-1) to [in=-90,out=90] (-0.75,0);
\filldraw[left color=white,right color=lightgray] (-0.5,0) ellipse (0.25 and 0.1);
\filldraw[left color=white,right color=lightgray] (0.5,0) ellipse (0.25 and 0.1);
\draw[dotted] (0.25,-1) arc (0:180:0.25 and 0.1);
\end{scope}
}

\newcommand\deltaC[2]{% Closed comultiplication
\begin{scope}[xshift=#1,yshift=#2]
\filldraw[left color=lightgray, right color=white] (-0.25,-1) to [out=90,in=180] (0,-0.66) to [in=90,out=0] (0.25,-1) to [out=-90,in=-90] (0.75,-1) to [in=-90,out=90] (0.25,0) to (-0.25,0) to [in=90,out=-90] (-0.75,-1) to [out=-90,in=-90] (-0.25,-1);
\filldraw[left color=white,right color=lightgray] (0,0) ellipse (0.25 and 0.1);
\draw[dotted] (-0.25,-1) arc (0:180:0.25 and 0.1);
\draw[dotted] (0.75,-1) arc (0:180:0.25 and 0.1);
\end{scope}
}

\newcommand\muA[2]{ % Open multiplication
\begin{scope}[xshift=#1,yshift=#2]
\draw (-0.75,0) -- (-0.25,0) to [out=-90,in=180] (0,-0.33) to [in=-90,out=0] (0.25,0) -- (0.75,0) to [in=90,out=-90] (0.25,-1);
\draw (-0.25,-1) -- (0.25,-1);
\draw (-0.75,0) to [in=90,out=-90] (-0.25,-1);
\end{scope}
}

\newcommand\pairA[2]{ % Open pairing
\begin{scope}[xshift=#1,yshift=#2]
\draw (-0.75,0) -- (-0.25,0) to [out=-90,in=180] (0,-0.33) to [in=-90,out=0] (0.25,0) -- (0.75,0) to [out=-90,in=0] (0,-0.83) to [out=180,in=-90] (-0.75,0);
\end{scope}
}

\newcommand\copairA[2]{ % Open copairing
\begin{scope}[xshift=#1,yshift=#2]
\draw (-0.75,0) -- (-0.25,0) to [out=90,in=180] (0,0.33) to [in=90,out=0] (0.25,0) -- (0.75,0) to [out=90,in=0] (0,0.83) to [out=180,in=90] (-0.75,0);
\end{scope}
}

\newcommand\deltaA[2]{ % Open comultiplication
\begin{scope}[xshift=#1,yshift=#2]
\draw (-0.75,-1) -- (-0.25,-1) to [out=90,in=180] (0,-0.66) to [in=90,out=0] (0.25,-1) -- (0.75,-1) to [in=-90,out=90] (0.25,0) -- (-0.25,0) to [in=90,out=-90] (-0.75,-1);
\end{scope}
}

\newcommand\zipper[2]{ % Zipper
\begin{scope}[xshift=#1,yshift=#2]
\draw (-0.25,-1) -- (0.25,-1);
\filldraw[right color=white,left color=lightgray] (-0.25,0) to (-0.25,-1) to [out=90,in=225] (0,-0.5) to [out=-45,in=90] (0.25,-1) to (0.25,0);
\filldraw[left color=white,right color=lightgray] (0,0) ellipse (0.25 and 0.1);
\end{scope}
}

\newcommand\cozipper[2]{ % Cozipper
\begin{scope}[xshift=#1,yshift=#2]
\draw (-0.25,0) -- (0.25,-0);
\filldraw[right color=white,left color=lightgray] (-0.25,-1) to (-0.25,0) to [out=-90,in=135] (0,-0.5) to [out=45,in=-90] (0.25,0) to (0.25,-1) to [in=-90,out=-90] (-0.25,-1);
\draw[dotted] (0.25,-1) arc (0:180:0.25 and 0.1);
\end{scope}
}

\newcommand\epsilonC[2]{ % Closed cap
\begin{scope}[xshift=#1,yshift=#2]
\filldraw[right color=white,left color=lightgray] (-0.25,0) to [out=-90,in=180] (0,-0.33) to [in=-90,out=0] (0.25,0);
\filldraw[left color=white,right color=lightgray] (0,0) ellipse (0.25 and 0.1);
\end{scope}
}

\newcommand\etaC[2] { % Closed cap
\begin{scope}[xshift=#1,yshift=#2]
\filldraw[right color=white,left color=lightgray] (-0.25,0) to [out=90,in=180] (0,0.33) to [in=90,out=0] (0.25,0) to [in=-90,out=-90] (-0.25,0);
\draw[dotted] (0.25,0) arc (0:180:0.25 and 0.1);
\end{scope}
}

\newcommand\epsilonA[2] {% Open cap
\begin{scope}[xshift=#1,yshift=#2]
\draw (-0.25,0) -- (0.25,0);
\draw (-0.25,0) to [out=-90,in=180] (0,-0.33) to [in=-90,out=0] (0.25,0);
\end{scope}
}

\newcommand\etaA[2] {% Open cap
\begin{scope}[xshift=#1,yshift=#2]
\draw (-0.25,0) -- (0.25,0);
\draw (-0.25,0) to [out=90,in=180] (0,0.33) to [in=90,out=0] (0.25,0);
\end{scope}
}

\begin{document}

\title{Entanglement entropy and edge modes in topological string theory II: The dual gauge theory story}
%\author[a]{William Donnelly,}
%\author[b]{Thomas Hartman,}
\author[a]{Yikun Jiang,}
\author[a]{Manki Kim,}% $^*$ 
\author[b]{and Gabriel Wong} %$^{**}$

%\affiliation[]{Perimeter Institute for Theoretical Physics, 31 Caroline St. N, N2L 2Y5, Waterloo ON, Canada}
\affiliation[a]{Department of Physics, Cornell University, Ithaca, New York, USA}
\affiliation[b]{Department of Physics, Fudan University, Shanghai, China}

%\emailAdd{wdonnelly@perimeterinstitute.ca}
%\emailAdd{hartman@cornell.edu}
\emailAdd{phys.yk.jiang@gmail.com}
\emailAdd{mk2427@cornell.edu}
\emailAdd{gabrielwon@gmail.com }

%\vspace*{6mm}
%\end{center}

%\end{spacing}

%\vskip 1cm
%\setcounter{tocdepth}{2}  
%\tableofcontents

%\begin{spacing}{1.3}
\abstract{This is the second in a two-part paper devoted to  studying entanglement entropy and edge modes in the A model topological string theory.  This theory enjoys a gauge-string (Gopakumar-Vafa) duality which is a topological analogue of AdS/CFT.  In part 1, we defined a notion of generalized entropy for the topological closed string theory on the resolved conifold. We provided a canonical interpretation of the generalized entropy in terms of the q-deformed entanglement entropy of the Hartle-Hawking state.  We found string edge modes transforming under a quantum group symmetry and interpreted them as entanglement branes.  In this work, we provide the dual Chern-Simons gauge theory description. Using Gopakumar-Vafa duality, we map the closed string theory Hartle-Hawking state to a Chern-Simons theory state containing a superposition of Wilson loops.  These Wilson loops are dual to closed string worldsheets that determine the partition function of the resolved conifold. We show that the \emph{undeformed} entanglement entropy due to cutting these Wilson loops reproduces the bulk generalized entropy and therefore captures the entanglement underlying the bulk spacetime. 
 Finally, we show that under the Gopakumar-Vafa duality, the bulk entanglement branes are mapped to a configuration of topological D-branes, and the non-local entanglement boundary condition in the bulk is mapped to a local boundary condition in the gauge theory dual. This suggests that the geometric transition underlying the gauge-string duality may also be responsible for the emergence of entanglement branes.
}  
\maketitle

\section{Introduction} 
\begin{figure}[h] 
\centering
\includegraphics[scale=.4]{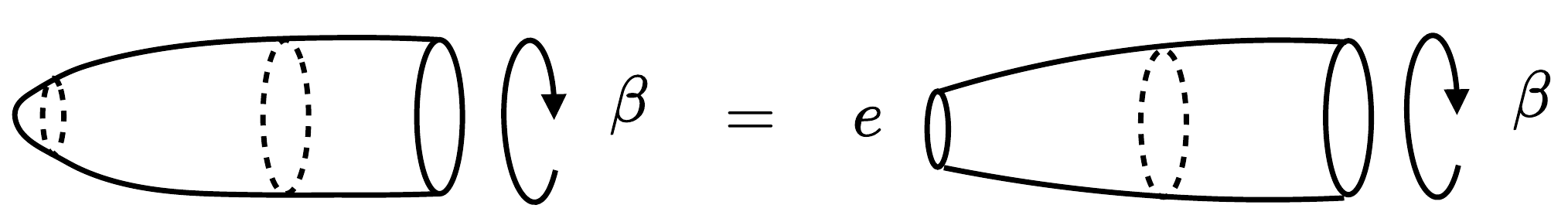}
\caption{The left figure shows the cigar geometry which is the saddle point that contributes the the area term in the  generalized entropy.  On the right we have removed a cap at the tip of the cigar and inserted a shrinkable boundary condition $e$. }\label{cigar}
\end{figure}

In the context of the AdS/CFT correspondence \cite{1999IJTP...38.1113M, 1998PhLB..428..105G, 1998AdTMP...2..253W}, the HRRT/generalized  entropy \cite{2006PhRvL..96r1602R, 2007JHEP...07..062H, 2013JHEP...11..074F, 2015JHEP...01..073E} formula provides the basis for our understanding of how spacetime emerges from quantum entanglement. It states that entanglement entropy of a boundary subregion in the strongly coupled regime is given by the generalized entropy of the semi-classical bulk theory: 
\begin{align} \label{QES}
   S_{\text{CFT}} =  S_{\text{gen}} = \frac{\langle{A}\rangle }{4G}+ S_{\text{bulk}} +\cdots
\end{align}
The generalized entropy $S_{\text{gen}}$ is defined via the  Euclidean gravity path integral $Z(\beta) $ on geometries for which the asymptotic boundary has a circle\footnote{This circle is non-contractible at asymptotic infinity but can shrink smoothly in the bulk.} of length $\beta$ \cite{2013JHEP...08..090L}  (see Fig. \ref{cigar}).   One then defines the generalized entropy as
\begin{align}\label{gen}
    S_{\text{gen}}&= (1-\beta \pd_{\beta})_{\beta=2\pi} \log Z(\beta) \\ 
    Z(\beta) &\sim  e^{-I_{\text{classical}}} Z_{\text{fluctuations}},
\end{align}
where $Z(\beta)$ is evaluated on a saddle point and $-I_{\text{classical}}$ is the on-shell action.
In order to interpret $S_{\text{gen}}$ in terms of a statistical mechanical entropy we must treat $Z(\beta)$
as a thermal partition function:
\begin{align} \label{trace}
    Z(\beta) = \tr e^{-\beta H }.
\end{align}
However, we do not have a general understanding of the bulk quantum gravity Hilbert space on which this trace would be defined.   As first discussed in \cite{PhysRevD.15.2752} and emphasized recently in \cite{2020arXiv201010539H},  the leading area term in \eqref{QES} only arises from saddles in which the circle shrinks smoothly in the interior, which gives the cigar geometry in the left of Fig. \ref{cigar}.  This represents an apparent obstruction to the trace interpretation \eqref{trace} from the viewpoint of effective field theory and obscures the bulk quantum mechanical origin of area term. This is an important puzzle to address because the area term,  which is the analogue of Bekenstein Hawking entropy, is expected to capture the entropy of the spacetime itself \cite{2010GReGr..42.2323V}.  

We can view equation \eqref{trace} as a constraint on the quantum gravity microstates, determined by the path integral that governs the low energy effective theory.  

\begin{figure}[h] 
\centering
\includegraphics[scale=.4]{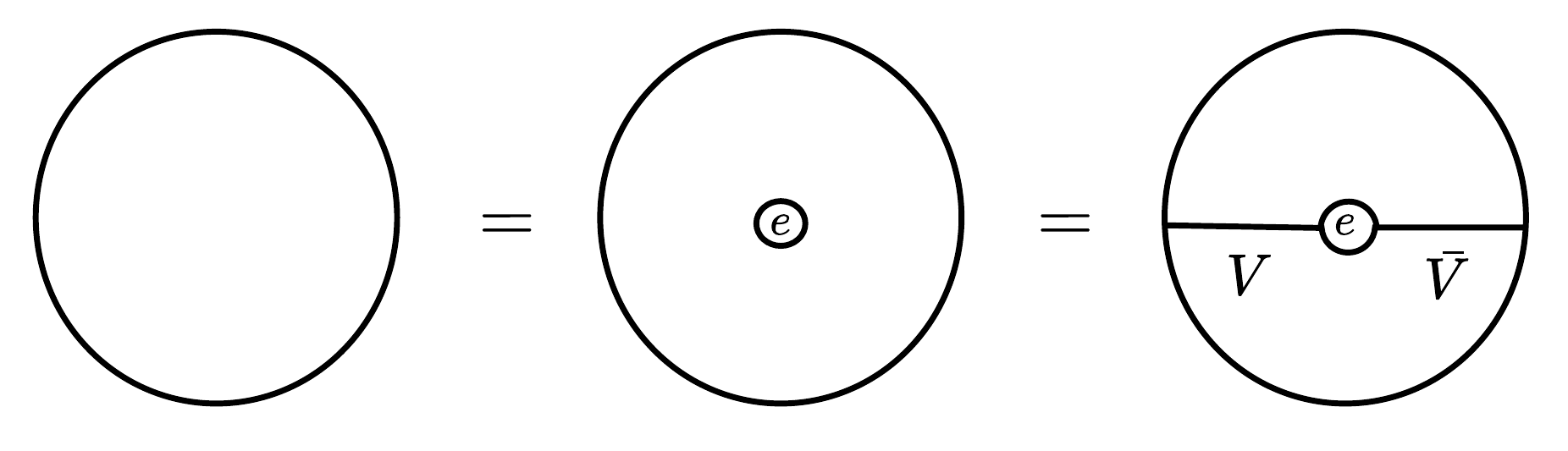}
\caption{ In this figure we have flattened out the cigar geometry into a disk.  On the right figure, we can view the lower half of the annulus as a path integral preparation of a factorized state with a shrinkable boundary condition at the entangling surface. Quantizing $Z(\beta)$ with respect to  the time variable around the origin shows that it can be viewed as the trace of a reduced density matrix on $V$.} \label{annulus}
\end{figure}

\begin{figure}
\centering 
\includegraphics[scale=1]{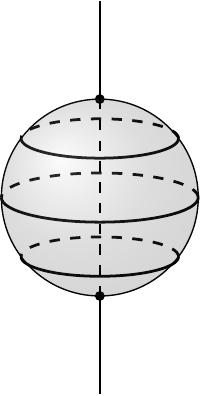}
\qquad \qquad
\includegraphics[scale=1]{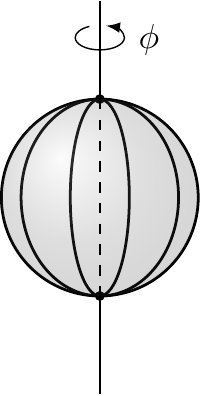}
\caption{ Susskind and Uglum considered the generalized entropy of perturbative closed strings in flat space, viewed as a limit of the cigar geometry.   Using off shell arguments, they computed generalized entropy by inserting a conical singularity in the background, corresponding to the tip of the cigar geometry.  In perturbative string theory, the area term comes from the sphere diagram which intersects the conical singularity.  Viewed in the open string channel, this is a one-loop open string diagram.  This interpretation amounts to an open-closed string duality which identifies Bekensten Hawking entropy as thermal entropy of open strings that end on the conical singularity.
Figure borrowed from Ref.~\cite{Donnelly:2016jet}.
} \label{fig:slicing}
\end{figure}

The idea is illustrated on the right of Fig. \ref{cigar}.  To interpret  $Z(\beta)$ on a cigar geometry as a trace we excise a small cap from the tip of the cigar and impose a ``shrinkable" boundary condition.   This boundary condition is defined so that the path integral on the excised geometry is the same as $Z(\beta)$.  It corresponds to inserting a boundary state given by the path integral on the small cap.  If the shrinkable boundary condition were local, we can immediately interpret $Z(\beta) $ as a thermal partition function by quantizing with respect to time variable around the circle.   Fig. \ref{annulus} suggests that the corresponding thermal density matrix can be viewed as the reduced density matrix $\rho_{V}$ on a subregion $V$ of a spatial slice\footnote{In gravitational path integral, there will be an extra complication due to the fact that we cannot fix the location of the `` stretched horizon" where we removed the small cap.  However we expect this construction remains valid provided that we sum over the location of the shrinkable boundary. }.   
If we could identify 
\begin{align} 
Z(\beta=2 \pi n) = \tr_{V} \rho_{V}^n,
\end{align}  then the generalized entropy \eqref{gen} would give the replica trick entanglement entropy of the subregion $V$.
Since string theory provides the UV completion of the bulk gravity theory in AdS/CFT, this suggest that the generalized entropy can be viewed as entanglement entropy of closed strings making up the spacetime. A worldsheet version of this proposal was first discussed by Susskind and Uglum \cite{Susskind:1994sm}.  As shown in Fig. \ref{fig:slicing}, from the worldsheet point of view, the quantization with respect to the modular time and the shrinkable boundary condition is equivalent to a form of open-closed string duality.

In \cite{Donnelly:2018ppr}, an explicit realization of these ideas was first obtained in two dimensional Yang Mills and its string theory dual, using the framework of  \emph{extended} topological quantum field theory (TQFT)\footnote{TQFT's have finite dimensional Hilbert spaces, where as area dependent QFT's such as 2D Yang Mills have infinite dimensional Hilbert spaces.  Nevertheless, they obey very similar sewing rules so we will use extended TQFT to refer to both types of theories in this paper.}.  Extended TQFT is a categorical reformulation of the path integral as a cobordism theory constrained by sewing relations.  In \cite{Donnelly:2018ppr}, the shrinkable boundary condition was interpreted as an additional sewing relation called the ``entanglement brane axiom."  In that theory, the shrinkable boundary condition is local and provides a constraint on the consistent factorization of the Hilbert space which requires the presence of edge modes localized to the entangling surface.  It was shown that the analogue of generalized entropy can indeed be interpreted as entanglement entropy of a subregion, and has a dominant edge mode contribution which plays the role of the area term.  In the string theory dual, these edge modes correspond to a large $N$ number of entanglement branes\footnote{Interestingly, the entanglement brane axiom requires the number of branes in that theory to be related to the closed string coupling as \cite{Donnelly:2016jet,Donnelly:2018ppr}: $N=\frac{1}{g_{s}}\,$. This is a direct example of how the entanglement brane axiom relates parameters of the low energy theory, i.e. the closed string coupling $g_{s}$, to high energy microstates given by the entanglement branes.}.

Unfortunately, in gravitational theories, the shrinkable boundary condition is  non-local due to a \emph{topological} feature of the gravity path integral on the small cap. As explain in \cite{Jafferis:2019wkd}, this is because the Gauss Bonnet theorem implies that reproducing the Einstein Hilbert action inside the cap requires a  non trivial  holonomy around the shrinkable boundary. This seems to create an additional obstacle to interpreting generalized entropy as a statistical entropy.   

\begin{figure}[h] 
\centering
\includegraphics[width=16cm,height=14cm,keepaspectratio]{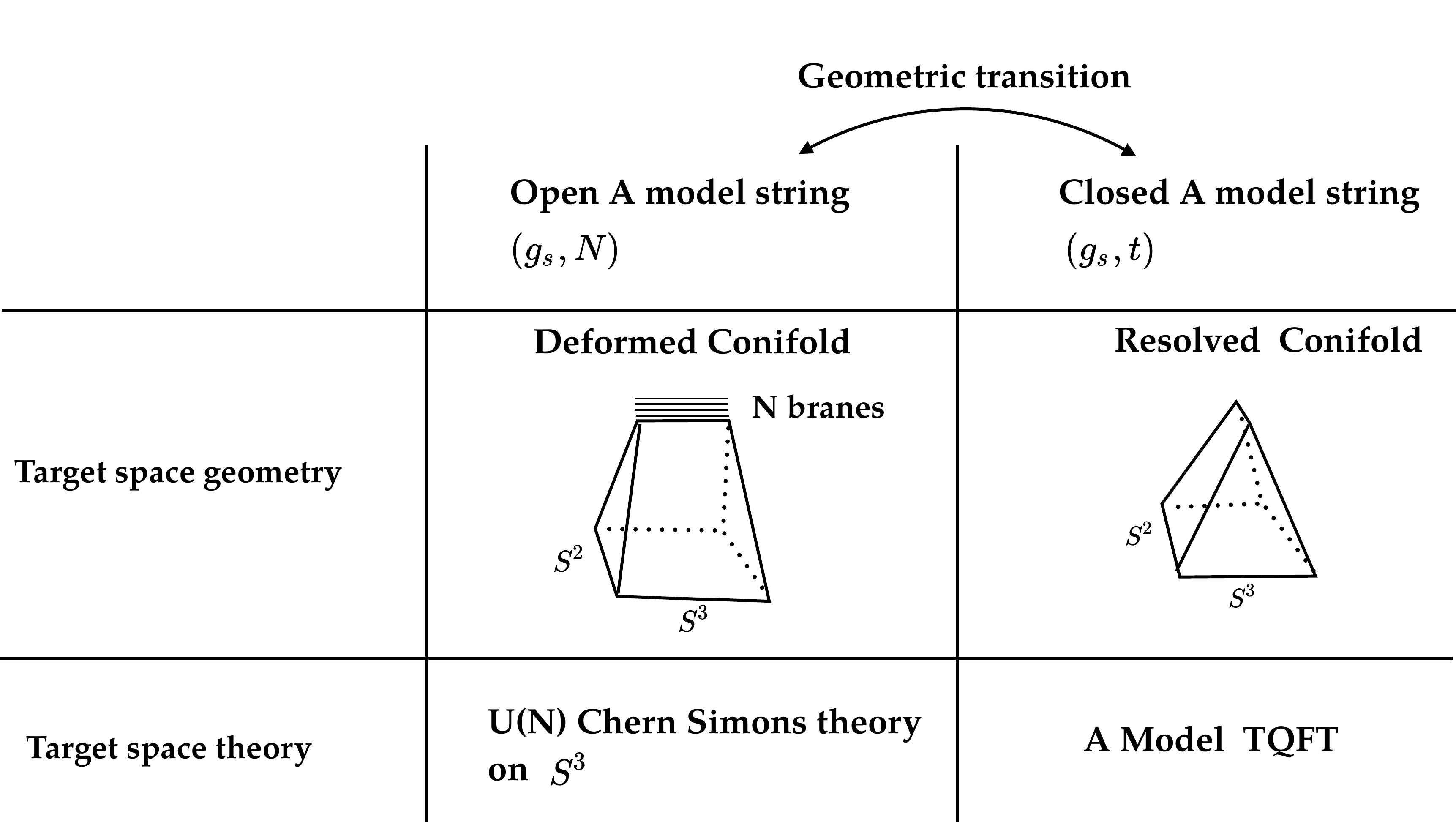}
\caption{Gopakumar-Vafa duality relates closed A-model string on the resolved conifold to the open A-model string on the deformed conifold }\label{chart}
\end{figure}

In \cite{2020arXiv201015737D}, we addressed these questions in the A model topological string theory using the extended TQFT framework developed in \cite{Donnelly:2018ppr}. We defined an analogue of  generalized entropy for closed strings on the resolved conifold geometry (see left of Fig. \ref{chart}) and showed that it has a canonical Hilbert space interpretation despite the presence of a non-local shrinkable boundary condition.  The analogue of the topological constraint in gravity is given by the Calabi-Yau condition\footnote{The A model string theory is well defined on any Kahler manifolds, so the Calabi Yau condition is a strong restriction.  The shrinkable boundary condition we obtained is specific to this sub-category of the target spaces for the topological string theory.}, which we imposed on the replica manifold so that the topology of the resolved conifold geometry is preserved as $\beta$ is varied in \eqref{gen}. The resulting boundary state corresponds to a ``Calabi-Yau" cap \cite{Bryan:2004iq, Aganagic:2004js}, and leads to string edge modes that obey anyonic statistics and transform under  the  quantum group  $U(\infty)_{q}$. As in \cite{Donnelly:2016jet, Donnelly:2018ppr}, these edge modes correspond a large $N$ number of entanglement branes which implements the entanglement cut on the closed strings.   Using a q-deformed version of the extended TQFT sewing relations,  we determined the factorization of the closed string Hilbert space and showed that the generalized entropy has a quantum mechanical description as a q-deformed entanglement entropy:
\begin{align} \label{q}
    S= -\tr_{q} \rho \log \rho = -\tr (D \rho \log \rho).
\end{align} Here $D$ is an operator called the Drinfeld element of  $U(\infty)_{q}$, whose insertion makes the quantum trace $\tr_{q}$ invariant under the quantum group symmetry. It can also be interpreted as a defect operator which creates the nontrivial bundle structure of the Calabi-Yau cap. The analogue of the area term in the generalized entropy is once again given by the edge mode contribution to the q-deformed entropy. Note that in  \cite{Jafferis:2019wkd}.  the same  formula \eqref{q} was obtained for the gravitational generalized entropy in JT gravity, with $D$ given by a defect operator which implements the topological constraint associated to the ``Einsten Hilbert"  cap.  

In this work we apply Gopakumar-Vafa (GV) duality \cite{Gopakumar:1998ki} to the A model topological string and give a dual calculation of the q-deformed entanglement entropy \eqref{q} from Chern-Simons gauge theory\footnote{Reference \cite{Hubeny:2019bje} also studied entanglement entropy in topological string theory using the dual Chern Simons gauge theory.  The idea of using the factorization map in Chern Simons theory to probe the entanglement structure in topological string theory via Gopakumar Vafa duality was originally suggested in \cite{Wong:2017pdm}.}    The GV duality is a topological analogue of AdS/CFT.   It is an open-closed string duality that relates bulk closed strings on the resolved conifold geometry to open strings on the deformed conifold geometry.  
This is illustrated in Fig. \ref{chart}.  Like AdS/CFT, the GV duality involves a geometric transition in which a large $N$ number of branes dissolve into fluxes.  On the deformed conifold, the branes wrap the Lagrangian submanifold $S^3$ at the tip and are replaced by flux passing through an $S^2$ on the resolved conifold across the geometric transition. 

The role of the boundary CFT is played by the large-$N$ limit of $U(N)$ Chern-Simons (CS) theory.  It is the worldvolume theory of the branes wrapping $S^3$ on the deformed conifold.  Remarkably, this is also the exact string field theory for open strings on the deformed conifold \cite{Witten:1992fb}.   The gauge coupling $g_{cs} =\frac{2 \pi }{k+N}$ and `t Hooft paramater $i g_{cs}N$ of the CS theory are related to the  closed string coupling $g_{s}$ and the Kahler modulus $t$ of the resolved conifold by
\begin{align}
    g_{s}&=g_{cs}= \frac{2 \pi }{k+N}\nn
    t&=ig_{s}N
\end{align}
Using the string field theory description, we can obtain the exact shrinkable boundary condition, edge modes, and entanglement entropy on both sides of the duality.  

As in AdS/CFT \cite{1998PhRvL..80.4859M, 2001EPJC...22..379R}, there is a local mapping between Wilson loops in the dual Chern-Simons gauge theory and worldsheets in the bulk closed string theory \cite{Ooguri:1999bv, Gomis:2006mv}. This is illustrated in Fig. \ref{fig:example1}, \ref{fig:example2}. The entanglement cut of the closed string worldsheets is therefore mapped to the entanglement cut of the Wilson loops.    We will reproduce the q-deformed entanglement entropy of the bulk closed string theory via a canonical calculation of the ``defect entropy'' \cite{2013arXiv1307.1132J, 2013PhRvD..88j6006J, Lewkowycz:2013laa} associated to Wilson loops.   The defect entropy is obtained from the \emph{undeformed} entanglement entropy by subtracting the entanglement entropy of the vacuum, thus capturing the entanglement due to the Wilson loops alone.      An analogous relation between the defect entropy of Wilson loops in the boundary gauge theory and bulk entropy of probe string worldsheets also holds in AdS/CFT \cite{Lewkowycz:2013laa}. However, our gauge theory calculation computes the entanglement entropy of a large superposition of Wilson loops.  These are dual to the worldsheets that determine the resolved conifold partition function, so our calculation captures the entropy that makes up the spacetime itself.

One important issue we will address using the GV duality is the nature of the entanglement branes, which were defined previously using the categorical language of extended TQFT.    While this provides a precise mathematical definition of a brane, its relation to the usual worldsheet definition as boundary conditions for the string  sigma model is rather obscure.    Here we will show  that the GV duality maps the entanglement branes to a configuration of D branes, which corresponds to Lagrangian boundary conditions for the topological  string.   In the string field theory description, the dual  brane configuration correspond to the CFT edge modes of the $U(N)$ Chern-Simons theory.  The shrinkable boundary condition is \emph{local} in the Chern-Simons theory, and the quantum group edge mode symmetry is replaced by the large $N$ Kac-Moody symmetry of the WZW model edge modes. This is a manifestation of the fact that quantum groups arise as a hidden symmetry in conformal field theories \cite{Witten:1988hf, Guadagnini:1989tj, Slingerland:2001aa}.  

As shown in Fig. \ref{fig:slicing}, the shrinkable boundary condition on the worldsheet implies a type of open-closed string duality.   This was manifest in the canonical calculation of  generalized entropy in \cite{2020arXiv201015737D},  in which a trace over the open string Hilbert space  (i.e. RHS of \eqref{trace})  reproduces \emph{closed} string amplitudes that determine the generalized entropy.  However the worldsheet mechanism behind this open-closed duality was not explained.   In this work we will find strong evidence that the open-closed duality responsible for the shrinkable boundary condition and the entanglement brane edge modes is related to the GV duality itself.  Remarkably the worldsheet mechanism behind GV duality is well understood and can be interpreted as a phase transition on the worldsheet corresponding to the condensation of vortices that represent the D branes \cite{Ooguri:2002gx}.  Our work seems to suggest that a similar worldsheet mechanism might be responsible for the emergence of entanglement branes.

Our paper is organized as follows. In section 2, we give an extended review of relevant results in our previous work on the generalized entropy of the A model closed string theory.  In particular we will review our construction of the closed string Hartle-Hawking state on the resolved conifold using the topological vertex formalism \cite{2005CMaPh.254..425A}.   We explain the construction of the string edge modes and the Drinfeld element, paying particular attention to the large $N$ limit  and regularization which is needed to define the shrinkable boundary condition. 
In section 3, we will give the dual Chern-Simons gauge theory calculation of the bulk generalized entropy, starting with a dual replica trick calculation.  The dual Hartle-Hawking state is given by a state on a torus containing a superposition of Wilson loops.  We explain the large $N$ limit which maps the entanglement edge modes of this state to the entanglement branes in string theory. 
In section 4, we will explain the duality between Wilson loops in Chern-Simons theory and worldsheets in topological string theory.   Moreover we will re-visit our discussion of Chern-Simons edge modes of the Hartle-Hawking state from the point of view of worldsheets on the deformed conifold.  We show that these edge mode correspond to a configuration of D branes that include dynamical branes on wrapping 3-spheres in the deformed geometry.   We will explain the precise sense in which GV duality relates these branes to the entanglement branes on the resolved conifold.

\section{Review of part 1}
\subsection{Generalized entropy for A model closed strings}
\paragraph{A model closed strings on the resolved conifold} 
The topological A model closed string theory is defined on target spaces which are six real dimensional Kahler manifolds \cite{Witten:1988xj}.   The perturbative string amplitudes can in principle be computed to all orders in the genus expansion and depend only on the Kahler modulus of the target space \cite{Witten:1992fb,Aganagic:2002qg,Aganagic:2003db,Klemm:1999gm,Dijkgraaf:2002fc,Aganagic:2003qj,Cota:2019cjx,Huang:2015sta}. The simplicity of this string theory is due to the localization of the worldsheet path integral to holomorphic instantons which wrap minimal-volume two cycles on the target space. 

In our previous work we considered the A model closed string theory on the resolved conifold geometry.  As depicted in the right of Fig. \ref{chart}, this geometry is obtained from a resolution of a cone over a $S^2 \times S^3 $ base.  The only minimal volume two-cycle is the $S^2$ at the tip whose (complexified) area determines the Kahler modulus, and the closed string instantons are arbitrary coverings of this sphere with winding number $n>0$. The exact resolved conifold partition function is given by\footnote{The free energy can also get contributions from constant maps, which are finite polynomials in t. For non-compact Calabi-Yau manifolds, they are ambiguous and not well-defined, and we naturally set them to zero \cite{Aganagic:2003db}. }
\begin{align}
    Z_{\text{res}}&=\exp\left( \sum_{n=1}^{\infty}  \frac{1}{ n (2\sin (\frac{ng_{s}}{2} ))^{2}} e^{-n t} \right)\nn
    &= \sum_{R}  (d_{q}(R))^{2} e^{-t l(R)}.
\end{align}
The first formula comes directly from the exponentiation of the free energy, corresponding to a sum over all connected string diagrams.  $e^{-nt}$ is the exponential of the worldsheet action for instantons wrapping the $S^2$  $n$ times, and the worldsheet genus has already been summed over \cite{Klemm:1999gm,Aganagic:2003db,Witten:1992fb,Gopakumar:1998ki,Dijkgraaf:2002fc,Aganagic:2003qj,Vafa:2004qa, 2008arXiv0809.3976M}.  The second formula can be obtained from the gluing of  topological vertices \cite{2005CMaPh.254..425A}, which are basic building blocks for A model amplitudes that satisfy gluing rules reminiscent of a cubic field theory in spacetime.  %Fig. \ref{tra} shows the representation of $Z_{\text{res}} $ in terms of topological vertices. \yikun{you mean toric diagrams?}

For the purposes of studying entanglement and edge modes, it will prove convenient to  work with the second formula. Here $R$ labels Young tableaux with an arbitrary number  of boxes denoted by  $l(R)$. The quantity $d_{q}(R)$ is the quantum dimensions of the symmetric group representation $R$. In term of the Young diagram, $d_q(R)$ is given by
\begin{align}\label{dqR} 
d_{q}(R) &= \prod_{\Box \in R}  \frac{i}{q^{ h(\Box)/2}-q^{- h(\Box)/2}}=\prod_{\Box  \in R} \frac{1}{2 \sin(\frac{h(\Box) g_s}{2})},
\end{align}
with $h(\Box)$ being the hook length.   
\paragraph{Generalized entropy in topological string theory}
We would like to define an analogue of generalized entropy for the A model by replicating the partiton function $Z_{\text{res}}$. The analogue of the cigar geometry is given by the minimal volume two-sphere where the string worldsheets wrap.  We will define the  replica manifold by making an opening angle of $\beta =2 \pi n$ around two antipodal points as shown in Fig. \ref{2spheres}.  %Naively, the corresponding replica entropy measures the entanglement between two subregions fibered over the eu 

In defining the replica manifold, it is important to note that global geometry of the resolved conifold is not a that of a direct product with a $S^2$ factor.  Instead it is a nontrivial rank 2 bundle over the sphere:
\begin{align}
     \mathcal{O}(-1) \oplus \mathcal{O}(-1) \rightarrow S^{2}
\end{align}
 Here $\mathcal{O}(-1)$ denotes the complex line bundle over the sphere with chern class $-1$. Moreover, the resolved conifold is a Calabi Yau manifold.  For rank 2 bundles of the form
 \begin{align}\label{vect}
     \mathcal{O}(k_{1}) \oplus \mathcal{O}(k_{2})  \rightarrow \mathcal{S}
\end{align}
 over a Riemann surface $\mathcal{S}$, the Calabi Yau condition translates into the relation 
 \begin{align}\label{CY}
     k_{1}+k_{2} = - \chi(\mathcal{S})
 \end{align}
 between the Chern classes and the Euler characteristics of $\mathcal{S}.$ 
 
 Since the bundle structure over the minimal $S^2$ is nontrivial, we need to specify what happens to the fiber directions when we replicate around the two antipodal points.  As discussed in \cite{2020arXiv201015737D}, a naive cyclic gluing of the resolved conifold replicas would lead to a vector bundle of the form 
 \begin{align}
    \mathcal{O}(-n) \oplus \mathcal{O}( -n ) \rightarrow S^{2}
 \end{align}
 While there is nothing apriori incorrect about this replica manifold, it does not provide a good candidate for the definition of  generalized entropy \cite{2013JHEP...08..090L}. This is because it violates the Calabi-Yau (CY) condition and changes the topology of the resolved conifold.   This implies that when analytically continuing to non-integer $n$, the replica partition function no longer has a geometric interpretation in terms of a target space where string worldsheets can propagate, since we can not define a bundle with non-integer Chern classes\footnote{Another reason for imposing the CY condition comes from mirror symmetry \cite{hori2003mirror, Witten:1991zz}.  In contrast to the A-model, the B-model is only well defined on Calabi-Yau manifolds. In order for the replica trick to  commute with mirror symmetry, the replica manifold must preserve the Calabi Yau condition. More comments related to the B-model are in the discussion section}.
 
However, if we impose the CY condition \eqref{CY} as a topological constraint on the replica manifold, we obtain a replica partition function which does have a geometric interpretation at all values of $n$, even when it is non integer.  This is because the Euler characteristic of the base sphere is invariant under replication, so the  CY condition forces the bundle structure to stay fixed as well.  This implies that the only effect of the replication is to rescale the area $t$ of the sphere\footnote{Note that the area $t$ is complex and includes the $B$ field flux, so we are replicating the flux as well. Also, when we increase the number of entangling points or consider other Riemann surfaces $\mathcal{S}$, the Euler characteristic of the base manifold will no longer be invariant under replication.  Nevertheless we can consistently impose the condition \eqref{CY} even though the geometric interpretation at non integer $n$ is obscured or may not exist. }.  Since the A model is only sensitive to the Kahler modulus given by this area, the replica partition function is simply given by rescaling $t$:
\begin{align}
    Z(n)_{\text{res}} = \sum_{R}  (d_{q}(R))^{2} e^{-n t l(R)}.
\end{align}
From this we can obtain the generalized entropy on the resolved conifold geometry by applying eq. \eqref{gen}
\begin{align}\label{qde}
    S_{\text{gen} } &= (1-n \pd_{n})_{n=1} \log Z_{\text{res}}(n)\nn
    &=\sum_R p(R)(-\ln{p(R)}+2 \ln d_{q} (R)),\quad
     p(R) = \frac{(d_{q}(R))^2 e^{- t l(R) }}{Z_{\text{res}}}.
\end{align}
This formula has exactly the same structure as the entanglement entropy of two dimensional non abelian gauge theory, with  $R$ playing the role of a representation label, $d_{q}(R)$ the associated dimension, and $p(R)$ a probablity factor.  As in 2DYM,  the $\sum_R 2 p(R) \log d_{q}(R)$ term plays the role of the area term. The factor of 2 counts the number of putative entangling surfaces given by branch points of the replica manifold. 
\subsection{The closed string Hilbert space, A model TQFT, and the Hartle-Hawking state}
As noted in the introduction, the A model string theory  has an exactly solvable string field theory, so we can apply the usual formulation of  entanglement entropy in terms of a second quantized theory of strings. For target space geometries which take the form of vector bundles like \eqref{vect}, the string field theory is given by a topological quantum field theory, which we will simply refer to as the A model TQFT \cite{Bryan:2004iq, Aganagic:2004js}.  We will use the TQFT formalism to define the closed string Hilbert space.

The A model TQFT is a map\footnote{The precise statement is that it is a functor from the category of 2-cobordisms with line bundles to the category of vector spaces} that assigns multi-string amplitudes to basic building blocks of spacetime that are represented as 2-cobordisms with line bundles.  These are target spaces of the form \eqref{vect} in which the Riemann surface $\mathcal{S}$ is viewed as Euclidean evolution from initial and final boundaries.  We represent such a cobordism by a decorated two-dimensional diagram (evolution from top to bottom) 
\begin{align}\label{cob}
   \mathtikz{\etaC{0}{0};
\node at (0,1/2){($k_{1},k_{2}$)}}, \quad  \mathtikz{\muC{0}{0};
\node at (0,1/2){($k_{1},k_{2})$}} ,\dots
\end{align}
In order to cut up the closed string amplitudes into these basic building blocks, we have to insert brane/anti branes at the in/out boundaries where the worldsheets can end.   The gluing of these cobordisms should then be viewed as the annihilation of these branes and anti branes.  

The D branes of the A model wrap three dimensional Lagrangian submanifolds.  Each diagram in eq.\eqref{cob} represent open string amplitudes consisting of worldsheets that end on these  Lagrangians,  which intersect $\mathcal{S}$ along its boundary circles.  The coupling of the worldsheet to the branes is given by multi-trace factors
\begin{align}\label{mtr}
     \prod_{i=1} &\tr (U^{i})^{k_{i}} ,\quad k_{j} >0, \nn
     \quad U &\in U(N)
\end{align}
where $U= P\exp \oint A$ is the holonomy on the brane, and $k_{j}$ labels the number of strings the wind $j$ times\footnote{$k_{j}>0$ reflects the fact that the A model is a chiral theory so the strings wind in a single direction, and around the boundary circles of $\mathcal{S}$.}.

In the large $N$ limit, we identify the multi-trace factors \eqref{mtr} with the winding basis of wavefunctions
\begin{align}\label{wind}
\braket{U|\vec{k}} =  \prod_{i=1} &\tr (U^{i})^{k_{i}}
\end{align}
that span a closed string Hilbert space $\mathcal{H}_{\text{closed}}$  assigned to each boundary of $\mathcal{S}$, with $U$ playing the role of a configuration space variable.  The Hilbert space is thus identified with class functions on $U(\infty)$. Note that when the holonomy $U$ is pulled back to the worldsheet, the wavefunctions \eqref{mtr} are \emph{functionals} of the string loops that make up the worldsheet boundary.

We will also make use of the  representation basis, related to \eqref{mtr} by the Frobenius relation
\begin{align} \label{frob}
    \braket{U|R}=\tr_{R}(U)&= \sum_{\vec{k} \subset S_{n}} \frac{\chi_{R}(\vec{k})}{z_{\vec{k}}}   \braket{U| \vec{k}},
\end{align}
where $R$ is a representation of $U(\infty)$ associated with Young diagram of $n$ boxes, and $\chi_{R}(\vec{k})$ is the symmetric group character for  $\vec{k}$, viewed as a conjugacy class in $S_{n}$.   In the large $N$ limit we can include states with an arbitrary number of boxes $n$.

Formally the A model TQFT assigns a tensor product of  $\mathcal{H}_{\text{closed}}$ to the disjoint union of in or out circles, and  linear maps to cobordisms that join these circles. 
The gluing of the cobordisms corresponds to the composition of linear maps, and is implemented by the haar integral on the closed string Hilbert space: 
\begin{align}
\int d U \, \tr_{R}(U) \tr_{R'}(U^{-1}) = \delta_{RR'}
\end{align}
\paragraph{Hartle-Hawking state }

Since our replica trick preserves the Calabi-Yau condition, we can also consistently restrict to this subset of vector bundles satisfying \eqref{CY}. The resulting TQFT forms a Frobenius Algebra \cite{Aganagic:2004js, Bryan:2004iq,2020arXiv201015737D}, and is generated by four basic cobordisms.
\begin{align}
 \mathtikz{\etaC{0}{0};
\node at (0,1/2){(0,-1)};} ,\quad 
 \mathtikz{\epsilonC{0}{0};
\node at (0,1/2){(-1,0)};} ,\quad 
\mathtikz{\muC{0}{0};
\node at (0,1/2){(0,1)};} 
,\quad  \mathtikz{\deltaC{0}{0};
\node at (0,1/2){(1,0)};} 
\end{align}
The resolved conifold is given by the overlap 
\begin{align}
        Z= \mathtikz{
\node at (0,1/2){(0,-1)};
\epsilonC{0}{0};
\etaC{0}{0};
\node at (0,-1/2){(-1,0)}}=\sum_{R}  (d_{q}(R))^{2} e^{-t l(R)}.
\end{align}

We define the Hartle-Hawking state to be the string amplitude on ``half" of the resolved conifold geometry:
\begin{align}\label{Hartle} 
 \ket{HH(t)} =  \mathtikz{ \etaC{0}{0};
\node at (0,1/2){(0,-1)}}  &= \sum_{R}(-i)^{l(R)} d_{q}(R) q^{\kappa_{R}/4} e^{-t l(R)} \ket{R} \nn
\kappa_{R}&=C_2(R)-N l(R)
\end{align}
where $C_2(R)$ is the eigenvalue of the quadratic Casimir operator in the representation $R$.
The other half of the resolve conifold geometry is given by the linear functional\footnote{ Note that the bra and ket states denote dual basis elements which are not related by a \emph{Hermitan} inner product. Instead they are related by an adjoint operation on the string amplitudes which maps branes to anti-branes \cite{2001hep.th....1218V, 2005CMaPh.254..425A}.}
\begin{align}
   \bra{HH^*(t)}=  \mathtikz{\epsilonC{0}{0};
\node at (0,1/2){(-1,0)}} =\sum_{R} i^{l(R)} d_{q}(R) q^{-\kappa_{R}/4} e^{-t l(R)} \bra{R} 
\end{align}
which is the string amplitude in the presence of anti branes on the Lagrangian that intersect $\mathcal{S}$.   

\begin{figure}[h]
\centering
\includegraphics[width=14cm,height=10cm,keepaspectratio]{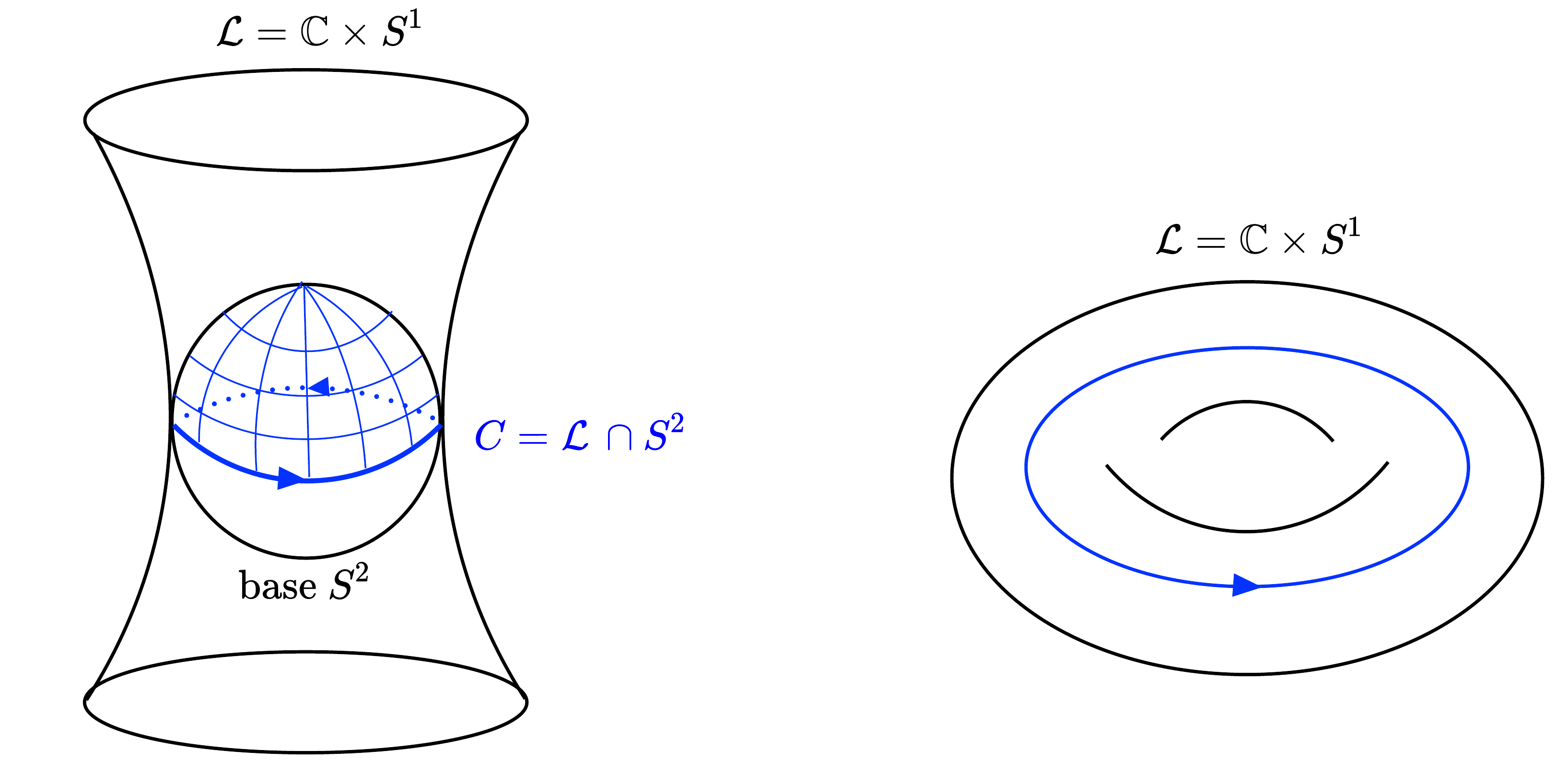}
\caption{The left figure shows worldsheet instantons ending on D-branes which cut the minimal volume $S^2$ of the resolved conifold along the equator.  The branes extend into the non compact fiber directions and wrap a Lagrangian submanifold with the topology $\mathbb{C} \times S^1$} \label{Sigma}  
\end{figure}

Fig. \ref{Sigma} shows the worldsheet instantons for the wavefunction  $\braket{U|HH}$ which end on branes that extend into the fiber directions as a hyperbola.  The topology of the corresponding Lagrangian submanifold is that of a non-compact solid torus $\mathbb{C}\times S^{1}$, and the winding basis \eqref{wind}  describe  string loops winding around the non contractible $S^{1}$.   The worldvolume theory on the branes is $U(\infty)$ Chern-Simons theory.

\subsection{Shrinkable boundary condition and the Calabi Yau Cap} 
\begin{figure}
\centering
\includegraphics[scale=.4]{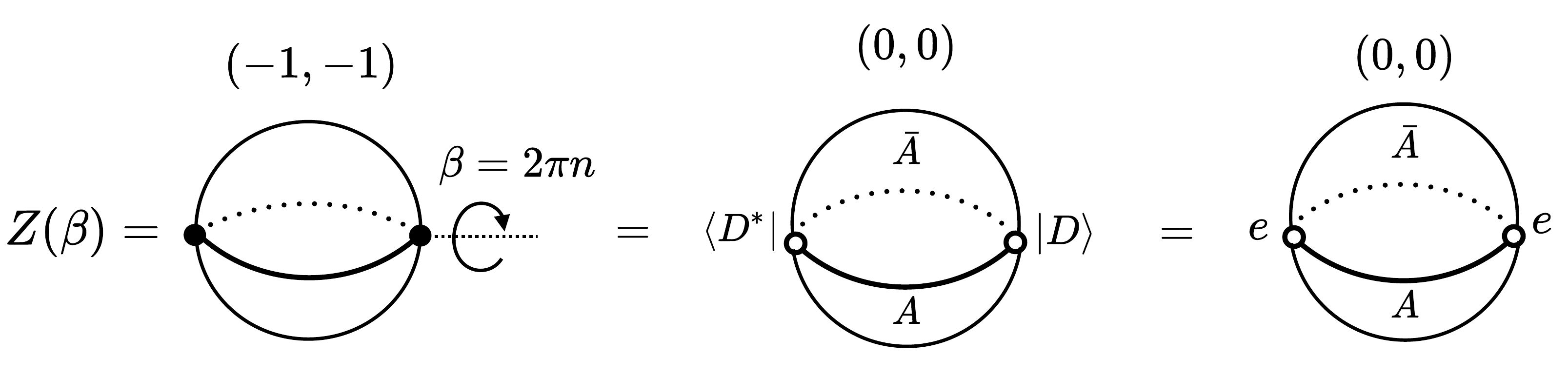}
\caption{The closed string channel description of entanglement branes on the resolved conifold geometry.   The total space is a six dimensional Calabi Yau manifold fibered over a sphere. We have shown only the base manifold and indicated the bundle structure with the Chern class labelling.} \label{2spheres}
\end{figure}

Having defined the closed string Hilbert space we can give a closed string channel description of the entanglement boundary state and shrinkable boundary condition.  Consider the partition function on the resolved conifold, viewed as a closed string amplitude between the entanglement boundary states:
\begin{align}
\label{DHD}
     Z_{\text{res}}(t)&= \braket{D^{*}|e^{-H_{\text{closed}}} |D} \nn
     H_{\text{closed}}&= t l(R) 
\end{align}
Here $ H_{\text{closed}}$ is the string field Hamiltonian. As shown in Fig. \ref{2spheres}, this corresponds to a decomposition of the base manifold into a cylinder and two small caps near the antipodal branch points associated with the entangling surfaces.  To satisfy the CY condition, the cylinder must have Chern class $(0,0)$, so the non trivial  topology is carried by the two ``Calabi Yau" caps \cite{Bryan:2004iq, Aganagic:2004js}.     These caps represent A model amplitudes with branes/anti branes and define a boundary states we call $\ket{D}$ and $\bra{D^*}$.    They are simply given by the states $\ket{HH}$ and $\bra{HH^*}$ with zero area $t=0$, and the corresponding wave functions are 
\begin{align}\label{BS}
    \braket{U|D}&= \sum_{R}(-i)^{l(R)} d_{q}(R) q^{\kappa_{R}/4}  \tr_{R}(U)\nn
    \braket{D^*|U} &= \sum_{R} i^{l(R)} d_{q}(R) q^{-\kappa_{R}/4}  \tr_{R}(U^{-1})
\end{align}
As discuss in the introduction, these boundary states determine the shrinkable boundary condition (see right of Fig. \ref{2spheres}).
The amplitudes of these wave functions capture the degeneracy factors in the partition function, which indicates the presence of q-deformed string edge modes.  

To understand this point it is useful to consider the analogous entanglement boundary state $\ket{\Omega}$,
obtained from the large $N$ limit of $U(N)$ 2DYM on a two dimensional cap \cite{Donnelly:2016jet}. This is a state in $\mathcal{H}_{\text{closed}}$  with wavefunction
\begin{align} \label{omega}
   \braket{U|\Omega} &=\sum_{R} \dim R \tr_{R}(U)\nn
    &= \delta(U,1)
\end{align}
In the second expression, we observed that the wavefunction for $\ket{\Omega}$ is a group theory Fourier transform of a delta function  which forces  $U=1$.  The triviality of this holonomy implies a local shirnkable boundary condition, corresponding to setting the gauge field component around the entangling surface to zero. 

In the large $N$ limit it was shown in \cite{Donnelly:2016jet} that the dimension factor $\dim R$ arises in the open string channel from edge modes transforming in the $R$ representation of $U(N)$, which were identified with the Chan-Paton factors labelling entanglement branes.

We can apply a similar analysis to the boundary state $\ket{D}$, keeping in mind that $U$ is now interpreted as a worldvolume holonomy on a three dimensional brane embedded inside a six dimensional Calabi-Yau manifold.  The wave function \eqref{BS} again gives a delta function on the group, but it now sets $U$ to a nontrivial group element
\begin{align} \label{dbs} 
\braket{U|D}&= \sum_{R}(-i)^{l(R)} d_{q}(R) q^{\kappa_{R}/4}  \tr_{R}(U)\nn
&=\delta(U,D)
\end{align} 
where $D$ is a diagonal $U(N)$  matrix of phases
\begin{align}
    D_{ij}&=\delta_{ij} q^{-i+\frac{1}{2}} \in U(N)\nn
    q&=e^{i g_{s}} 
\end{align}
The non-triviality of holonomy $D$ implies that there is no way to enforce it as a \emph{local} boundary condition on the worldvolume gauge field. Thus the shrinkable boundary condition for the A model is non-local, just as in Einstein-Hilbert gravity. 

If we analytically continue the string coupling $g_{s}$ to give it an imaginary part, we can identify the overlap $\braket{R|D}$  as a suitably regularized trace:
\begin{align} \label{traceD}
\braket{R|D}=\lim_{N\to \infty} \tr_{R}(D)= (-i)^{l(R)} d_{q}(R) q^{\kappa_{R}/4} 
\end{align} 
Note that in addition to the degeneracy factor $d_{q}(R)$, this overlap contains phases with nontrivial information about the fiber bundles structure in the transverse directions.

A similar formula holds for the coupling to the boundary state $\bra{D^*}$\footnote{The fact that this happens to be the complex conjugate of $\braket{R|D}$ is an accidental feature of the state when $t$ is real. We emphasize once again that we are not applying a Hermitian adjoint to obtain $\bra{D^*}$ from $\ket{D}$, but are instead using the branes to anti brane mapping defined in \cite{Aganagic:2003db} \cite{2001hep.th....1218V}. }
\begin{align}
    \braket{D^*|R}=\lim_{N\to \infty} \tr_{R}(D^{-1})=  (i)^{l(R)} d_{q}(R) q^{-\kappa_{R}/4}
\end{align}
We can thus write
\begin{align}\label{Ztr}
    Z_{\text{res}} &= \sum_{R} \braket{D^{*}|R} e^{-t l(R)} \braket{R|D}\nn
    &= \sum_{R} \tr_{R}(D)\tr_{R}(D^*) e^{-t l(R)} =\sum_{R} d_{q}(R)^2 e^{-t l(R)}
\end{align}
We see that the degeneracy factor $d_{q}(R)^2$ arises from the coupling of the closed string states $\ket{R}$ to the entanglement boundary state $\ket{D}$ \cite{Aganagic:2003db}.  Notice that the phases in \eqref{traceD} have cancelled out to give a real, positive degeneracy factor. 

Equation \eqref{Ztr} suggests that we can obtain a thermal interpretation of $Z_{\text{res}}$ if we associate a degenerate edge mode Hilbert space $V_{R} \times V_{\bar{R}}$ with the superselection label $R$, so that
\begin{align}
     Z_{\text{res}} = \sum_{R} \tr_{R}(D)\tr_{\bar{R}}(D) e^{-t l(R)} = \sum_{R} \tr_{R\times \bar{R} } (D e^{-H_{\text{open}}} ) \label{ampt}
\end{align}
where $H_{\text{open}}$ is an open string Hamiltonian with the same eigenvalues as $H_{\text{closed}} $, and $D$ should be viewed as a choice of measure on the degenerate open string microstates\footnote{We thank Laurent Freidel for suggesting this interpretation.}.  Notice that this measure is quite nontrivial\footnote{It is also a complex measure, as indicated in equation \eqref{traceD}}, since tracing over $U(N)$ indices of $D_{ij}$ gives \emph{symmetric} group dimensions. 
\begin{align}
    \lim_{N \to \infty} \tr_{R\otimes \bar{R}} (D) =d_{q}(R)^2
\end{align}
As we explain below, $D$ is the Drinfeld element of a quantum group and should be absorbed into the trace to define the \emph{quantum} trace.  This is a modification of the trace which makes it invariant under the adjoint action of the quantum group symmetry on the open string microstates.  

\subsection{Factorization and the q-deformed entanglement entropy}
In this section we identify the open string microstates whose thermal partition function is given by $Z_{\text{res}} $ and provide the statistical interpretation for its generalized entropy.   More precisely we determine a factorization map
\begin{align}\label{cfact0}
\mathtikz{ \zipper{0cm}{0cm}\deltaA{0cm}{-1cm} \draw (0cm,-1.9cm) node {\footnotesize $e$};\draw (0cm,-.8cm) node {\footnotesize $e$}} :    \mathcal{H}_{\text{closed}} \to   \mathcal{H}_{\text{open}}\otimes \mathcal{H}_{\text{open}
}
\end{align}
from the the closed string Hilbert space into an extended Hilbert space of open strings.  We apply the factorization map to the Hartle-Hawking state and obtain its reduced density matrix 
\begin{align}
    \rho= e^{- H_{\text{open}}}
\end{align}by 
doing a quantum partial trace over half of the closed string.
The corresponding q-deformed entanglement entropy 
\begin{align}
    S= - \tr (D \rho \log \rho) 
\end{align}
agrees with the generalized entropy \eqref{qde} and provides the statistical interpretation we were after.
\paragraph{Factorization as an extension of the TQFT} 
In terms of elementary cobordisms, equation \eqref{cfact0} is the composition of two elementary factorization maps
\begin{align} \label{fact}
\mathtikz{\zipper{0cm}{0cm}\draw (0cm,-0.8cm) node {\footnotesize $e$} } ,\quad \mathtikz{ \deltaA{0cm}{0cm}\draw (0cm,-0.9cm) node {\footnotesize $e$}} 
\end{align}
which describe the \emph{extension} of the closed TQFT into an open-closed TQFT that includes cobordisms with corners. These corners are the boundaries of the initial and final slice and carry labels which specify objects in the category of $D$ branes. The extended TQFT assigns an open string Hilbert space to such labelled intervals and linear maps to cobordisms that connect them.
These cobordisms satisfy sewing relations which provide local constraints on the factorization maps. 

In our setup, corners correspond to \emph{entanglement} branes that represent string  edge modes satisfying the shrinkability condition.  Denoting these branes by the label $e$, the shrinkablitiy condition is formulated as the sewing relation
    \begin{align} \label{EbraneCY}
     \mathtikz{\node at (0,.6){(0,-1)}; \etaC{0cm}{0cm} } = \mathtikz{  \etaA{0cm}{0cm} \cozipper{0cm}{0cm}
\draw (0cm,0.5cm) node {\footnotesize $e$};
},
\end{align}
called the entanglement brane axiom \cite{Donnelly:2018ppr}. 

When combined with the sewing relations of the open-closed TQFT, it can be shown that the entanglement brane axiom implies all holes labelled by $e$ can be closed.
\begin{equation}\label{holes}
\mathtikz{ \cozipper{0cm}{0cm} \zipper{0cm}{1cm} \draw (0cm,0.2cm) node {\footnotesize $e$}} 
= \mathtikz{ \idC{0cm}{0cm} },
\qquad
\mathtikz{ \muA{0cm}{0cm} \deltaA{0cm}{1cm}\draw (0cm,0cm) node {\footnotesize $e$} } = \mathtikz{ \idA{0cm}{0cm} } ,\quad \mathtikz{\node at (0,.6){(0,-1)}; \node at (0,-.6){(-1,0)};\epsilonC{0}{0}
\etaC{0}{0}}  = \mathtikz{\pairA{0}{0}\copairA{0}{0}\draw (0cm,0cm) node {\footnotesize $e$} ; \draw (0cm,1cm) node {\footnotesize $e$}} , \cdots
\end{equation}
The first two relation implies that the factorization maps do not change the state. The third is the cobordism description of eq.\eqref{ampt} which identifies the resolved conifold partition function as a thermal partition function. 

In fact, this partition function is precisely the \emph{categorical} trace of the un-normalized reduced density matrix for the Hartle-Hawking state \eqref{Hartle}, obtained by tracing out half of the closed string.   To see how this works in the cobordism language, we first factorize $\ket{HH}$ using \eqref{cfact}:
\begin{align}\label{HHF}
    \ket{HH} &\to \mathtikz{ \node at (0,1.5){(0,-1)};\deltaA{0cm}{0cm} \zipper{0cm}{1cm} \etaC{0cm}{1cm} }
= \mathtikz{ \deltaA{0cm}{0cm} \etaA{0cm}{0cm} } 
= \mathtikz{ \copairA{0cm}{0cm} } \nn
    \bra{HH^*} &\to \mathtikz{\node at (0,-2.7){(-1,0)}; \epsilonC{0cm}{-2cm} \muA{0cm}{0cm} \cozipper{0cm}{-1cm} }
= \mathtikz{  \epsilonA{0cm}{-1cm} \muA{0cm}{0cm} } 
= \mathtikz{ \pairA{0cm}{0cm} }
\end{align}

\begin{equation}
    \tilde{ \rho }=  \ket{HH}\bra{HH^*}= \mathtikz{ \node at (0,.7){(-1,0)}; \node at (0,-.75){(0,-1)}; \epsilonC{0cm}{.4cm}\etaC{0cm}{-.4cm}} \to \mathtikz{\pairA{0cm}{1cm} \copairA{0cm}{-1cm}  } .
\end{equation}

The corresponding reduced density matrix is given by the categorical partial trace 
\begin{align} \label{red}
    \tilde{\rho}_{A} =\tr_{B}\tilde{\rho}  = \mathtikz{\pairA{0cm}{1cm} \copairA{1cm}{1cm} \copairA{0cm}{-1cm} \pairA{1cm}{-1cm} \idA{1.5cm}{0cm}\idA{1.5cm}{1cm} } =\mathtikz{\idA{0cm}{0cm}\idA{0cm}{1cm}}=e^{-H_{\text{open}} }.
\end{align}
This partial trace operation is defined by the half annulli which turns input into output intervals and glues them together.  Such a trace operation can be defined abstractly from the cobordism theory; we will see that for the A model, it coincides with the \emph{quantum} partial trace.  In a purely topological theory where the Hamiltonian is strictly zero, the resulting strip in \eqref{red} is a trivial evolution operator which is equal to the identity operator.   However, the A model has a non-trivial Hamiltonian due to its dependence on the Kahler modulus of the target space, so the strip is an open string propagator which that depends on the complexified area and modular energies.

Applying the categorical partial trace on the remaining subregion $A$ gives the  quantum trace of the reduced denstiy matrix.
\begin{align}\label{qtrace} 
\tr_{q,A} ( \tilde{\rho}_{A}) =\mathtikz{\epsilonC{0}{0}
\etaC{0}{0}\draw (0cm,0.6cm) node {(0,-1)};\draw (0cm,-0.6cm) node {(-1,0)};}  = \mathtikz{\pairA{0}{0}\copairA{0}{0}\draw (0cm,0cm) node {\footnotesize $e$};\draw (0cm,1cm) node {\footnotesize $e$};},
 \end{align}
 where we have applied the entanglement brane axiom in the last line.  
\paragraph{The Open string Hilbert space and the factorization map} 
In \cite{2020arXiv201015737D}, we defined the open string Hilbert space $\mathcal{H}_{\text{open}} $ in terms of a noncommutative algebra of functions on the quantum group $U(\infty)_{q}$.  This Hilbert space is spanned by a basis of open string wavefunctions given by representation matrix elements of $U(\infty)_{q}$:
\begin{align}
    \mathcal{H}_{\text{open}}&=  \lim_{N \to \infty}\text{span} \{ R_{ij}(U)= \braket{U|Rij},\, U \in U(N)_{q},\, \, i,j =1,\cdots \dim R  \}\nn
    &=\otimes_{R} V_{R} \otimes V_{\bar{R} }
\end{align}

The matrix indices  of $R_{ij}$ correspond to edge mode degrees of freedom which transform under the quantum group as a representation space $V_{R} \otimes V_{\bar{R}}$. Given this definition of $\mathcal{H}_{\text{open}} $, the embedding \eqref{cfact0} of the closed string Hilbert space into the extended Hilbert space of open strings is given by

\begin{align} \label{cfact}
\mathtikz{ \zipper{0cm}{0cm}\deltaA{0cm}{-1cm} \draw (0cm,-1.9cm) node {\footnotesize $e$};\draw (0cm,-0.8cm) node {\footnotesize $e$};} : 
\ket{R} \rightarrow  \sum_{ijk} (D^{-1})^{R}_{ij} \ket{R jk}\ket{R k i }
\end{align}
We showed previously that this satisfies the entanglement brane axiom.  Here $D_{R}$ is  the quantum group representation of the Drinfeld element $D$, which we will explain below.  

\begin{figure}[h]
\centering
\includegraphics[scale=.4]{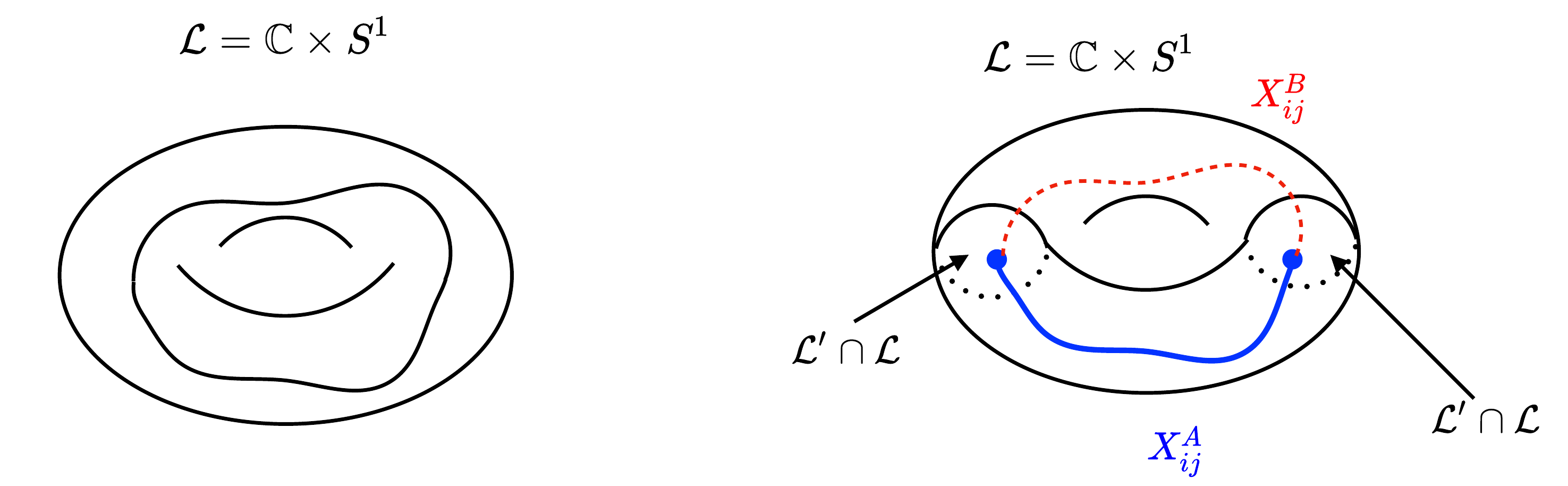}
\caption{ On the left, we show a closed string loop $X(\sigma)$ inside the non- compact Lagrangian manifold $\mathcal{L}= \mathbb{C} \times S^{1} $ where we put probe D branes.  On the right $\mathcal{L}$ is split into subregions by the entanglement branes on $\mathcal{L}'$ .  The factorization map embeds the closed string $X(\sigma)$ into open string configuration $X_{ij}^{A}(\sigma),X_{ij}^{B}(\sigma)  $ which are glued together along the entanglement branes} \label{stringsplit}
\end{figure}

We can understand the mapping \eqref{cfact} intuitively as follows.    Fig. \ref{stringsplit} shows the closed string loops in the Lagrangian submanifold $\mathcal{L}$ where the closed string states are defined as a function of the worldvolume holonomy $U$.  To cut the closed string loops into open strings, we introduce a large $N$ number of entanglement branes which intersect $\mathcal{L}$ as shown along two open disks. This introduces new sectors of open strings inside complementary subregions of $\mathcal{L}$ that end on the entanglement branes.  We denote these open strings configurations by 
\begin{align}
    X^{A}_{ij},X^{\bar{A}}_{ij}, \quad i,j =1,\dots N \gg 1
\end{align} where $i,j$ labels the branes and define corresponding Wilson lines
\begin{align}
    U^{A}_{ij}&= P \left(\exp \oint X_{ij} ^{A*}A \right)\nn
     U^{\bar{A}}_{ij}&= P \left(\exp \oint X_{ij} ^{\bar{A}*}A \right)
\end{align}
The factorization of the closed string states would naively follow by splitting the configuration space holonmoy $U$ into the subregion Wilson lines 
\begin{align}
   U= U^{A} U^{\bar{A}}
\end{align}
This would give the factorization map
\begin{align} \label{Ufact}
\tr_{R}(U) &\to \tr_{R}(U^{A} U^{\bar{A}}) = \sum_{ij} R( U^{A})_{ij}R( U^{\bar{A}})_{ji}\nn
\ket{R} &\to \sum_{ij} \ket{Rij}\ket{Rji} 
\end{align} 
where $R_{ij}(U^{A})$ is a representation matrix element, viewed as a wavefunction in a subregion Hilbert space. This factorization map preserves a diagonal part of the $U(N)\times U(N)$ edge mode symmetry which acts on the subregion wavefunctions by conjugation
\begin{align}\label{edge}
    U^{A} \to g U^{A} g^{-1}\nn
    U^{\bar{A}} \to g U^{\bar{A}} g^{-1}
\end{align}
Unfortunately, this fails to satisfy the entanglement brane axiom for the A model.  This is because the factorization map gives a reduced density matrix in the $R$
sector of the form
\begin{align}
\rho_{R} = \sum_{i,j} \ket{Rij}\bra{Rij}
\end{align} 
which has a $(\dim R)^{2}$ degeneracy due to the $U(N)$ edge mode symmetry. We can view this as a choice of measure on the edge mode Hilbert space which is compatible with the degeneracy factors of 2DYM \eqref{omega}, but incompatible with the q-deformed symmetric group dimensions \eqref{traceD} of the A model.

The symmetry viewpoint naturally suggests a modification of \eqref{Ufact}, which does give the correct factorization.   The presence of the q-deformed dimension factors implies that the edge mode symmetry is also q-deformed.  Thus we should treat the holonomies  $U,~ U^{A},~ U^{\bar{A}}$ as elements of the  quantum group $U(N)_{q}$.   The trace function which is invariant under the adjoint action of the quantum group is given by the \emph{quantum} trace:
\begin{align}
    \tr_{q,R}(U)&=\tr_{R}(uU)\nn
    u&= \delta_{ij} q^{-i + (N+1)/2} 
\end{align}
where $u$ is the Drinfeld element of $U(N)_{q}$. This is an object defined purely from quantum group data and reproduces the quantum dimension of a $U(N)_{q}$ rep:
\begin{align}
 \tr_{q,R}(1)=\tr_{R}(u)=\dim_{q}(R)
\end{align}
This defines a q-deformed measure on the edge mode Hilbert space.  To get the precise edge mode measure for the A model partition function we need to define a large $N$ limit of $u$ which captures symmetry group quantum dimensions and the phases in \eqref{traceD}.
\paragraph{The most important formula of this paper}
The key to finding this limit  is to observe that the shrinkable holonomy $D$ is a renormalized version of the $U(N)_{q}$ Drinfeld element.  According to equation \eqref{dbs}, it gives the correct edge mode measure for the A model provided we use a suitable regularization of the trace as $N \to \infty$. Explicitly we have
\begin{align} \label{Du}
    D&= q^{-N/2} u\nn 
    \lim_{N \to \infty} \tr_{R}(D) &= (-i)^{l(R)} d_{q}(R) q^{\kappa_{R}/4} 
\end{align}
Moreover, the proportionality of $D$ and $u$ implies that the quantum trace defined with $D$ or $D^{-1}$ is also invariant under the conjugation action of the quantum group $U(\infty)_{q}$.  

Equation \eqref{Du} can also be interpreted as a particular large $N$ limit \eqref{limit} of the $U(N)$ quantum dimension $\dim_{q}R$, which was previously applied to derive the topological vertex from Chern-Simons link invariants \cite{Aganagic:2002qg, Aganagic:2003db}.   In section 3, we discuss this limit from the point of view of the Chern-Simons dual, and in section 4 we give a string theory interpretation in terms of the geometric transition.  The upshot is that equation \eqref{Du} is a form of open-closed string duality that intimately related to GV duality itself.

The above discussion suggests the correct factorization map can be obtained by promoting the closed string wavefunctions to \emph{quantum} characters\cite{de_Haro_2007, 2020arXiv201015737D}
\begin{align}
    \tr_{R}(U) \to \tr_{R}(DU)
\end{align}
and then applying the splitting $U=U^{A}U^{\bar{A}}$
\begin{align}
 \tr_{R}(DU)  \to \tr_{R}(DU^{A} U^{\bar{A}} ) =\sum_{ijk} D^{R}_{ij} R_{ jk}(U^{A})R_{ k i }(U^{\bar{A}}), 
\end{align}
The open string wavefunctions $R_{ jk}(U^{A})$ now transform under a the $U(\infty)_{q}$ version of the edge mode symmetry  \eqref{edge}.   
It was shown in \cite{2020arXiv201015737D} that this factorization map satisfies the entanglement brane axiom and sewing relations of a properly q-deformed extended TQFT.    By applying this factorization map to the Hartle-Hawking state, we can compute the reduced density matrix on the open string Hilbert space and compute its entanglement entropy.   In the corbodism computation, the partial trace operations on each subregion as defined by the half annulus diagrams are automatically quantum traces which preserve the  edge mode symmetry.  As a result the entanglement entropy is q-deformed \cite{Couvreur_2017, Quella:2020aa}: 
\begin{align}
    S= -\tr_{q} (\rho \log \rho) = - \tr (D \rho \log \rho ) 
\end{align}
An explicit computation shows that the q-deformed entropy matches precisely with the generalized entropy \eqref{qde}, with the leading ``area term" arising from the entropy of edge modes. 

\subsection{Quantum group symmetry, defect operator and non-local boundary conditions}
Since the quantum group $U(N)_{q}$ is the symmetry of anyons, its presence implies that the string edge modes are anyons with  nontrivial braiding.  This can be understood via the large $N$ duality with Chern-Simons gauge theory,  since the string worldsheets are mapped to Wilson lines representing worldlines of anyons.  We will present this mapping in section 4.

Here we give an heuristic explanation in the bulk closed string theory for why quantum group symmetry emerges from the non local shrinkable boundary condition.   We noted earlier that the boundary state $\ket{D}$ defined by the Calabi Yau cap produces the shrinkable boundary condition which sets the worldvolume holonomy to 
$U=D$ along the stretched entangling surface.  

\begin{figure}[h]
\centering
\includegraphics[scale=.4]{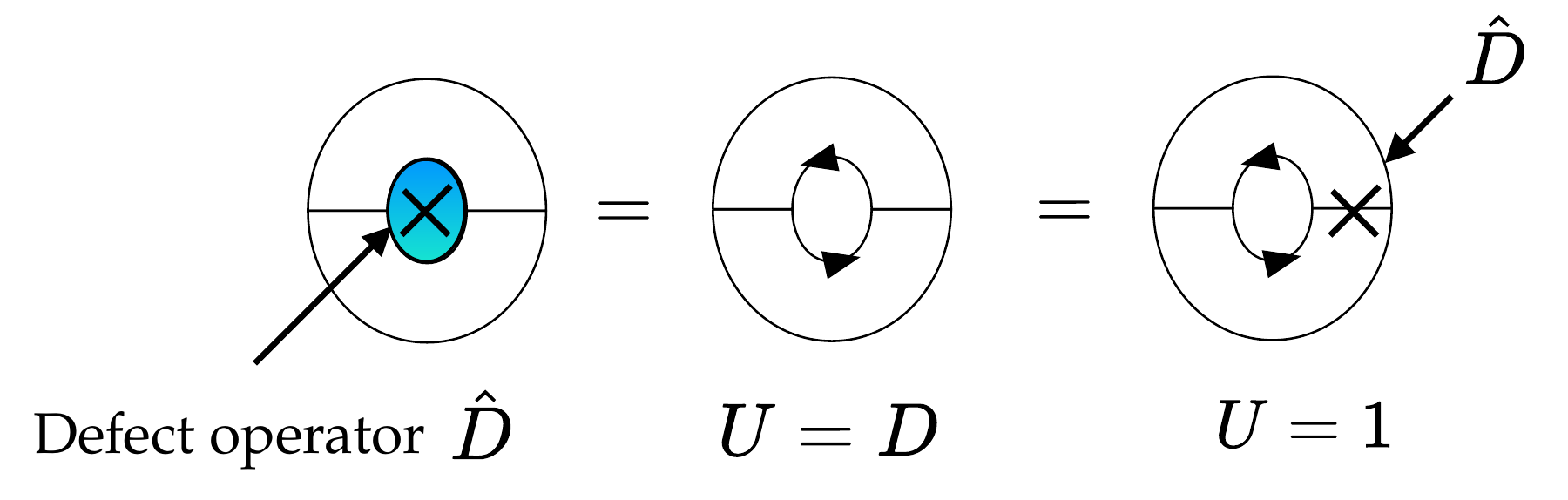}
\caption{On the left,the blue disk corresponds to a $(0,0)$ cap with trivial bundle structure.  To obtain the Calabi Yau cap, a defect operator is inserted to implement the nontrivial topology.  In the first equality, we have integrated over the cap to obtain a non local shrinkable boundary condition $U=D$.  An equivalent A model amplitude on the disk can be obtained by moving the defect operator outside the hole and putting a local boundary condition on its boundary.  } \label{defecthole}
\end{figure}

We could define a new holonomy basis 
\begin{align}
    \ket{U} \to \ket{U}'=\sum_{R} \tr_{R}(DU) \ket{R} 
\end{align}
so that
\begin{align}
    \ket{D}= \ket{U=1}'
\end{align}
In terms of this new holonomy variable for the configuration space, it would seem that the boundary condition is local as in the 2DYM example.  However the new wavefunction $\tr_{R}(DU) $ is no longer invariant under 
\begin{align}
    U \to g U g^{-1} , \quad g \in U(N)
\end{align}
because $D$ is not in the center of $U(N)$ so it doesn't commute with a general group element $g$.  However, the new wavefunction  $\tr_{R}(DU) $ is invariant under the adjoint action of \emph{quantum group} elements $g \in U(N)_{q}$.  Thus by insisting on quantizing in the open string channel with a local boundary condition, we see the emergence of a q-deformed edge mode symmetry ! 
\paragraph{Defect operator and the Calabi Yau cap}
As noted earlier, the Drinfeld element $D$ also has an interpretation as a defect operator which is associated with the nontrivial topology of the Calabi Yau cap.  Naively, the operator associated with the boundary state $\ket{D}$ for this cap is just the identity.  However as in the discussion above, if we compare the Calabi Yau cap with $(0,-1)$ Chern classes to a trivial cap with $(0,0)$ Chern classes, we find that the difference can by accounted for by the insertion of a defect operator (see figure \ref{defecthole}) .  This operator creates poles in the local sections of the bundles which leads to a nontrivial Chern class, and was shown to be equivalent to insertions of the Drinfeld element of the quantum group in the trace over the open string Hilbert space\cite{2020arXiv201015737D}. We thus have the equivalences
\begin{align}
\text{Defect operator}  \leftrightarrow \text{Non-local shrinkable boundary condition}
 \leftrightarrow \text{Quantum group symmetry}
\end{align} 

\section{Chern-Simons dual of the Hartle-Hawking state and the entanglement entropy}
\label{section:CS} 
In the previous section, we defined a factorization of the closed string Hilbert space $\mathcal{H}_{\text{closed}}$ in terms of an extension of the A-model closed TQFT.  We formulated the closed TQFT in terms of the representation category of quantum groups, and derived an extension compatible with the quantum group symmetry as well as the E-brane axiom.  This naturally led to a q-deformed notion of entanglement entropy consistent with the presence of edge modes transforming under the quantum group $U(\infty)_{q}$.  In this section, we provide additional evidence supporting this definition of closed string entanglement entropy by appealing to the dual Chern-Simons theory.

As shown in Fig. \eqref{chart}, the dual Chern-Simons description can be obtained by first applying a geometric transition which maps closed strings on the resolved conifold to open strings on the deformed conifold $T^{*}S^{3}$, with a large $N$ number of branes wrapping $S^{3}$ \cite{Gopakumar:1998ki}.   The string field theory for these open strings is then given by  $U(N)$ Chern-Simons theory on $S^3$.   The equivalence between Chern-Simons theory on $S^3$ and open topological string theory on $T^{*}S^{3}$ holds even at finite $N$ and can be understood as follows \cite{Witten:1992fb}. 
The A model open string  theory contains only zero mode degrees of freedom and localizes to holomorphic instantons that must wrap a minimal volume two dimensional manifold with boundaries ending on $S^3$.  In the deformed conifold geometry, the only such minimal volume manifolds are points on $S^3$.  These point-like worldsheets are degenerate instantons and act like particles charged under $U(N)$. We thus expect that this theory should be a topological field theory on the $S^{3}$, which can be shown to be Chern-Simons gauge theory.   From the point of view of the original closed string theory on the resolved conifold, the Chern-Simons gauge theory can be thought of as living on the $S^3$ at infinity. \cite{2002math......1219T, Ooguri:1999bv, Gomis:2006mv} This is reminiscent of AdS/CFT.

The relation between Chern-Simons theory and the gravitational dual closed string theory extends to more general geometries with multiple $S^2$ resolutions and to backgrounds with branes wrapped on Lagrangian manifolds \cite{Ooguri:1999bv, Gomis:2006mv}.  The duality provides a local mapping between branes on the resolved conifold and Wilson loops in Chern-Simons theory. It is essential for obtaining the gauge theory dual of the Hartle-Hawking state and  relates the entanglement cuts on the two sides.  

We will give a full derivation of this duality in section 4.  In this section, our immediate goal is to present the Chern-Simons dual of Hartle-Hawking state \eqref{Hartle} and compute its standard and undeformed entanglement entropy using the extended Hilbert space factorization into left and right moving WZW model edge modes \cite{Wen:2016snr, Wong:2017pdm, Das:2015oha}.
We will find a precise matching between the vacuum subtracted defect entropy \cite{2013arXiv1307.1132J, 2013PhRvD..88j6006J, Lewkowycz:2013laa} in Chern-Simons theory and the $q$-deformed entanglement entropy we calculated above in the dual string theory. We also present a Chern-Simons dual to the replica trick calculation of generalized entropy \cite {Dong:2008ft, Fliss:2020cos} and explain the construction of the generating functional for Wilson loops which plays an essential role in the duality map on the branes.

\subsection{Review of Chern-Simons theory} 
We begin by summarizing some important well-known results from Chern-Simons theory. More details can be found for example in  \cite{Witten:1988hf, Elitzur:1989nr, Marino:2004uf}. 
Consider Chern-Simons theory with gauge group $G$ on a manifold $M$ with a boundary $\Sigma=\pd M$. The action is given by
\begin{equation}
S_{CS,M}(A)=\frac{ik}{4\pi} \int_M \text{Tr}\left(A\wedge dA+\frac{2}{3}A\wedge A\wedge A \right).
\end{equation}
where the integer $k$ is the level determining the central extension of the lie algebra.
\paragraph{The Chern-Simons Hilbert space on a Torus}
The path integral on $M$ defines a state in the Hilbert space $\mathcal{H}(\Sigma)$ on $\Sigma$. This is given by the wave functional 
\begin{equation}
    \Psi_M(A_{\Sigma})=\langle A_{\Sigma}|\Psi\rangle=\int_{A|_{\Sigma}=A_{\Sigma}} \mathcal{D}A e^{iS},\label{eqn:CS wave fun}
\end{equation} where $A_{\Sigma}$ is the boundary value of the gauge field.

In \cite{Witten:1988hf}, it was shown that $\mathcal{H}(\Sigma)$ is isomorphic to the space of conformal blocks of a WZW model on $\Sigma$ with gauge group $G.$ In particular, $\mathcal{H}(S^2)$ is one dimensional and $\mathcal{H}(T^2)$ is spanned by irreducible representations of the affine Kac-Moody algebra. We will focus on the torus since the Chern-Simons dual of the Hartle-Hawking state $\ket{HH}$ belongs to $\mathcal{H}(T^2)$.

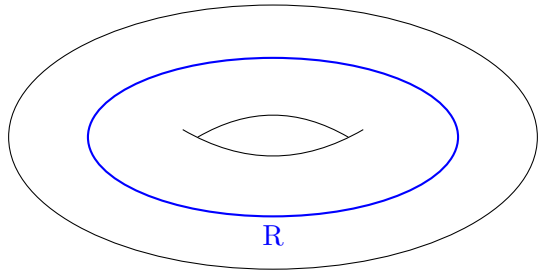
\begin{figure}[h]
\centering
\begin{tikzpicture}
  \draw (-1,0) to[bend left] (1,0);
  \draw (-1.2,.1) to[bend right] (1.2,.1);
  \draw[rotate=0] (0,0) ellipse (100pt and 50pt);
  \draw[blue,thick] (0,0) ellipse (70pt and 30pt) ;
  \node[blue] at (0,-1.3) {R};
\end{tikzpicture}
\caption{ The state $\ket{R_{CS}}$ is defined by the path integral on a solid torus with a Wilson loop operator inserted.}
\label{fig:Rtorus}
\end{figure}

The basis elements $\ket{R}_{CS}$ for $\mathcal{H}(T^2)$ are obtained by performing the Chern-Simons path integral on a solid torus $D^2\times S^1$ with an insertion of the Wilson loop operator
\begin{align}
    W_{R} =\tr_{R} P \exp \oint_{C} A,
\end{align}
where $R$ labels the representation and $C$ is the non-contractible cycle of the solid torus. 
This is shown in figure \ref{fig:Rtorus}. The trivial representation corresponds to the vacuum state $|0\rangle\in\mathcal{H}(T^2)$ with no Wilson loops inserted. 

We can superpose the states $\ket{R}_{CS}.$ Since no local operator can connect states with different representation labels, each representation labels a superselection sector corresponding to an anyon of type $R$.  
\paragraph{Chern-Simons partition functions from Heegaard splitting}

Now we can consider partition functions of Chern-Simons theory on a manifold $M$ without a boundary by gluing the aforementioned building blocks.  Consider a Heegaard splitting  of $M$
into two manifolds with boundaries $M_1$ and $M_2$ that are glued together by a nontrivial diffeomorphism  $f:\partial M_1\rightarrow \partial M_2.$  Then there is a linear map $U_{f}$
such that
\begin{align}
    Z(M)=\langle \Psi_{M_2}|U_f|\Psi_{M_1}\rangle,\label{eqn:CS partition}
\end{align}
where $\Psi_{M_1}$ and $\Psi_{M_2}$ are states assigned to $M_{1}$ and $M_{2}$ via \eqref{eqn:CS wave fun}, and $U_{f}$ forms a representation of the diffeomorphisms that define the gluing. 

In particular, under the Heegard splitting, $S^3$ is decomposed into two solid tori $\Bbb{T}^3_i=D^2_i\times S^1_i$ for $i=1,2.$ which are glued together with an $S$ transformation that exchanges the A and B cycles.  Thus we have
\begin{equation}
    Z(S^3)=\langle 0|S|0\rangle
    \label{eqn:CS partition on S3}
\end{equation}
The value of the matrix element \eqref{eqn:CS partition on S3} is fixed by a normalization for the vacuum state $|0\rangle$, which we choose to be

\begin{equation}
   \langle 0|0\rangle= Z(S^2\times S^1)=1.
    \label{eqn:CS normalization}
\end{equation}
Notice that this normalization is equivalent to a choice of path integral measure.
This defines the S matrix element
\begin{equation}
    Z(S^3)=S_{00}
\end{equation}

The same Heegard splitting can be applied to compute expectation values of Wilson loops  $S^{3}$.  By gluing the state $\ket{R}_{CS} \in \mathcal{H}(T^2)$ to the the vacuum state $\ket{0} \in \mathcal{H}(T^2)$ with an S transformation, we find 
\begin{align} 
\braket{\tr_{R}(U)}_{S^{3} } &=\braket{0|S|R}_{CS}=S_{0R}\nn
&= S_{00} \dim_{q}(R) 
\end{align} 
In the second equality we introduced the quantum dimension $\dim_{q}(R)$ of $U(N)$.  It is the \emph{normalized} expectation value of the unknot in $S^3$ and give the effective dimension of the topological Hilbert space for the anyon labelled by $R$.

\subsection{Generating functional for Wilson loops and the $\Omega$ state}
Here we define the generating functional for Wilson loop operators in Chern-Simons theory \cite{Ooguri:1999bv}, which plays an essential role in the duality between Wilson loops and worldsheets in topological string theory.

This can be obtained from the Ooguri-Vafa operator \cite{Ooguri:1999bv}
\begin{align}\label{OGV}
\exp \left( \sum_{n=1}^{\infty} \frac{1}{n}\tr U^n \tr V^n\right) = \sum_{R} \tr_{R}(U) \tr_{R} (V) 
\end{align}
where $U=\exp(\oint_{\gamma} A)$, and $\gamma$ is an unknot in $S^3$. This can be derived from integrating out a massless bi-fundamental field which couples to both the source and dynamical gauge fields \cite{Ooguri:1999bv}.
Treating $V$ as a source, we insert this operator into the path integral to obtain the generating functional
\begin{align} \label{loopfn}
Z(V)=\int \mathcal{D}A e^{i S_{CS}(A)+\sum_n \frac{1}{n}\tr U^n \tr V^n} =\sum_{R} \braket{\tr_{R}(U)}_{S^3} \tr_{R}(V) 
\end{align}
for Wilson loops in an arbitrary representation $R$. 

It will be useful to view the generating functional \eqref{loopfn} as a wavefunction for a state $\ket{\Omega}$ on the torus Hilbert space, defined by
\begin{align}
    |\Omega\rangle &=\sum_R S_{0R} \ket{R}_{CS} \nn
&=S_{00}  \sum_R \dim_{q}(R)  \ket{R}_{CS}
\end{align}
If we introduce the coherent state basis $|V\rangle\in\mathcal{H}(T^2) $ with wavefunctionals
\begin{align}
\langle V|R\rangle=\tr_R V,
\end{align} 
then the wave function of $\ket{\Omega}$ in this basis can be identified with the generating functional $Z(V)$  
\begin{align}
Z(V)&= \sum_{R}  S_{00} \dim_{q}(R)  \tr_{R}(V)= \braket{V|\Omega}
\end{align} 

\subsection{Hartle-Hawking state in Chern-Simons theory}
Consider Chern-Simons theory at level $k$ with gauge group $U(N)$, which corresponds to the $q$ parameter 
\begin{align} 
q= \exp (\frac{2\pi i}{k+N}).
\end{align}    
For any $t \in \mathbb{C}$, define the following state on the torus 
\begin{align}\label{Omt}
\ket{\Omega(t)} &=  S_{00} \sum_{R}  \dim_{q}(R) e^{-\frac{t}{2}l(R)} \ket{R}_{CS},
\end{align} 
which consists of a superposition of Wilson loops, and reduces to the state $\ket{\Omega}$ at $t=0$.  We claim that the Hartle-Hawking state $\ket{HH(t)}$ for string theory on the resolved conifold is dual  to a suitable large $N$ limit of  $\ket{\Omega(t)}$.   In the coherent state basis, the wavefunction for this state is
\begin{align}
\braket{V| \Omega(t)} = \braket{ e^{-t/2} V|\Omega}= S_{00} \sum_{R}  \dim_{q}(R) e^{-\frac{t}{2}l(R)} \tr_{R}(V) 
\end{align}
As in the case of $t=0$ we can identify this expression as a generating wavefunctional obtained by a path integral on $S^3$: 
\begin{align}
\braket{V|\Omega(t)} = Z(V,t) &:=\int \mathcal{D}A e^{i S_{CS}(A)+\sum_n  \frac{e^{-nt/2} }{n}\tr U^n \tr V^n}\nn
 &=\sum_{R} \braket{\tr_{R}(U)}_{S^3}  e^{- l(R) t/2} \tr_{R}(V) 
\end{align}
where we have applied a generalization of \eqref{OGV} by replacing $V \to e^{-t/2} V$. $Z(V,t)$ can again be obtained from integrating out a massive bi-fundamental field coupling the source and dynamical gauge fields \cite{Ooguri:1999bv, 2004CMaPh.247..467A, Aganagic:2003db}.

To obtain the appropriate large $N$ limit of eq \eqref{Omt}, we need to specify  the large N limit of the states $\ket{R_{CS}}$.  To do this, first define the states $\ket{k_{CS}}$ whose wavefunctions are obtained by inserting Wilson loops in the winding basis \eqref{wind}
\begin{align}
    \braket{A_{T^2}|\vec{k}_{CS}} &=\int_{A|_{T^2}=A_{T^2}} \mathcal{D}A e^{iS} \prod_{n=1}^{\infty} \tr (U^n)^{k_{n}}\nn
    U&=\exp(\oint_{S^{1} }A) 
\end{align}
$\ket{k_{CS}}$ is well defined in the large $N$ limit, and $\ket{R_{CS}}$ is defined by it's relation to $\ket{k_{CS}}$ via the Frobenius relation \eqref{frob} In this limit $R$ is a representation label for $U(\infty)$ corresponding to a Young tableaux.

The large $N$ limit of  the amplitudes $\braket{R|\Omega(t) } $ is more subtle.    Following \cite{2005CMaPh.254..425A, Marino:2004uf} , we first take the large $N$ limit while fixing the  't Hooft coupling $t'$ 
\begin{align}
N &\to \infty \nn
t'&=\frac{ 2 \pi i N }{k+N}=\text{constant} ,
\end{align} 
which is the same as holding the ratio $\frac{k}{N}$ constant. We then analytically continue $t'$ to a real number and then take $t' \to \infty$.  In this limit we can expand the quantum dimensions as
\begin{align} \label{limit}
\dim_q (R)=(-i)^{l(R)}d_q(R) q^{Nl(R)/2}q^{\kappa(R)/4}+\mathcal{O}(q^{-l(R)N/2}),
\end{align}
where $d_{q}(R)$ is defines as in \eqref{dqR} with 
\begin{align}
g_{s}&= \frac{2\pi}{k+N}
\end{align}

Notice that we needed to analytically continue $t'$ so that $q^{Nl(R)/2}= e^{\frac{t'}{2} l(R) }$ has a divergent norm  $t' \to \infty$; the first term of \eqref{limit} can then be considered large relative to the rest.  Remarkably, we can absorb this divergence into the Boltzman factor $e^{-\frac{t}{2} l(R) }$ because they both have the same  exponent dependence on $l(R)$.  More precisely, we will apply a shift to the ``coupling" $t$ in the Bolztman factor
 \begin{align} \label{shift}
    e^{-\frac{t}{2}l(R)}&\to e^{-\frac{t+t'}{2}l(R)} 
 \end{align}
 and identify the leading dependence of $\dim_{q}(R)$ on $N$  as
 \begin{align} \label{qdiv}
     q^{Nl(R)/2}=e^{ t' l(R)/2}
 \end{align}
The state
 \begin{align}
     \ket{HH_{CS}(t)} := \lim_{t' \to \infty} \lim_{N \to \infty} \ket{\Omega (t+t') } =\lim_{t' \to \infty}  \lim_{N \to \infty} S_{00} \sum_{R}  \dim_{q}(R) e^{-\frac{t+t'}{2}l(R)} \ket{R}_{CS}
 \end{align}
 is then well defined since the divergent term \eqref{qdiv} has been cancelled.   Note that $S_{00}(t')$ diverges as $t' \to \infty$, but this just gives the usual infinite normalization which arises from the path integral measure.  Using \eqref{limit}, we find that the putative Chern-Simons dual of the Hartle-Hawking state $\ket{HH(t)}$ is: 
 \begin{align}\label{putative}
     \ket{HH_{CS} (t) } = S_{00} \sum_{R} (-i)^{l(R)}d_q(R) q^{\kappa(R)/4}  e^{-\frac{t}{2}l(R)} \ket{R_{CS}}
 \end{align}
In the dual closed string theory on the resolved conifold, the shift  $t\rightarrow t+t'$ in \eqref{shift} is performed to correctly parametrize the K\"ahler cone in the presence of the B-flux on the $S^2$ on which the worldsheet ends \cite{Diaconescu:2002sf}.  Notice that the this shift by $t'$ is equivalent the introducing the relative factor between the Drinfeld elements $D$ and $u$:
\begin{align}
    D=q^{-N/2} u
\end{align}
In the large $N$ limit, this choice of $D$ effectively absorbs the divergence in the quantum dimension as defined by $u$ and assigns a finite dimension to the edge mode Hilbert space in the $R$ sector.    

To see that we have obtained the correct dual to the Hartle-Hawking state, we must show that the identification 
\begin{align}\label{R}
    \ket{R_{CS}} \to \ket{R}
\end{align}
preserves locality in some sense. Otherwise equation \eqref{R} is just a linear map between basis elements which is rather trivial, since all vector spaces of the same dimensionality are isomorphic.  The preservation of locality is captured most precisely by showing that local Hilbert space factorization of CS theory is mapped to the factorization in the string theory. The sharpest expression of the local nature of this mapping between $\ket{HH(t)}$ and $\ket{HH_{CS}(t)}$ is obtained by identifying the Wilson loops in CS theory to the boundaries of worldsheets in the string theory: we will explain this duality in the next section \cite{1998PhRvL..80.4859M, 2001EPJC...22..379R, Ooguri:1999bv, Gomis:2006mv}. We will also show the preservation of locality by checking that the entanglement entropy computed with either factorization agrees.  

\subsection{Matching with the dual partition function and emergence of the bulk geometry } 
 Consider the density matrix for $\ket{HH_{CS}}$.  In the following it is important to observe that a TQFT is not endowed with a canonical choice of a Hermitian inner product. Indeed, the sesquilinear property of such an inner product is incompatible with the holomorphic nature of the A model.  Therefore in the dual Chern-Simons theory, we will define the density matrix and the partial trace without reference to a Hermitian inner product. While this departs from the conventions of \cite{Wen:2016snr, Wong:2017pdm, Dong:2008ft, Fliss:2020cos},  this is consistent with the usual representation of the reduced density matrix as a Euclidean path integral with a cut along the subregion.  It also agrees with conventions in defining entanglement entropy in non Hermitian systems where a positive definite Hermitian inner product is not readily available. 

We define the density matrix for the Hartle-Hawking state by
\begin{align}\label{density}
  \rho&= \ket{HH_{CS} }\bra{HH_{CS}^{*} } \nn
   \bra{HH^{*}_{CS}} &:= \sum_R (i)^{l(R)} d_{q} (R) q^{-\kappa_R/4}  e^{-l(R)(t)/2} \bra{R_{CS}}
\end{align}
where we deonte by $\bra{R_{CS}}$ the basis dependent dual of $\ket{R_{CS}} $, obtained by doing the path integral on a torus of opposite orientation with the insertion of the Wilson loop in the conjugate representation $\bar{R} $. By definition the dual basis satisfies 
\begin{align} \label{RR'}
    \braket{R_{CS}|R'_{CS}}  = \delta_{RR'},
\end{align}
but note that this makes no reference a Hermitian inner product\footnote{The usual Hermitian inner product  agrees with \eqref{RR'} on a basis.  However its sesquilinear property is not consistent with the holomorphic nature of the A model and its Chern-Simons dual.}. Instead equation \eqref{RR'} arises from evaluating the Wilson loops expectation values in the $S^{2} \times S^{1}$ geometry obtained from  gluing the two tori with opposite orientation.  In particular $ \bra{HH^{*}_{CS}}$ is not related to $\ket{HH_{CS}}$ by an anti-linear map.

In the string theory, $ \bra{HH^{*}_{CS}}$ corresponds to the the linear functional that is related to the geometric state $\ket{HH_{CS}}$ by flipping orientation and mapping branes to anti-branes \cite{2001hep.th....1218V, 2005CMaPh.254..425A}.  In the Chern-Simons theory, we will take $ \bra{HH^{*}_{CS}}$ as part of the  choice of a density matrix $\rho$ which determines the expectation value of operators via
\begin{align}
    \braket{O}= \tr (\rho O ) 
\end{align}
The trace of the density matrix agrees with the A model partition function on the resolved conifold: 
\begin{align}
    Z=\tr (\rho)  = \sum_{R}  (d_{q}(R))^{2} e^{-t l(R)} = Z_{\text{top}} 
\end{align}
Just like the A model, this is a holomorphic quantity and is not a real norm. 
It is important to note that from the point of view of the  Chern-Simons theory, the parameters $t$  carry no geometric interpretation; it merely specifies a particular superposition of Wilson loops.  Remarkably, in the dual closed string theory, a geometry has emerged in which $t$ becomes the Kahler modulus of the target space.   Note that this Kahler modulus is not the one that arise from applying a geometric transition to branes wrapping $S^3$ in Chern-Simons theory. Instead it arises from a particular superposition of Wilson loops.

\subsection{Entropy from geometrical replica trick in Chern-Simons theory}
Here we perform the replica trick calculation of entanglement entropy for the putative Chern-Simons dual \eqref{putative} to the Hartle-Hawking state and show that they match on the two sides of the duality. More precisely, we show that the q-deformed entanglement entropy in the closed string theory coincides with the undeformed defect entropy in Chern-Simons theory in the large N-limit.  The defect entropy is the difference between the entanglement entropy and the state independent ground state entanglement entropy, and it measures the entanglement entropy due to cutting the Wilson loops \cite{2013arXiv1307.1132J, 2013PhRvD..88j6006J, Lewkowycz:2013laa}.
We will sidestep the question of how to factorize of the Chern-Simons Hilbert space by applying the geometric replica trick via surgery methods as in  \cite{Dong:2008ft}.  

As pointed out earlier,  our calculation of the entanglement entropy of a generic state
\begin{align}
    \ket{\psi}= \sum_{R} \psi(R) \ket{R} 
\end{align}
 differs from \cite{Dong:2008ft} in the choice of inner product: since reference \cite{Dong:2008ft} uses a Hermitian inner product which defines an anti-linear adjoint operation, their density will involve complex conjugation of the amplitudes $\psi(R)$, whereas ours do not.  

\begin{figure}[h]
\centering
\begin{tikzpicture}
  \draw (-1,0) to[bend left] (1,0);
  \draw (-1.2,.1) to[bend right] (1.2,.1);
  \draw[rotate=0] (0,0) ellipse (100pt and 50pt);
  \draw[blue,thick] (0,0) ellipse (70pt and 30pt) ;
  \node[blue] at (0,-1.3) {R};
  \node at (-2,0){$A$};
  \node at (2,0){$\bar{A}$};
  \draw[dashed] (0,-50pt) to[bend left] (0,-0.26);
  \draw (0,-50pt) to[bend right] (0,-0.26);
  \draw (0,50pt) to[bend left] (0,0.27);
  \draw[dashed] (0,50pt) to[bend right] (0,0.27);
\end{tikzpicture}
\caption{Separated solid torus with a Wilson loop operator inserted.}
\label{fig:torus with wilson2 main}
\end{figure}
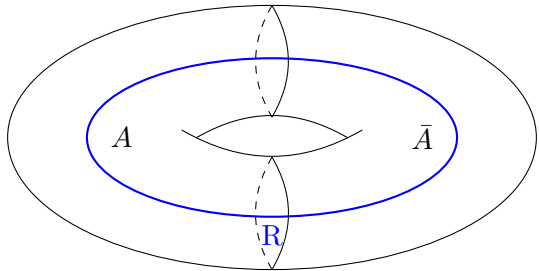

 However apart from determining the coefficients in the density matrix $\rho$, the choice of inner product has no bearing on the construction of the replica manifold which computes the replica partition functions 
 \begin{align}
     Z(n)= \tr_{A} \rho^{n} _{A} ,
 \end{align}
 since neither the partial trace nor matrix multiplication implicit in the replica trick uses the inner product.  This point was recently emphazised in \cite{Dupic_2018}.  As a result, we can borrow the results of \cite{Dong:2008ft} directly.   First note that in when computing $Z(n)$ there are no ``cross term" corresponding to replica manifolds with insertions of Wilson loop $R$ and $R'$ which are not conjugate to each other.  

The reduced density matrix defined by the replica manifold will therefore decompose into superselection sectors labelled by $R$.  This means we can apply the result of  \eqref{Dong:2008ft} for each state $|R_{CS}\rangle$ separately and sum up the results.   For each $R$, we partition the torus in Fig. \ref{fig:torus with wilson2 main} defining $\ket{R_{CS}}$ as well as the dual torus corresponding to $\bra{R_{CS}}$.  We glue together the $\bar{A}$ to obtain a reduced density matrix, and then construct the replica manifold by the usual cyclic gluing.  The reference \eqref{Dong:2008ft} showed that the partition function for the $n$th replica is
\begin{equation}\label{eqn:surgery4}
    Z(n)= \tr{\rho_A(R)^n}= Z(S;R_i)^{1-n} Z(S;\overline{R}_i)^{1-n}.
\end{equation}
The normalization factor is
\begin{equation}\label{eqn:surgery3}
   Z(1)=\tr{\rho_A(R)}= \langle R_i| R_i \rangle = \frac{Z(S^3;R_i)^2 Z(S^3; \overline{R}_i)^2}{Z(S^3;R_i)^2 Z(S^3;\overline{R}_i)^2}=1.
\end{equation}
Using the large N identity 
\begin{equation}
\lim_{N \to \infty}    Z(S^3;R)Z(S^3;\overline{R})=S_{00}^{2} d_q(R)^2.
\end{equation}
we obtain the replica trick entropy
\begin{align}
     S_{R}=-\frac{\partial}{\partial n} \frac{Z_n}{Z_1^n}|_{n=1}= 2 \log S_{00} d_{q}(R).
\end{align}

Now we apply the replica trick the putative dual to the Hartle-Hawking state \eqref{putative} 
\begin{equation}
    |HH_{CS}(t)\rangle =S_{00} \sum_R (-i)^{l(R)}d_q(R)q^{\kappa(R)/4}e^{-\frac{t}{2}l(R)}|R_{CS}\rangle.
\end{equation}
As noted earlier, the reduced density matrix
\begin{equation}
    \rho_A=\tr_{\overline{A}}|HH_{CS}(t)\rangle \langle HH_{CS}(t)^*|,
\end{equation}
breaks into superselection sectors labelled by $R$, so the replica partition function is just 
\begin{align}\label{eqn:replica}
    Z_n=&\tr_A(\rho_A^n)=\sum_R(S_{00}d_q(R))^{2-2n} (S_{00} d_q(R) e^{-\frac{t}{2}l(R)})^{2n}  \nn
    =& \sum_R S_{00}^2 d_q(R)^2 e^{-nt l(R)}.
\end{align}
It is very interesting to note that the only effect of the replication is the replication of the complexified area $t$. This means that from the string theory point of view, the topology of the target space is not changed, and the  Calabi-Yau condition is preserved.   This was precisely the topological constraint we imposed in our definition of generalized entropy on the resolved conifold \cite{2020arXiv201015737D}.
We can thus view our Chern-Simons replica manifold as the in gauge theory dual to the replica manfiold we constructed for the resolved conifold in \cite{2020arXiv201015737D}.

Given \eqref{eqn:replica}, we find that
\be \label{eqn:CS renyi}
\frac{Z_n}{Z_1^n}=\frac{\tr_A {\rho_A^n}}{\left(\tr_A {\rho_A}\right)^n}=S_{00}^{2-2n}\sum_R d_q(R)^2 e^{-ntl(R)},
\ee
where we defined $Z\equiv \sum_R d_q(R)^2 e^{-tl(R)}.$. The total entanglement entropy is 
\begin{equation}
    S_{tot}=-\frac{\partial}{\partial n} \frac{Z_n}{Z_1^n}|_{n=1}=\sum_R p(R)(-\ln{p(R)}+2 \ln d_q(R))+2\ln(S_{00})\label{eqn:CS replica answer}
\end{equation}
where  $p(R)=\frac{d_q (R)^2 e^{-t l(R)}}{Z}$

As noted earlier, we will only be interested in the entropy due to cutting the Wilson loops, since these are dual to the closed string worldsheets.   This is captured by the defect entropy which subtracts the state independent contribution 
\be
S_0=2\ln(S_{00}) \label{eqn:CS replica constant}
\ee 
which is the analogue of the background extremal entropy  in \cite{Jafferis:2019wkd, 2018arXiv180706575L, 2013arXiv1307.1132J, 2013PhRvD..88j6006J, Lewkowycz:2013laa}.
The defect entropy is therefore
\begin{equation}
    S=S_{tot}-S_0=\sum_R p(R)(-\ln{p(R)}+2 \ln d_q R).
\end{equation}
which matches with the generalized entropy computed in the closed string theory.

\subsection{Factorization and edge modes in the dual Chern-Simons theory}
In this section we give the Chern-Simons dual of the factorization map \eqref{cfact} and edge modes for the closed string theory on the resolved conifold.   These can be obtained from a suitable large N limit of the factorization map developed in \cite{Wen:2016snr, Wong:2017pdm}, where an explicit description of the extended Hilbert space for Chern-Simons theory was given in terms of CFT edge modes.   We will apply this factorization map to the state $\ket{\Omega(t)}$ and give a canonical calculation of entanglement entropy which agrees with the results of the previous section.   We find that the quantum group edge mode symmetry of the closed string theory is described in the dual gauge theory by the large N limit of Kac-Moody symmetry of the CFT edge modes.
 
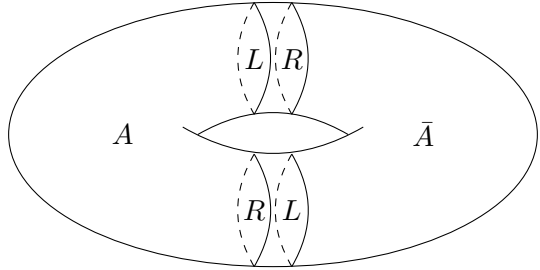
\begin{figure}[h] 
\centering
\begin{tikzpicture}
  \draw (-1,0) to[bend left] (1,0);
  \draw (-1.2,.1) to[bend right] (1.2,.1);
  \draw[rotate=0] (0,0) ellipse (100pt and 50pt);
  \node at (-2,0){$A$};
  \node at (2,0){$\bar{A}$};
  \draw[dashed] (-0.25,-50pt) to[bend left] (-0.25,-0.26);
  \draw (-0.25,-50pt) to[bend right] (-0.25,-0.26);
  \draw (-0.25,50pt) to[bend left] (-0.25,0.27);
  \draw[dashed] (-0.25,50pt) to[bend right] (-0.25,0.27);
  \draw[dashed] (0.25,-50pt) to[bend left] (0.25,-0.26);
  \draw (0.25,-50pt) to[bend right] (0.25,-0.26);
  \draw (0.25,50pt) to[bend left] (0.25,0.27);
  \draw[dashed] (0.25,50pt) to[bend right] (0.25,0.27);
   \node at (-0.25,1){$L$};
  \node at (0.25,1){$R$};
    \node at (0.25,-1){$L$};
       \node at (-0.25,-1){$R$};
\end{tikzpicture}
\caption{Geometric entanglement entropy in Chern-Simons theory is equivalent to left-right entanglement entropy in the WZW model. L and R in the diagram represents left and right moving chiralities for the WZW models.}
\label{fig:LREE}
\end{figure}

\paragraph{Entanglement cut and the factorization map at finite N}
The entanglement cut which we apply to the state $\ket{\Omega(t)}$ is shown in Fig. \ref{fig:LREE}.  The surface of the torus is partitioned into two disconnected subregions $A$ and $\bar{A}$,separated by a cylindical region of size $\epsilon$.  This is a UV regulator which we will send to zero at the end of the calculation.  To define the subregion Hilbert space $\mathcal{H}_{A}$,$\mathcal{H}_{\bar{A}}$, we choose a ``shrinkable" entanglement boundary condition which breaks the topological invariance\footnote{More specifically, the boundary condition introduces a choice of complex structure which defines the chiral edge modes \cite{Fliss:2020cos}.} and introduces CFT edge modes along $\pd A$ and $\pd \bar{A}$ \cite{Wen:2016snr, Wong:2017pdm}.  This shrinkable boundary condition is \emph{local} and corresponds to setting the component of the gauge field in the angular direction around the entangling surface to zero.  For $U(N)_{k}$ Chern-Simons theory, the edge modes correspond to chiral $U(N)_{k}$ WZW models at the boundaries of $A$ with opposite chiralities.

The subregion Hilbert space can be expressed as 
\begin{align}
    \mathcal{H}_{A} &=  \mathcal{H}^{L}_{WZW} \otimes   \mathcal{H}^{R}_{WZW} \nn
    \mathcal{H}_{\bar{A}} 
    &=  \mathcal{H}^{R}_{WZW} \otimes   \mathcal{H}^{L}_{WZW} 
\end{align}

Due to the Gauss Law constraint, the Chern-Simons Hilbert space $\mathcal{H}(T^2)$ on the torus does not naively factorize into a tensor product of subregion Hilbert spaces.  Instead, the factorization should be viewed as a mapping 
\begin{align}
    \mathcal{H}(T^2)\to \mathcal{H}_{A} \otimes \mathcal{H}_{\bar{A} } 
\end{align}
that embedds the physical Hilbert space into the extended Hilbert space\footnote{Equivalently, we can view the physical Hilbert space as a fusion product
\begin{align}
    \mathcal{H}(T^2)=\mathcal{H}_{A} \otimes_{G} \mathcal{H}_{\bar{A} }
\end{align}
in which we impose quotient relation determined by a quantum gluing condition. }; due to the holographic nature of Chern-Simons theory, the extended Hilbert space consists entirely of edge modes.  The factorization map on each basis element is given by 
\begin{align} \label{Ishi}
    \ket{R_{CS}} &\to |R\rangle \rangle_{1} |R\rangle \rangle_{2}\nn
   |R\rangle \rangle_{1} &= \frac{ e^{\frac{-8\pi \epsilon}{l} (\bar{L}_{0}-\frac{c}{24})} }{\sqrt{n_R}} \sum_{N=0}^{\infty} \sum_{j=1}^{d_R(N)} \ket{R,N,j}_L \overline{\ket{R,N,j}}_R ,
\end{align}
where $1,2$ labells the two entangling surfaces, and $|R\rangle \rangle $ is a normalized Ishibashi state that satisfy the Gauss law constraint  across the entangling surface.  The integers $N,j$ label descendents, and $d_R(N)$ is a degeneracy for each level $N$.   As explained in \cite{Wong:2017pdm}, the factorization map is implemented by the Euclidean path integral with a ``brick wall" regularization.  If we flip a ket into a bra in \eqref{Ishi} using a CPT conjugation\footnote{This is refered to as the state-channel duality}, the normalized Ishiabashi state becomes the modular operator which implements the ``half modular flow" from $ \mathcal{H}_{A}$ to $\mathcal{H}_{\bar{A}}$.  Finally the normalization factor 
\begin{align}
    n(R)=\chi_R (e^{\frac{-8 \pi \epsilon}{l}}) =\tr_{R}  e^{\frac{-8\pi \epsilon}{l} (\bar{L}_{0}-\frac{c}{24})}
\end{align}
is the charactor of the integrable representation $R$ of the Kacs Moody algebra for $U(N)$ at level $k$, and $l$ is the length of entangling surface. Observe that the brick wall regulator $\epsilon$ is needed to render the normalization finite, as $\epsilon \to 0$ corresponds to an infinite temperature limit.    

For fixed $\ket{R_{CS}}$, the reduced density matrix is obtained by factorizing the density matrix \eqref{density} and tracing over $\bar{A}$.  This corresponds to tracing out a chiral half of the left-right entangled Ishibashi states $|R \rangle \rangle$, which gives the reduced density matrix

\be
\rho^{R}=\rho^{R}_{1} \otimes  \rho^{\bar{R}}_{2},
\ee
with
\begin{align}\label{CSfact}
    \rho^{R}_{1}&=\frac{1}{\bar{n}_R} \sum_{N_1,j_1} e^{\frac{-8\pi \epsilon}{l} (\bar{L}_{0}-\frac{c}{24})} \ket{R,N_1,j_1}_L \bra{R,N_1,j_1}_L, \nn 
\rho^{\bar{R}}_{2}&=\frac{1}{n_R} \sum_{N_2,j_2} e^{\frac{-8\pi \epsilon}{l} (L_{0}-\frac{c}{24})} \ket{\overline{R,N_2,j_2}}_R  \bra{\overline{R,N_2,j_2}}_R 
\end{align}
Note that edge modes at the two  boundaries of a subregion have opposite chiralities, and combine to form a diagonal CFT. The entanglement Hamiltonian\footnote{ Our entanglement Hamiltonian is slightly different the usual definition of the modular Hamiltonian since it doesn't contained the constant term due to $Z_{A}$ in the denominator  } is identified with the non-chiral WZW Hamiltonian: 
\begin{align}
    H_{A} &= \frac{-8\pi \epsilon}{l} (L_{0}+\bar{L}_{0} -\frac{c}{12})\\ \label{CFTH}
    \rho_{A}&=\frac{e^{-H_{A}}}{Z_{A}}
\end{align}
where $Z_{A}= \bar{n}_{R}n_{R}$ is the partition function of the CFT edge modes.

The entanglement entropy can be obtained directly without appealing to the replica trick:

\begin{align}
    S=- \tr \rho_{A} \log \rho_{A}= \tr_{A}(\rho_{A} H_{A})+ \log Z_{A}
\end{align}
As $\epsilon \to 0$, the ``modular energy term" vanishes and the entropy is identified with the free energy.  For a fixed $R$ sector we have
\begin{align}
  S_{R}= \log Z_{A}= \log \chi_{R}(e^{\frac{-8\pi \epsilon}{l}}) \chi_{\bar{R}} (e^{\frac{-8\pi \epsilon}{l}})
\end{align} 
In the $\epsilon \to 0$ limit, the reduced density becomes maximally mixed, so the entropy is the logarithm of the number of states in a suitable sense.   In fact this interpretation becomes sharper when we consider only the entropy due to cutting the Wilson loop.   This is obtained by subtracting the vacuum entropy for $R=0$, which gives the defect entropy $S_{\text{defect}}(R)$. This gives a counting of the degeneracy according to 
\begin{align}\label{dR}
 e^{S_{\text{defect}}(R)+ S_{\text{defect}}(\bar{R})}=   \left( \lim_{\epsilon \rightarrow 0} \frac{ \tr_{R} e^ {-2 \pi \epsilon H_{\text{CFT}} }}{\tr_{R=0} e^ {-2 \pi \epsilon H_{\text{CFT}}}} \right) \left( \lim_{\epsilon \rightarrow 0} \frac{ \tr_{\bar{R}} e^ {-2 \pi \epsilon \bar{H}_{\text{CFT}} }}{\tr_{\bar{R}=0} e^ {-2 \pi \epsilon \bar{H}_{\text{CFT}}}}   \right)  =\dim_{q}(R) \dim_{q}(\bar{R} )
\end{align}
The ratio of CFT partition functions define the ``regularized dimension" of a representations $R/\bar{R}$ of the chiral algebra, which is one way to \emph{define} the quantum dimensions.    In practice the $\epsilon \to 0$ limit is computed by first applying a modular transformation to  $\chi_{R}$ 
\begin{align}
 \chi_R(e^{-\frac{8\pi  \epsilon}{l}})=\sum_{R'} S_{R R'} \chi_R(e^{-\frac{\pi l}{2\pi  \epsilon}}) \to   S_{R 0} e^{\frac{\pi c l}{48  \epsilon}}= S_{00} \dim_q(R) e^{\frac{\pi c l}{48  \epsilon}}
\end{align}
At finite $N$ we have $\dim_{q}(\bar{R})= \dim_{q}(R)$, so from the point of view of the defect entropy, we have an effective degeneracy of $\dim_{q}(R)^{2}$ for each superselection sector labelled by $R$.

In the section 2, we observed that the quantum dimensions $\dim_{q}(R)$ can be viewed as a choice of measure determined by the Drinfeld element of $U(N)_{q}$ and the corresponding quantum trace \eqref{qtrace}. Here we see an alternative definition of this measure  via the ratio of CFT partition functions.  We will see below that in the large $N$ limit, these correspond to the two alternative descriptions of the edge modes in the string theory and the Chern-Simons dual.  

By linearity, we can apply the factorization map \eqref{CSfact} to the state $\ket{\Omega(t)} $
\begin{equation}
    \ket{\Omega(t)}= S_{00}\sum_R \dim_{q}(R)   e^{-l(R)t/2} | R \rangle_{CS}
    \end{equation}
which gives
\begin{equation}
 \ket{\Omega(t)}=S_{00}\sum_R \dim_{q}(R)   e^{-l(R)t/2} \ket{R} \rangle_1 \ket{R}\rangle_2,
\end{equation}

The (normalized)  reduced density matrix now consists of a sum over superselection sectors labelled by $R$ 
\begin{align}
    \rho_{A} &= \sum_{R} P(R) \rho^{R}_{1} \otimes \rho^{\bar{R}}_{2} \nn
    P(R) &= \frac{ \dim_{q}(R)^{2} e^{ - t l(R) } }{Z} 
\end{align}
This takes the form of a ``thermo-mixed double" state \cite{2020arXiv200313117V} in which the two edge modes CFT's are classically correlated due to the Wilson loop threading the torus.

This reduced density matrix takes a form which is directly analogous to that of a nonabelian gauge theory. We can make this manifest by writing the normalized density matrices $\rho_{i}^{R} $ explicitly as a maximally mixed state in the $R$ sector.   
\begin{align}
    \rho_{1}^{R} =   \frac{ e^{ \frac{ - 8 \pi \epsilon}{l} (L_{0}-\frac{c}{24})  }}{\chi_{R}(e^{ \frac{ - 8 \pi \epsilon}{l}) }}    \to  \frac{ \mathbf{1}_{R} }{\dim_{q}R S_{00} e^{\frac{\pi c l}{48  \epsilon}} } 
\end{align}
and similarly for $\rho_{2}^{\bar{R}} $. 
Then we have
\begin{align} \label{red}
    \rho_{A} =\sum_{R} P(R) \frac{ \mathbf{1}_{R\otimes \bar{R}}} {|\dim_{q}R|^{2} |S_{00} e^{\frac{\pi c l}{48  \epsilon}} |^{2}}
\end{align}
so we can identify $P(R)$ as a probability factor for being in the $R\otimes \bar{R}$ sector, where the density matrix is just proportional to the identity.  The analogy become exact when we subtract off the vacuum entropy which gets rid of the contribution from $S_{00} e^{  \frac{\pi c l}{48 \epsilon}}$.
Written in this form, we see that the entanglement Hamiltonian should be identified (up to a state independent constant) with the operator 
\begin{align}
    H_{A} = t l(R) 
\end{align}
The CFT edge modes Hamiltonian \eqref{CFTH} merely plays an intermediary role in regularizing the trace, in order to determine a finite degeneracy factor for fixed $R$.

We now repeat the edge mode calculation of defect entropy in the large N limit, and show that it matches with string edge mode calculation involving entanglement branes.   In particular we 
want to identify the entanglement spectrum and degeneracy on both sides.  
In the string theory, the edge mode Hilbert space breaks up into superselection sectors  $R\otimes \bar{R}$ labelled by a young tableaux.  In  the large N limit, each sector has an infinite number of states with modular eigenvalue $t l(R)$.   The degeneracy in each sector is obtained by applying a regularized trace
\begin{align}
    \tr_{R\otimes \bar{R}}( D) = \tr_{R}(D) \tr_{\bar{R}}(D) = (d_{q}(R))^{2}
\end{align} where $D$ is related to the Drinfield element $u$ of $U(N)_{q}$ by
\begin{align}
    D_{ij}= q^{-N/2} u_{ij}=\delta_{ij} q^{-i+\frac{1}{2}}
\end{align}

The regularization involves a continuation
\begin{align}
     q \to q e^{\epsilon_{\text{string}} }
\end{align}
which makes the trace converge.  We interpret the trace as a sum over entanglement branes. 
 
In the Chern-Simons dual, we find  a similar structure from the large N limit of the WZW model edge mode CFT \cite{2011JHEP...04..113K}.   The primary states of the $U(N)_{k}$ WZW model are labelled by a finite number of integrable representations.  However in the large $N$ limit, the truncation is lifted and we can associate each a chiral/antichiral primary to each Young Tableaux.    The conformal dimensions for these chiral primaries are given by
\begin{align}
    \Delta (R)=  \frac{C_{2}(R)}{2(k+N)},
\end{align}
where $C_{2}(R)$ is the quadratic Casimir.  The large $N$ limit at fixed $t'$ gives 
\begin{align}
  \Delta_R(t')=\frac{1}{4\pi i}l(R) t'+\mathcal{O}(\frac{1}{N}) .
\end{align}
This determines the large $N$ spectrum which defines the WZW model propagator, and the associated normalized Ishibashi state \eqref{Ishi}. Notice that we had to introduce $UV$ regulator $\epsilon$ to define this propagator, which regularizes the trace over the CFT edge modes.   This is the CFT analogue of the string theory regulator $\epsilon_{\text{string}}$.

Given these definitions we can factorize each state $\ket{R_{CS}}$ as in \eqref{Ishi} for the finite N case.   Applying this factorization to $\ket{HH_{CS}(t) }$ and $\bra{HH_{CS}(t)^*} $ and doing the partial trace gives the reduced density matrix 
\begin{align}
    \rho_{A}&= \sum_{R} p(R) \rho^{R}_{1} \otimes \rho^{\bar{R}}_{2} ,\quad 
    p(R) = \frac{ d_{q}(R)^{2} e^{ - t l(R) } }{Z} \nn
    \rho_{1}^{R} &= \frac{ e^{ \frac{ - 8 \pi \epsilon}{l} (L_{0}-\frac{c}{24})  }}{\chi_{R}(e^{ \frac{ - 8 \pi \epsilon}{l}) }} ,\quad \rho_{2}^{R} = \frac{ e^{ \frac{ - 8 \pi \epsilon}{l} (\bar{L}_{0}-\frac{c}{24})  }}{\chi_{\bar{R}}(e^{ \frac{ - 8 \pi \epsilon}{l}) }}
\end{align}

where $Z$ is the resolved conifold partition function.   
As in the finite $N$ case the density matrix in the fixed $R$ sector is maximally mixed when we take $\epsilon \to 0$, and the entanglement entropy in each sector just computes the $\log $ of the degeneracy.   The entanglement Hamiltonian is then given by 
\begin{align}
    H_{A}= t l(R)
\end{align}
which is the same as in the string theory 

We can obtain this degeneracy factor for fixed $R$ by first keeping $N$ and $t'$ finite and taking the $\epsilon \to 0$.  In this limit, we can  formally write 
\begin{align} \label{Red}
    \rho_{A} = \sum_{R} p(R)  \frac{ e^{ \frac{ - 8 \pi \epsilon}{l} (L_{0}+\bar{L}_{0}-\frac{c}{12})  }}{|\chi_{R}(e^{ \frac{ - 8 \pi \epsilon}{l}) }|^{2}} 
    \sim \sum_{R} p(R) \frac{ \mathbf{1}_{R\otimes \bar{R}}} {|\dim_{q}R|^{2} |S_{00}  e^{\frac{\pi c l}{48  \epsilon}} |^{2}}
\end{align}

If we now take the large $N$ limit of the quantum dimensions using \eqref{limit}, then
\begin{align}
     \rho_{A} \sim \sum_{R} p(R) \frac{ \mathbf{1}_{R\otimes \bar{R}}} {d_{q}(R)^{2} |S_{00} e^{\frac{\pi c l}{48  \epsilon}} |^{2}}
\end{align}
which shows that after the vacuum subtraction there is an effective degeneracy $d_{q}(R)^2$ just as in the string theory.  Note that in obtaining this degeneracy, it was crucial to take into account the opposite chiralities of the two edges, which have complex conjugate characters. 
Explicitly, using this order of limits the entanglement entropy of $\ket{HH_{CS}}$ is given by 

\begin{align}
   S&= \lim_{t' \to \infty} \lim_{N \to \infty} \sum_{R} ( - p(R) \ln p(R))+  p(R) \ln{|\dim_q R|^{2} } +\frac{\pi c l}{12 \epsilon}+2\ln S_{00} \nn
   &=  \sum_{R} ( - p(R) \ln p(R))+  p(R) \ln{(d_{q} (R))^{2} } +\frac{\pi c l}{12 \epsilon}+2\ln S_{00}
\end{align}
If we subtract the ``extremal entropy", which is the ground state entanglement entropy in the absence of Wilson loops:
\be \label{ext} 
S_{ext}=\frac{\pi c l}{12 \epsilon}+2\ln S_{00}
\ee
we obtain the defect entropy 
\begin{align}
   S_{\text{defect}}&= S-S_{ext}=\sum_{R} ( - p(R) \ln p(R))+  p(R) \ln{(d_{q} (R))^{2} }
\end{align}
which again agrees with the q-deformed entropy of the closed string theory.

In Chern-Simons theory, the area term  in \eqref{ext} originates from UV divergences in field theories, which can also be obtained from careful treatment of the replica trick calculation \cite{Fliss:2020cos}.  The area term is important when applying CS theory as a low energy effective field theory,  since it is required for the positivity of the entanglement entropy. However, our definition of generalized entropy in string theory does not include this term, since we are only capturing the entanglement purely due to cutting the worldsheets, which is dual to cutting the Wilson loops.  As a result we obtain a manifestly positive entropy without including the area term.

\section{Large N expansion of Wilson loops and dual string worldsheets}
In the previous section we applied a large $N$ limit to the state \eqref{putative} in Chern-Simons theory, and showed that its factorization leads to an entanglement entropy consistent with the q-deformed entropy of the Hartle-Hawking state in the closed string theory.  This bolsters our claim that the q-deformed entropy should be viewed as the topological string analogue of generalized entropy in AdS/CFT.   Here we will explain the matching of the states and the dual entropy calculations from the point of view of the large N duality between Wilson loops and string worldsheets \cite{Ooguri:1999bv,Gomis:2006mv}.  

\paragraph{Toric diagrams and geometric transitions.}
Toric diagrams provide a useful representation for topological string amplitudes which gives a precise description of the geometric transition between the resolved and deformed conifold.  They capture the duality in the presence of branes in a simple graphical language.   We give a very brief description here and defer a more detailed explanation to the Appendix. 

Toric manifolds such as the resolved and deformed conifold can be characterized as a $T^{2} \times \mathbb{R}$ fibration over $\Bbb{R}^3.$   The toric diagrams specify the degeneracy locus of this fibration where a cycle of $T^2$ shrinks.  It turns out this locus lives in a $R^2$ subspace of the base, and we can specify this locus by edges on a graph. The orientation of the edges determines which cycle degenerates on $T^2$.

For example $\mathbb{C}^3$ is given by a trivalent graph with a single vertex as shown as the third diagram in Fig. \ref{DualHH}.   The topological vertex is given by adding branes on this graph, labelled by arrows.   In particular,the Hartle-Hawking state correspond to adding one stack of branes along an edge, as shown in the right of figure \ref{DualHH}.

Gluing two topological vertices with branes and anti branes inserted gives the resolved conifold geometry as shown in the right of  Fig. \eqref{gtrans}.
Note that the inner edge describes a sphere with Kahler modulus $t'$, which can by described by a cycle which expands from a point and then shrinks.    The deformed conifold geometry  $T^{*}S^3$ is given by the left diagram in figure \eqref{gtrans}.  The dotted line describes the base $S^3$; this can be understood via the Heegaard splitting in which $S^3$ is described as two solid torus glued together with an $S$ transformation.  The dotted line captures this geometry as a foliation of 2- tori $T^2$ which begins with a pinched $A$ cycle and ends with a pinched $B$ cycle.

In terms of toric diagrams, the geometric transition is captured precisely by the equality in \eqref{gtrans}, in which the dotted line representing the three-sphere wrapped by a large $N$ number of branes is replaced by a two sphere with flux $t'= i g_{s}N $

\begin{equation} \label{gtrans}
\tikz[baseline=.1ex]{\draw[thick](0,0.1)--(0,2)node[right]{(0,1)};
\draw[thick](0,-0.1)--(0,-2)node[right]{z=0};
\draw[thick](0,0)--(2,0)node[below]{(1,0)};
\draw[thick](0,0)--(-2,0)node[below]{z=-a};
\draw[dashed](-1,0)--(0,1)node[left]{$S^3$};
}=\tikz[baseline=.1ex]{
\draw[thick](1/2,1/2)--(1/2,3/2)node[right]{(0,1)};
\draw[thick](1/2,1/2)--(3/2,1/2)node[below]{(1,0)};
\draw[thick](-1/2,-1/2)--(0,0)node[align=left,   above]{$t'$};
\draw[thick](0,0)--(1/2,1/2);
\draw[thick](-1/2,-1/2)--(-3/2,-1/2)node[above]{(-1,0)};
\draw[thick](-1/2,-1/2)--(-1/2,-3/2)node[below]{(0,-1)};
}.
\end{equation}

\subsection{Mapping Wilson loops to worldsheets on the deformed conifold}
\paragraph{Worldsheet description at finite N} 
We first consider the worldsheet description of  $\ket{\Omega(t)}$ from the point of view of the open string theory on the deformed conifold geometry.  This is valid even at finite $N$, since Chern-Simons theory is the exact string field theory for strings on the deformed conifold. We start with the case  $t=0$.  Note that while $\ket{\Omega(0) }$ is defined on the torus, its wavefunction in the coherent state basis is given by the generating functional \eqref{loopfn} for Wilson loops on $S^3$. 

To obtain a worldsheet description of the Wilson loops, we use the Frobenious relation to change to the winding basis.
\begin{align}
    W_{\vec{k}}(U)  &:= \prod_{n} (\text{tr}U^n)^{k_{n}}
\end{align}
In this basis, $\ket{\Omega}$ is given by  
\begin{align}
    \ket{\Omega} &= S_{00} \sum_{\vec{k} } \frac{\dim_{q}(\vec{k})}{z_{\vec{k}}} \ket{\vec{k}} \nn
    \dim_{q}(\vec{k}) &= \frac{1}{S_{00}}\braket{ W_{\vec{k}} }_{S^{3}} ,
\end{align}
where 
\begin{align}
\langle{\vec{k}} | \vec{k}' \rangle&=\delta_{\vec{k},\vec{k}'} z_{\vec{k}} \nn
z_{\vec{k}}&=\Pi_{j} k_j! j^{k_j}
\end{align}
The combinatorial factor $z_{\vec{k}}$ reflects the different ways to glue together the Wilson loops in the bra and ket state. 

The generating functional \eqref{loopfn} is given by 
\begin{align}\label{zvk}
   Z(V)&=\int \mathcal{D}A e^{i S_{CS}(A)+\sum_n \frac{1}{n}\tr U^n \tr V^n} \nn
&= \sum_{\vec{k}} \frac{1}{z_{\vec{k}}} \braket{ W_{\vec{k}}(U)}_{S^3} \braket{V| \vec{k}},
\end{align}

where we applied the identity
\begin{align}
    \exp \left( \sum_{n=1}^{\infty} \frac{1}{n}\tr U^n \tr V^n\right) =  1 + \sum_{\vec{k}}  \frac{1}{z_{\vec{k}}} W_{\vec{k}}(U) W_{\vec{k}}(V),
\end{align}
used $U$ to denote the holonomy of the dynamical gauge field $A$, and $V$  to denote the holonomy of the source.  We used the notation $\braket{V|k}$ instead of $W_{\vec{k}}(V)$ in \eqref{zvk} to distinguish the source Wilson loop, which should be viewed as a state  in  $\mathcal{H}(T^2) $.
 \begin{figure}
\centering
\includegraphics[scale=.4]{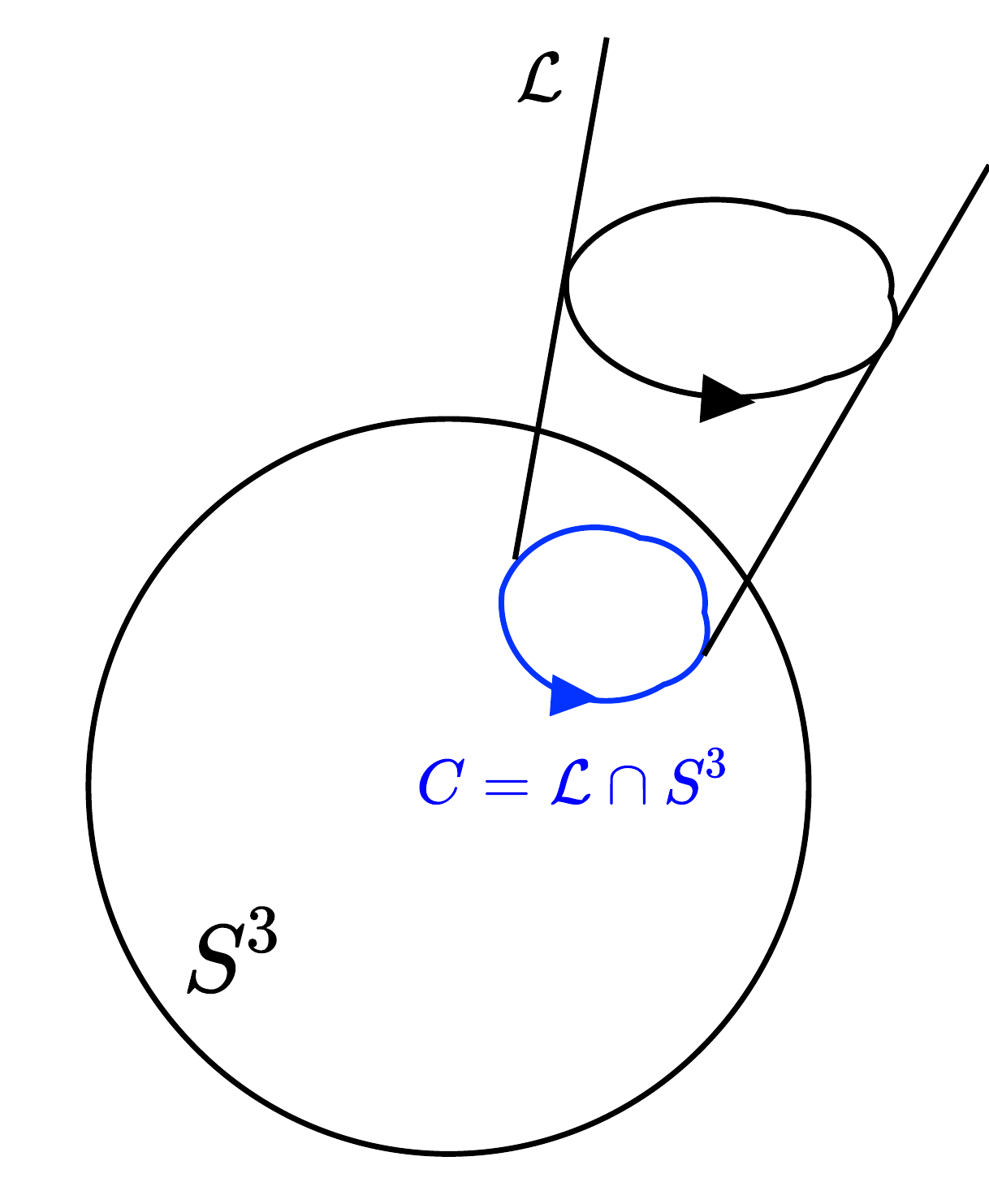}
\caption{The generating functional for Wilson loops can be viewed as the open string amplitude on the deformed conifold with dynamical branes wrapping $S^3$ and probe branes wrapping $\mathcal{L}$.  They intersect along a knot $C$ on $S^3$,colored in blue. } \label{IntBrane}
\end{figure}  

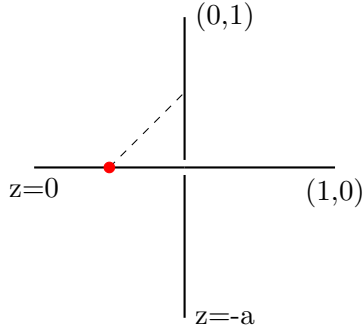
\begin{figure}[h]
\centering
\begin{tikzpicture}
\draw[thick](0,0.1)--(0,2)node[right]{(0,1)};
\draw[thick](0,-0.1)--(0,-2)node[right]{z=-a};
\draw[thick](0,0)--(2,0)node[below]{(1,0)};
\draw[thick](0,0)--(-2,0)node[below]{z=0};
\draw[dashed](-1,0)--(0,1);
\filldraw[red] (-1,0) circle (2pt);
\end{tikzpicture}
\caption{Toric diagram for the deformed conifold with probe D-branes on $\mathcal{L}$ intersecting the $S^3.$ For the simplicitiy, we have not specified the frame of the D-brane and the probe brane is depicted as a red dot.}
\label{toric:deformed conifold2}
\end{figure}

As we have noted earlier,  $Z(V) =\braket{V|\Omega} $ is the coherent state wavefunction for $\ket{\Omega}$.  When expressed in the winding basis,  each term in \eqref{zvk}  labelled by $\vec{k}$ has a string theory interpretation on the deformed conifold in terms of open string worldsheets ending on a configuration of intersecting D branes. As shown in Fig. \ref{IntBrane}, this configuration consists of non-compact, probe branes on a Lagrangian submanifold  $\mathcal{L}$ which intersects a large N number of \emph{dynamical } branes on $S^{3}$ along the knot $C$ \cite{Ooguri:1999bv}. It was shown in \cite{Ooguri:1999bv} that $Z(V)$ is the spacetime effective field theory obtained by integrating out the strings ending on the intersection of these branes. 
More precisely, each term in \eqref{zvk} labeled by $\vec{k}$ corresponds to open string worldsheets that end on the intersection of the D branes, with one set of boundaries on dynamical branes coupled to the holonomy $U$, and the other set of boundaries on probe branes with holonomy $V$.  The winding pattern $\vec{k}$ of the Wilson loop variables is identified with the winding of the open string endpoint around the knot $C$.   The toric diagram for $Z(V,t)$ is given in Fig. \ref{toric:deformed conifold2}.

The worldsheets stretched between the two sets of branes with winding numbers $\vec{k}$ should be viewed as worldsheet instantons, i.e. a classical backgrounds in the string path integral \cite{Gomis:2006mv}.  When quantizing strings around these backgrounds there is a sector of open string worldsheets living on $S^3$ which ends on the winding boundary of the worldsheet instanton.   These can be identified with the ribbon diagrams of the Chern-Simons path integral which produces the expectation value  $\braket{ W_{\vec{k}}(U)}_{S^3} = S_{00} \dim_{q}(\vec{k}) $.   These ribbon diagrams can be seen seen explicitly by expanding $\dim_{q}(k)$ in small string coupling
\begin{align}
g_{s} =\frac{ 2 \pi  }{k+N}   .    
\end{align}
This gives
\begin{align}\label{dimqk}
\dim_{q}(\vec{k})&= \prod_j \left(\frac{ \sin((j N g_s/2))}{ \sin(j g_s/2)}\right)^{k_j}=\Pi_j (\frac{q^{j N/2}-q^{-j N/2}}{q^{j /2}-q^{-j /2}})^{ k_j} \nn
&\to \prod_{j} \left( N+\frac{j^2 g_s^2}{24} (N-N^3)+\mathcal{O}(g_s^4)\right)^{k_{j}} 
\end{align} 
At zero string coupling, there are no interactions between the instanton worldsheets and the ribbon diagrams, so we just get a factor of $N$ per boundary due to the Chan Paton factors running in a loop.  However turning on the string coupling introduces corrections where the fatgraphs interact with the winding boundary of the instanton. 
For example, the first subleading term proportional to $N-N^3$ comes from the well known ``theta" diagrams. Remarkably the all order corrections sum up into a $q^{j}$ deformed number:
\begin{align}
    [N]_{q^j}=\frac{q^{j N/2}-q^{-j N/2}}{q^{j /2}-q^{-j /2}}
\end{align}  This shows explicitly how the open string interactions obey a hidden quantum group symmetry which dictates the final form of the target space amplitude.  We have performed a series expansion in $g_{s}$ to make this open string interactions explicit, but it is important to observe that the series has to be summed to obtain the quantum group symmetry. 

The factor $\braket{V| \vec{k}}$ in \eqref{zvk} arises from the opposite boundaries of the worldsheet instantons that end on the probe branes on $\mathcal{L}$.  Since these branes are non-compact, the worldvolume gauge field is non-dynmaical and  $\braket{V| \vec{k}}$ comes from the coupling of the worldsheet boundary to the background gauge field.  These are identified with the coherent state wavefunctionals of the Chern-Simons theory. 
 
\begin{figure}[h]
\centering
\begin{tikzpicture}
\draw[thick](0,0.1)--(0,2)node[right]{(0,1)};
\draw[thick](0,-0.1)--(0,-2)node[right]{z=0};
\draw[thick](0,0)--(2,0)node[below]{(1,0)};
\draw[thick](0,0)--(-2,0)node[below]{z=-a};
\filldraw[red] (1,0) circle (2pt);
\draw[dashed](-1,0)--(0,1);
\draw[red] (-1,0.3) .. controls (-0.2,0) and (0.2,0) .. (1,0.3);
\draw[red] (-1,-0.3) .. controls (-0.2,-0) and (0.2,-0) .. (1,-0.3);
\draw[red] (-1,0) ellipse (0.2 and 0.3);
\draw[red] (1,0) ellipse (0.2 and 0.3);
\end{tikzpicture}
\caption{Toric diagram for the deformed conifold with M D-branes on $\mathcal{M}$. Note that we have not specified the frame of the probe D-brane on $\mathcal{M}.$}
\label{toric:deformed conifold3}
\end{figure}
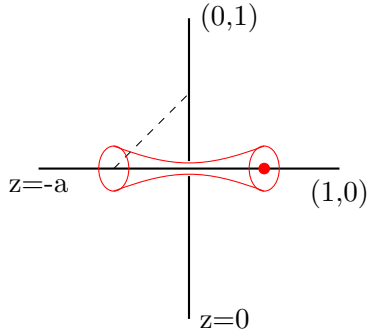

\begin{figure} [h]
\centering
\includegraphics[scale=.3]{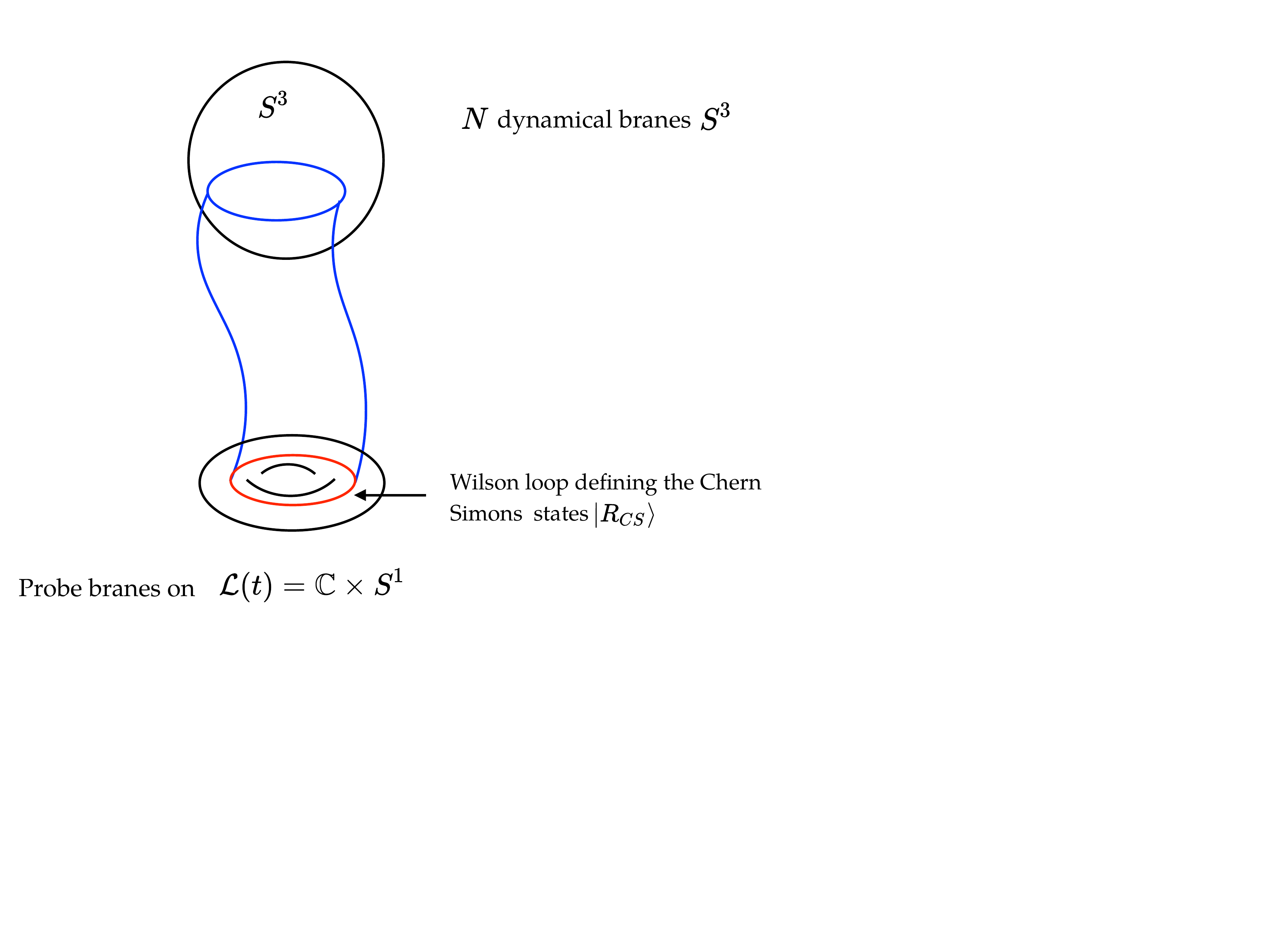}
\caption{Displacing the probe branes away from $S^3$ gives rise to open string instantons of finite area, stretched between the branes.   } 
\label{stretch}
\end{figure}

The worldsheet description given above generalizes to the state $\ket{\Omega(t)}$.   Making $t$ non zero corresponds to displacing the probe branes away from $S^3$, so that the stretched worldsheet instantons now have (complexified) area $t$.   In terms of the winding basis, the wavefunctional for $\ket{\Omega(t)}$ is
\begin{align}
    \braket{V|\Omega(t)} &= Z(V,t)\nn
    &= S_{00} \sum_{\vec{k}}
\dim_{q}(\vec{k}) e^{-tl(\vec{k}) /2}  \prod_{n} \tr (V ^n)^{k_{n}},
\end{align}
where the Boltzman factors $e^{-tl(k) /2}  $ originates from the exponential of the worldsheet action for the stretched worldsheets. Fig. \ref{toric:deformed conifold3} shows the toric diagram for $\braket{V|\Omega(t)}$.
\paragraph{ The large $N$, $t'$  limit and shift of the the worldsheet area}
Just as in the representation basis, the large $N$ and $t'$ limit of $\dim_{q}(k)$ has a divergent factor
\be
\dim_q(\vec{k}) \to q^{N l(\vec{k})/2} (-i)^{\sum_j k_j} d_q(\vec{k})+\mathcal{O}(q^{-N l(\vec{k}) /2})
\ee
which should be absorbed into a shift of the coupling $t$
\begin{align}
e^{- t l(\vec{k}) /2} \to  e^{- (t+t') l(\vec{k}) /2} 
\end{align}  
In terms of the string theory, the shift is interpreted as a modification of the worldsheet area of the stretched instantons to account for a nontrivial $B$ field flux.  
Applying  this limit, we can write
\begin{align}
\lim_{t'\to \infty}  \lim_{N\to \infty}  \braket{V|\Omega(t+t')}&=S_{00}\sum_{\vec{k}}
q^{N l(\vec{k})/2} (-i)^{\sum_j k_j} d_q(\vec{k}) e^{-(t+t') l(\vec{k})/2}\braket{V|k_{CS}} \nn  
&= S_{00} \sum_{\vec{k}}
 (-i)^{\sum_j k_j} d_q(\vec{k}) e^{-t l(\vec{k})/2} \braket{V|k_{CS}} 
\end{align}
The worldsheet description of each Wilson loop insertion in this wavefunction is given in Fig. \ref{stretch}.   

\subsection{Applying the geometric transition to the resolved conifold geometry}
\begin{figure}[h] 
\centering
\includegraphics[scale=.5]{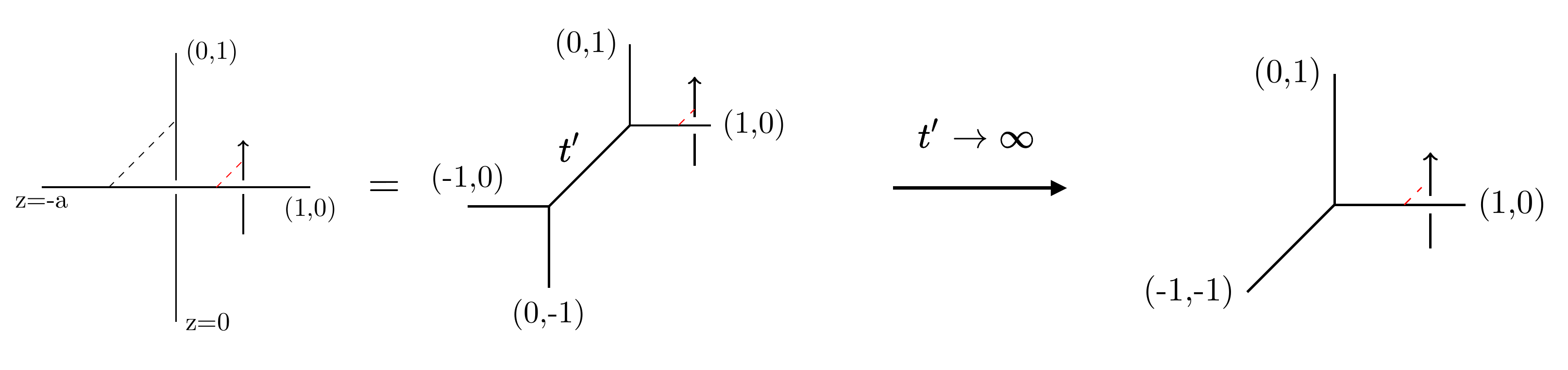}
\caption{The toric diagrams show the result of the geometric transition applied to the deformed reosolved with probe branes displaced from $S^3$.  Under the transition the $S^3$ is replaced by a sphere of size $t'$.   On the right, we show the $t'$ limit in which that sphere $t'$ is sent to infinity.  This results in ``half" the resolved conifold geometry with probe branes inserted, which defines the Hartle-Hawking state. It is important to note that we have specified the framing for the probe D-brane for the HH state, which is determined by the direction of the arrow \cite{Aganagic:2003db}.}
\label{DualHH}
\end{figure}

\begin{figure}[h]
\centering
\includegraphics[scale=.4]{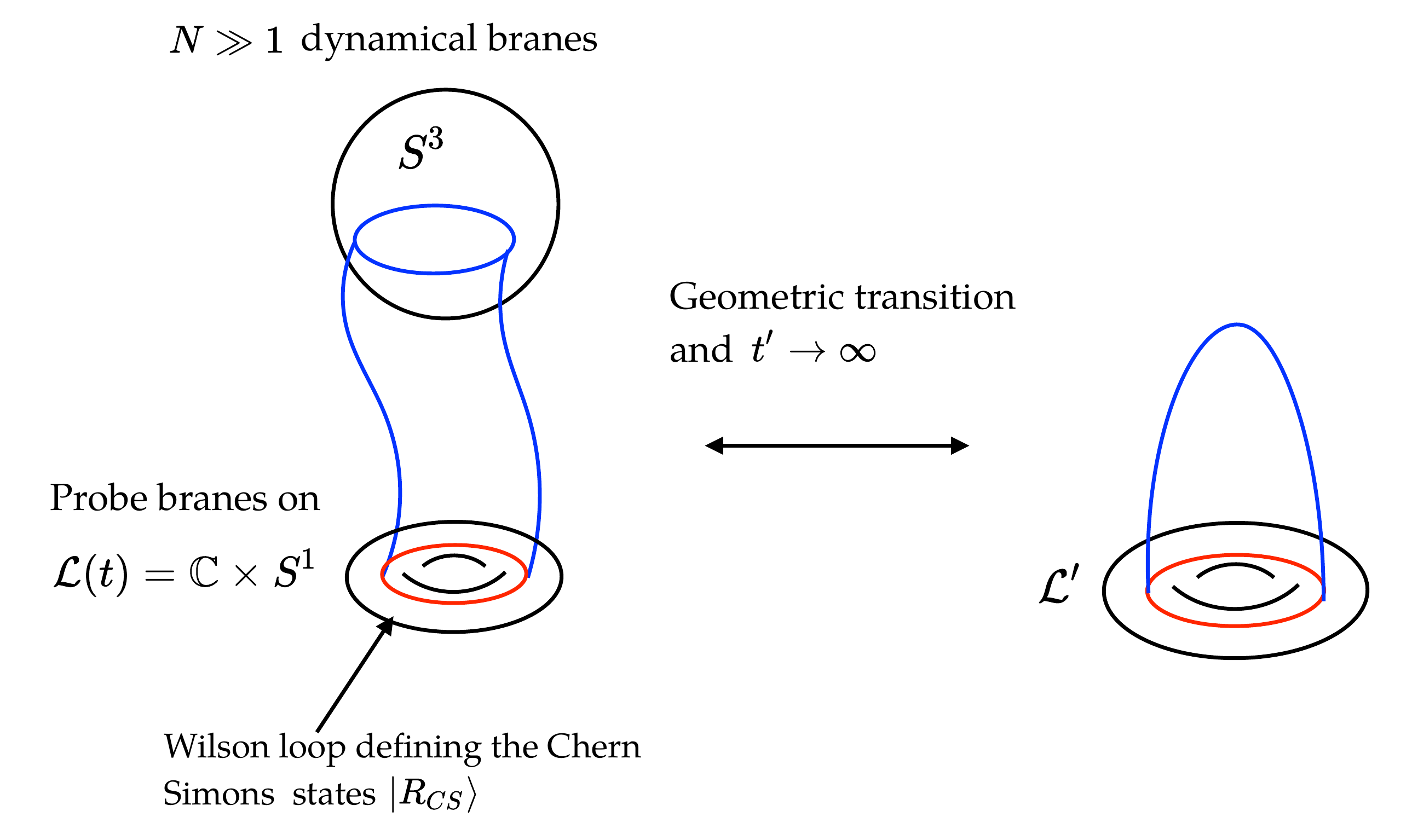}
\caption{The geometric transition, combined with the $t' \to \infty $ limit maps the state on the deformed conifold (left figure) to the Hartle-Hawking state on the resolved conifold. } \label{transition}
\end{figure}

Having described the large $N$ limit of the Chern-Simons state  $\ket{\Omega(t)}$ in terms of string theory on the deformed conifold, we now apply the geometric transition to show that it is mapped to the Hartle-Hawking state $\ket{HH(t)}$. We have to consider the geometric transition of the brane configuration in Fig. \ref{stretch}.  This is illustrated in Fig. \ref{transition};  In the large N limit,
\begin{align}
    \lim_{N \to \infty}  \ket{\Omega(t+t') }
\end{align}
the geometric transition replace the branes wrapping a three-sphere  into  B field flux $t'=i g_{s} N$ threading a two sphere  of size $t'$.   The area of the two-sphere is then ``sent to infinity" by taking the limit $t' \to  \infty $.  A precise geometric description of this limit is described in terms of toric diagrams in Fig. \ref{DualHH}.

As shown in Fig. \ref{transition},
under the geometric transition, the worldsheet boundaries ending on the $S^3$ closes up.  On the other hand, the displaced probe branes $\mathcal{L}$ on the deformed conifold are mapped to probe branes $\mathcal{L'}$ on the resolved conifold, which cut through the equator of the base $S^2$.  These are precisely the probe branes which defines the Hartle-Hawking state on the resolved conifold. We can therefore identify the Chern-Simons wavefunction
\begin{align}
  \braket{V|HH_{CS}(t)}&=  \lim_{t' \to \infty } \lim_{N \to \infty}  \ket{\Omega(t+t') } = S_{00}\sum_{\vec{k}}
 (-i)^{\sum_j k_j} d_q(\vec{k}) e^{t l(\vec{k}) /2}  \prod_{n} \tr (V ^n)^{k_{n}} 
\end{align}
with the string amplitude that defines the Hartle-Hawking state (neglecting an overall normalization).   

Similarly, the overlap $\braket{HH_{CS}(t)^*|HH_{CS}(t)}$ can be identified with the resolved conifold partition function by applying the geometric transition as shown in Fig. \ref{toricdual}, and then taking the $t' \to \infty $ limit.  As stated earlier, notice that the geometric transition is being applied to the spheres with Kahler Modulus $t'$, rather than the inner sphere with Kahlar modulus $t$.

\subsection{Dual description of the entanglement brane}

In section 2, we observed that the entanglement brane boundary state $\ket{D}$ in the closed string theory can be identified with the Hartle-Hawking state $\ket{HH(t=0) }$ when the Kahler modulus is set to zero.  Geometrically,  this boundary state describes a ``Calabi Yau cap",  which corresponds to a non-local boundary condition that leads to a q-deformation of the edge mode symmetry. Since we have identified the dual description of $\ket{HH(t=0)}$ on the deformed conifold, we can also identify the dual of the entanglement brane boundary state;  this is given by configuration of \emph{dynamical} D branes wrapping the $S^3$ and probe branes on a deformed conifold geometry.

\paragraph{E-brane boundary state and the shrinkable boundary condition on the deformed conifold geometry}

\begin{figure}[h]
\centering
\includegraphics[scale=.4]{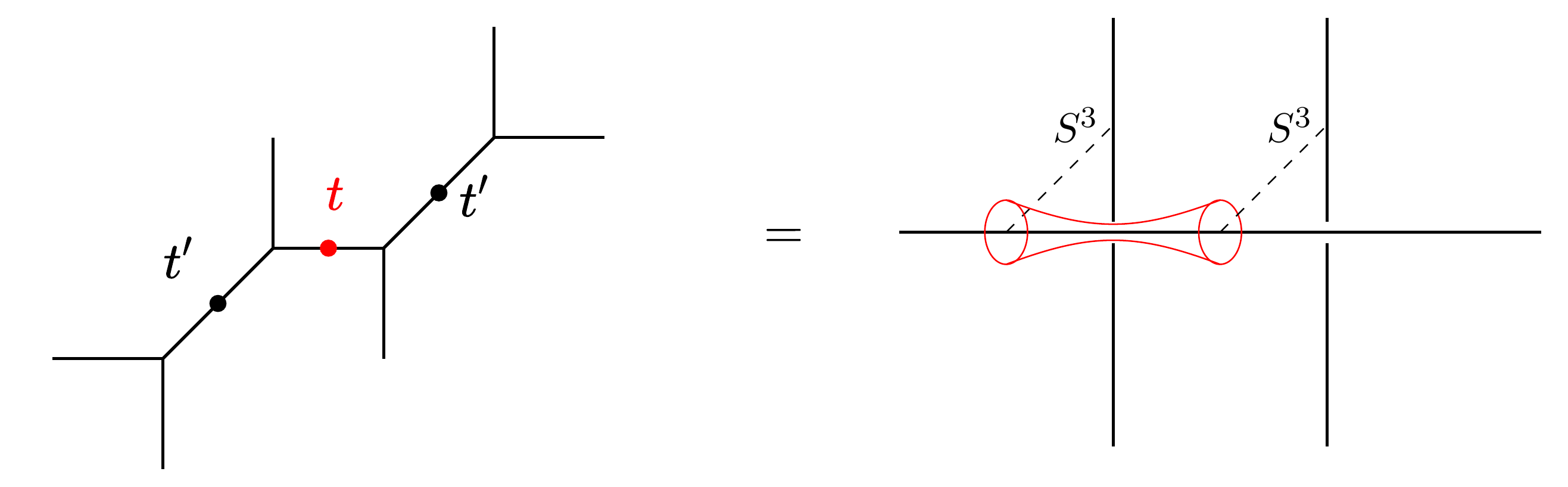}
\caption{The dual thermal interpretation of the resolved conifold partition function is obtained by first introducing two $S^2$ with Kahler paremeter $t'$ (left figure) and then applying a geometric transition to the deformed geometry on the right with branes on two $S^3$.  The worldsheets stretched between these branes describe open string loop diagrams .} \label{toricdual}
\end{figure}

We want to give a dual thermal description of the E brane boundary state $\ket{D}$ in terms of a shrinkable boundary condition on the deformed conifold geometry. This is obtained essentially by running the duality transformation described in the previous section backwards.  We begin with a closed string channel interpretation of two stacks of E branes on the resolved conifold geometry:
\begin{align}\label{DHD}
     Z_{\text{resolved}}(t)&= \braket{D^{*}|e^{-H_{\text{closed} }} |D} \nn
     &= \braket{HH^{*}(t=0)|e^{-H_{\text{closed}} } |HH(t=0)}\nn
    H_{\text{closed}}&= l(R) t
\end{align}
This is an amplitude between two E brane boundary states on the resolved conifold.  To obtain a dual thermal interpretation, we have to first introduce two $S^2$'s with Kahler parameter $t'= ig_{s} N$  as shown in the left diagram of Fig. \ref{toricdual}.  For large $t'$ this effectively introduces a constant factor of $S_{00}(t')^{2}$ into $ Z_{\text{resolved}}(t)$.   We then apply a geometric transition to obtain a deformed geometry where the fluxes are replaced by a large $N$ number of branes wrapping two $S^3$'s. In terms of the overlap \eqref{DHD}, this corresponds to mapping:
\begin{align}
     S_{00}\ket{HH(t=0)} = \ket{HH_{CS}(t=0)} \to   \lim_{N \to \infty}  \ket{\Omega(t') }\nn 
    \bra{HH^*(t=0)}S_{00}= \bra{HH_{CS}^{*}(t=0)} \to  \lim_{N \to \infty}  \bra{\Omega(t')^{*} } 
\end{align}
while keeping the same string theory Hamiltonian.    

In the previous section, we developed a string worldsheet description of $  \ket{\Omega(t')}$ in terms of configurations of D branes wrapping $S^3$ and probe branes on a Lagrangian $\mathcal{L}$.    The linear functional $\bra{\Omega(t')^{*} }$ corresponds a similar D brane configuration where we change probe branes on $\mathcal{L}$ to anti branes.   The overlap then corresponds to annhilation of the probe branes, giving the the string theory  partition function on a deformed geometry with dynamical branes on two $S^3$'s:
\begin{align} \label{Zdef}
   Z_{\text{deformed}}(t,t')&=   \lim_{N\to \infty}  \braket{\Omega(t')^{*}|e^{- H_{\text{closed}} }| \Omega(t') } \nn
   &=   \lim_{N \to \infty} \sum_{\vec{k}} \frac{1}{z_{\vec{k}}} \langle \Omega(t')^{*} | e^{-t l(\vec{k})/2 }|\vec{k} \rangle \langle \vec{k}|e^{-t l(\vec{k})/2 } | \Omega(t')  \rangle \nn
   &=  \lim_{N \to \infty}  (S_{00}(t'))^{2} \sum_{\vec{k}} \frac{1}{z_{\vec{k}}}  \dim_{q}(\vec{k}) \overline{\dim_{q}(\vec{k})} e^{-t l(\vec{k}) }
\end{align}

\begin{figure}[h]
\centering
\includegraphics[scale=.4]{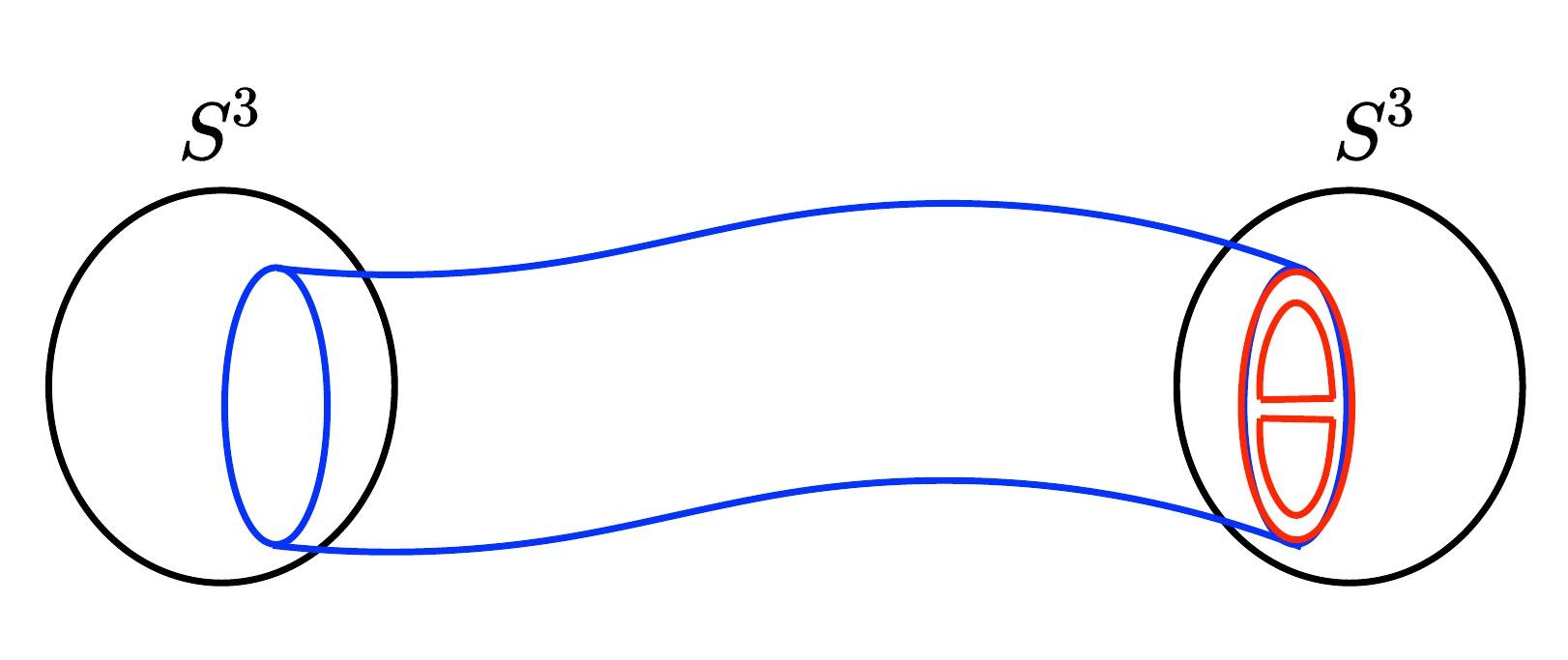}
\caption{ Annihilation of the probe branes gives rise to the partition function $ Z_{\text{deformed}}(t,t')$, which describes world worldsheet instantons stretched between D branes on the two $S^3$'s. The quantum dimensions which lead to the q-deformation of the edge modes arise from the ribbon diagrams (red) interacting with the boundary of the instanton worldsheets.  We have shown one term in the ribbon diagram expansion corresponding to the ``theta" diagram. } \label{thermal}
\end{figure}

Here we have identified the coupling of the closed strings to the entanglement boundary state to be $\braket{\vec{k}|\Omega(t')} = S_{00} \dim_{q}(\vec{k})$.     As shown in Fig. \ref{thermal}  each term labelled by $\vec{k}$ describes nondegenerate worldsheet instantons stretched between dynamical branes on the two $S^3$. 
In the open string channel, these are viewed as loop diagrams describing a thermal ensemble of open strings.    Thus, in the large $N$ limit, the D-branes on $S^3$'s give the shrinkable boundary condition in the dual geometry.   Note that the shrinkable boundary condition in the deformed geometry is \emph{local} in the sense that D-branes define local boundary conditions on the worldsheet BCFT. 

Let's consider in more detail what these worldsheets look like at finite $N$ in order to understand the worldsheet description of the shrinkable boundary condition.   We thus consider the overlap
\begin{align} \label{finiteN}
\sum_{\vec{k}} \frac{1}{z_{\vec{k}}} \langle \Omega^{*}(t') | e^{-t l(\vec{k})/2 }|\vec{k} \rangle \langle \vec{k}|e^{-t l(\vec{k})/2 } | \Omega(t')  \rangle = (S_{00})^{2} \sum_{\vec{k}} \frac{1}{z_{\vec{k}}} \dim_{q}(\vec{k}) \overline{\dim_{q}(\vec{k})} e^{-t l(\vec{k}) },
\end{align} 
where the quantum dimensions should be viewed as a function of the open string parameters:
\begin{align}
    \dim_{q}(\vec{k})= \dim_{q}(\vec{k})(N, g_{s}) 
 \end{align}
As we saw in eq \eqref{dimqk}, by expanding $\dim_{q}(k)$ in small $g_{s}$, we obtain the ribbon diagrams of the worldvolume Chern-Simons theory on the D branes which arise from the quantization of the open strings in the background of the instantons.   Thus, the thermal open string partition function describes a modification of the usual  one-loop diagrams in which the boundary of the winding worldsheet instantons interact with ribbon diagrams living on the $D$ branes (see Figure \ref{thermal} ).     These interactions on the branes are crucial to the shrinkability of the D brane boundary condition, since it is the full summation over these ribbon diagrams which leads to a q-deformation of the usual $N^2$ degeneracy factor for one-loop open strings
\begin{align}
N^{2} \to ([N]_{q_{j}})^{2} .
\end{align} 
Here $j$ is the winding number of the worldsheet instanton around the thermal circle.   This q-deformation is needed to reproduce wavefunctions of the E brane boundary state, and is therefore essential to the shrinkability condition. 

As we noted earlier, applying the large $N$ geometric transition and sending $t' \to \infty $ closes up the holes on the open string worldsheets due to the D branes, recovering the closed string worldsheets on the resolved conifold.    The sigma model description of this transition is well known.    The novelty of this set up is that these D branes are related to the entanglement branes which are inserted to define an entanglement cut in the string theory. 

\paragraph{Replica trick and the thermal partition function for open strings}
An open string Hilbert space description of  $Z_{\text{deformed}}(t,t')$ can be given explicitly in term of the unnormalized reduced density matrix of the large $N$ Chern-Simons theory \eqref{Red}:
\begin{align}
     Z_{\text{deformed}}(t,t')&= \tr_{A} \tilde{\rho}_{A} \nn 
     \tilde{\rho}_{A}&= \lim_{\epsilon \to 0}  \sum_{R} (S_{00}(t'))^{2}  d_{q}(R))^{2} e^{-t l(R)}  \frac{ e^{ \frac{ - 8 \pi \epsilon}{l} (L_{0}+\bar{L}_{0}-\frac{c}{12})  }}{|\chi_{R}(e^{ \frac{ - 8 \pi \epsilon}{l}) }|^{2}}    \nn
    &\sim   \sum_{R} e^{-t l(R)}  1_{R\otimes \bar{R}} 
\end{align}
In the last expression, have denoted by $1_{R\otimes \bar{R}}$ a maximally mixed state in the sector $R\otimes \bar{R}$ , which has a degeneracy factor of 
\begin{align}
\chi_{R} \chi_{\bar{R}}  \to  (d_{q}(R))^{2} S_{00}^{2} (t') e^{ \frac{cl}{\epsilon}} 
\end{align}
as $\epsilon \to 0$. 
 Notice also that $\tilde{\rho}_{A}$ can be identified with the reduced density of open strings on the deformed conifold, obtained from cutting the stretched open string instantons defining $\ket{\Omega}$.    This is because we have previously identified the worldsheet instanton description of the wavefunction $ \braket{V|\Omega}=Z(V)$, so we can lift the entanglement cut of Wilson loops in $\ket{\Omega} $ directly to the entanglement cut of the worldsheet instantons.  
\paragraph{The $S_{00}^{2} $ factor }
In the Chern-Simons theory description $Z_{\text{deformed}}$ , the $S_{00}(t')^{2} $ factor arises from the measure in the path integral, and $t'$ is viewed as the 't Hooft parameter. This constant sets the normalization of the partition functions \footnote{Interestingly, this choice matches the normalization chosen for the q-deformed Yang Mills theory \cite{2005NuPhB.715..304A}}. In the string theory description, this constant has a geometric origin. It comes from the two $S^3$'s in the deformed conifold geometry, and two $S^2$ in the resolved geometry after the transition. 

It should be noted that multiplying the partition function or the reduced density matrix by a constant has no physical consequences and does not change the entanglement entropy.  We see this explicitly in the replica trick, which computes the ratio
\begin{align}
\frac{Z(n)}{Z(1)^n} = \frac{ \tr_{A} \tilde{\rho}^{n}_{A}} { (\tr_{A} \tilde{\rho}_{A})^{n}} 
\end{align}
so any rescaling of  the un-normalized reduced density matrix $\tilde{\rho}_{A}$ would cancel.  However, a change in the measure for the path integral does change the entanglement entropy, because the measure is \emph{not} replicated in $Z(n)$.  This is why we obtain an extra $\log S_{00}(t')^{2}$  in the entanglement entropy computed in Chern Simons.

In the string theory we can understand the distinction between an overall constant and a choice of measure geometrically.   When we replicate the deformed geometry in figure  \ref{2spheres}, we do \emph{not} duplicate the $S^3$'s.  Instead the replica manifold is obtained simply by rescaling the Kahler modulus $t$ by a factor of $n$, which rescales area of the stretched worldsheet instantons by the same factor.   This is the string theory analogue of treating  $S_{00}(t')^{2}$ as a measure rather than an overall constant, and leads to the $\log S_{00}(t')^{2}$ term in the entanglement entropy.  Notice that not duplicating the $S^3$'s also manifestly preserves the Calabi Yau condition, consistent with the constraint imposed on the dual replica manifold for the resolved conifold geometry.   

Now, we explain why we added an additional layer to the story by including and subtracting the contributions from $S^{3}$. In the original closed string calculation,  we computed the generalized entropy due to cutting non-degenerate worldsheet instantons with finite area.  For the Hartle Hawking state corresponding to "half"  of the resolved conifold, these instantons wrap half of the minimal volume $S^2$ and end on probe branes intersecting the equator.  However there is no way to cut the deformed conifold  $T^{*}S^{3}$  in half and also obtain non-degenerate instantons.    To obtain a dual description on a deformed geometry with non-degenerate instantons, we had to first introduce the $S^3$ to allow strings that stretch between them and then subtract the contribution from ribbon diagrams that do not connect to the instantons. These give precisely the $S_{00}(t')^{2}$ we subtracted.

%The aim of this paper is to calculate the extra entanglement entropy due to the inclusion of Wilson loops. However, we first computed the entanglement entropy of the Chern-Simons theory on $S^3$ in which the Wilson loop is inserted. Then later, we subtracted the background contribution of the entropy $2\ln (S_{00}),$ rather than computing the entropy of the Wilson loops in isolation. The reason for considering Wilson loops on $S^3$ is as follows. The very building block of the computation we used extensively in Chern-Simons theory is the generating functional or the HH state. Without the inclusion of the $S^3$ background, the generating functional $Z(V)$ hardly makes any sense. As a result, the entanglement entropy computation in the absence of the $S^3$ background is vacuous \cite{2013arXiv1307.1132J}. Hence, we first computed the EE as a whole and subtracted the background contribution. From the string theory point of view, the necessity of the background $S^3$'s is also manifest because without more than one stack of D-branes one does not have non-trivial non-degenerate instantons. 

\section{Discussion}
\begin{figure}[h]
    \centering
    \subfloat{{\includegraphics[width=7cm]{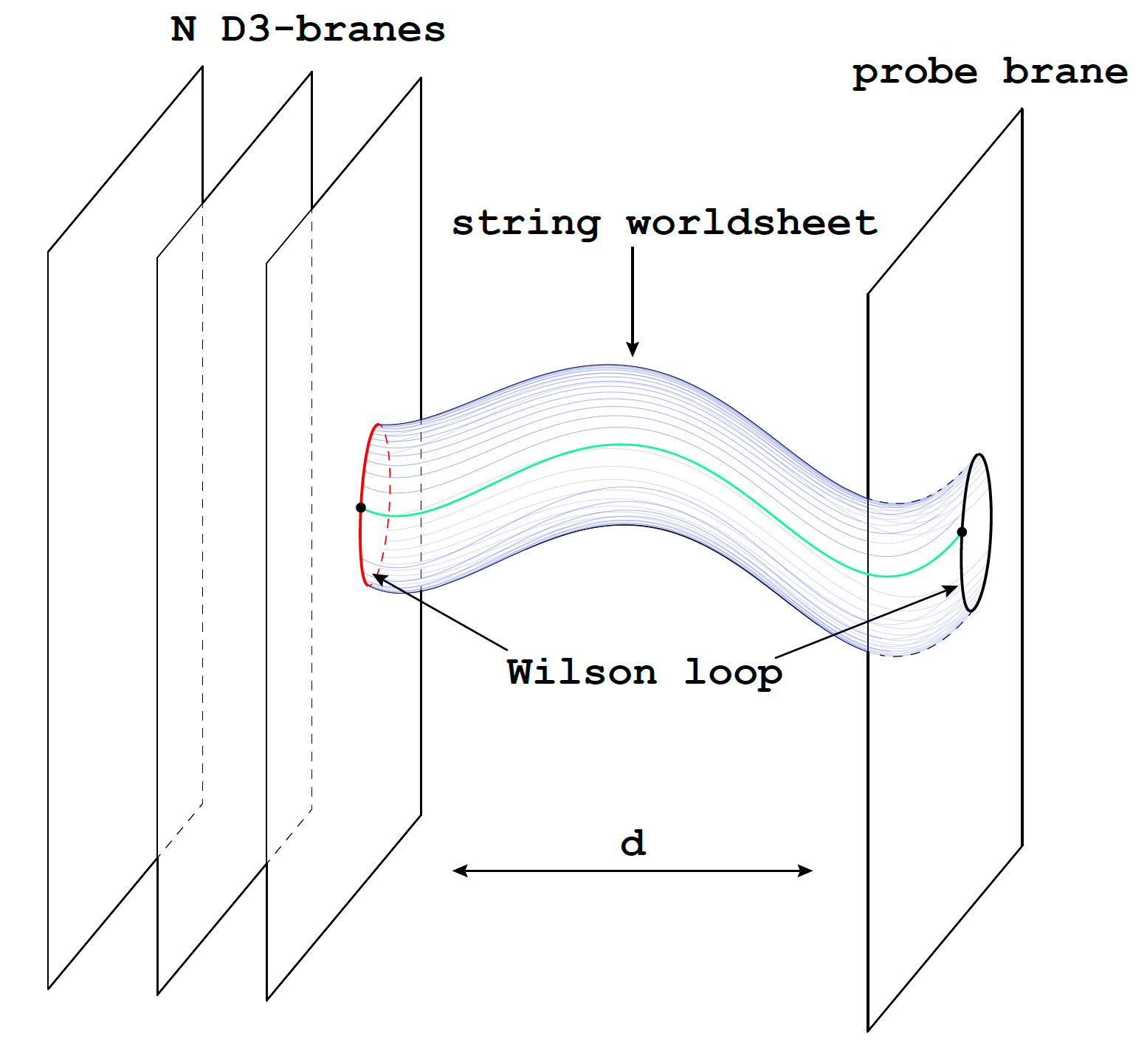} }}%
    \qquad
    \subfloat{{\includegraphics[width=7cm]{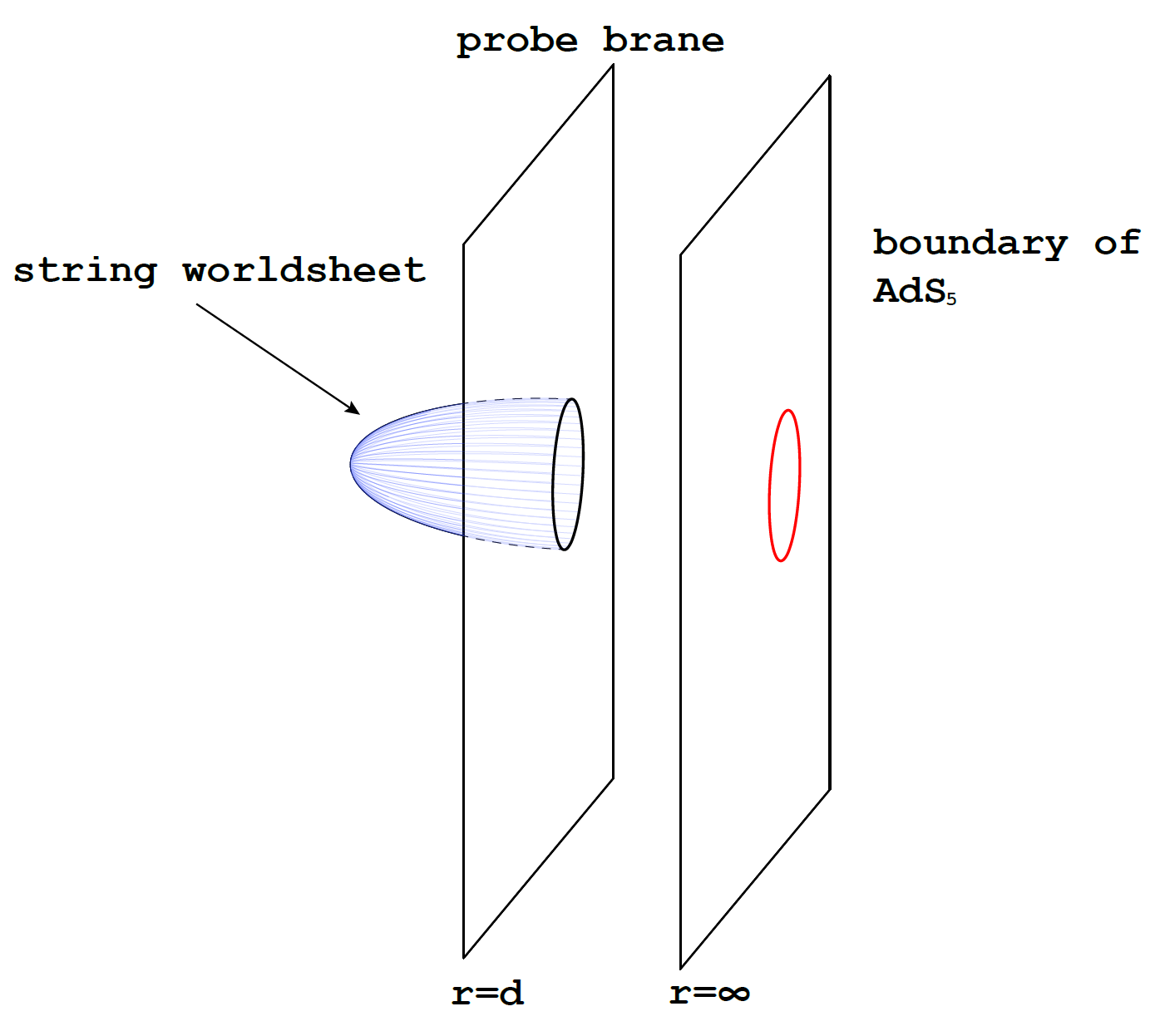} }}%
    \caption{Duality between Wilson loop and worldsheet in AdS/CFT. On the left figure, we showed the open string frame, where displacing a probe brane away from a stack of D branes leads to stretched worldsheets ending on Wilson loops.   On the right we applied a geometric transition to obtain the closed string $AdS \times S^5$ geometry where the worldsheet has only one boundary ending on the probe brane.  This is a direct analogue of the duality between Wilson loops and worldsheets in topological string theory  }%
    \label{fig:example1}
\end{figure}

\begin{figure}[h]
    \centering
    \subfloat{{\includegraphics[width=7cm]{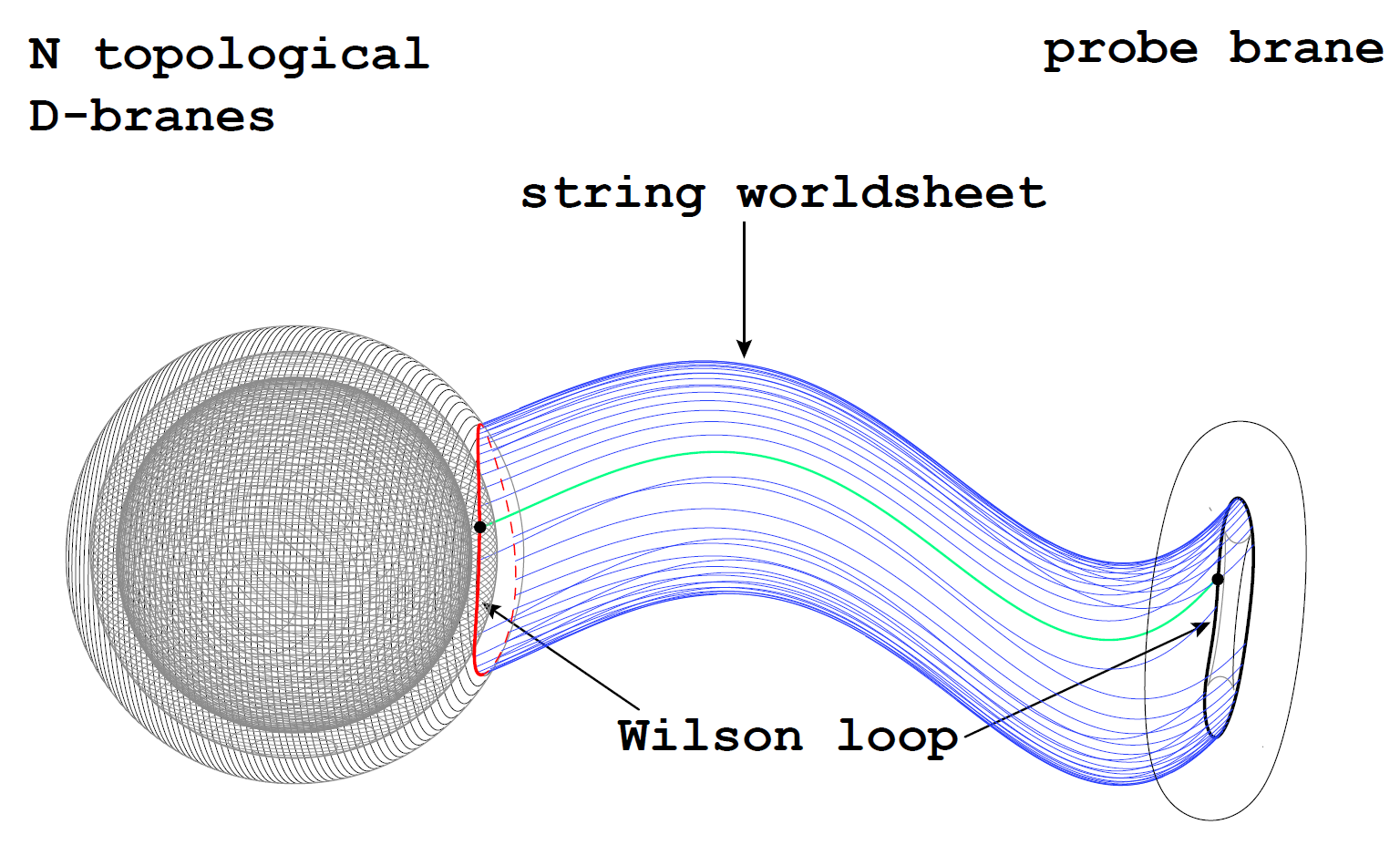} }}%
    \qquad
    \subfloat{{\includegraphics[width=7cm]{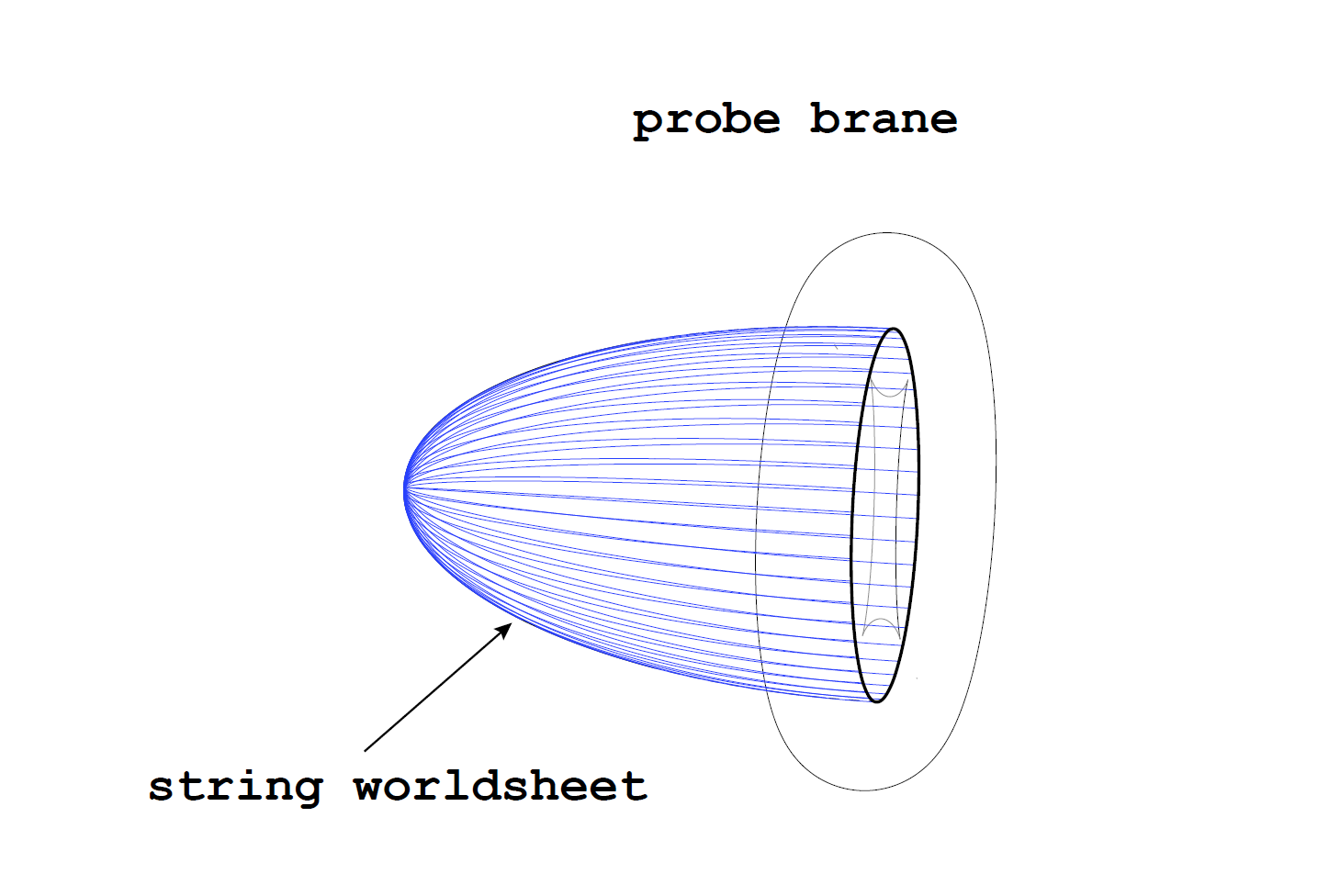} }}%
    \caption{Duality between Wilson loop and worldsheet in topological string. There is an $S^3$ at asymptotic infinity that has been omitted from the picture. \cite{2002math......1219T, Gomis:2006mv}} %
    \label{fig:example2}
\end{figure}

In this work we provided a dual gauge theory description of generalized entropy for closed topological strings on the resolved conifold.  This was obtained by applying a geometric transition to the brane configurations for Hartle Hawking state,  as shown \ref{fig:example2}. The duality map on the branes has a direct analogue in AdS/CFT, where the anyons are replaced by heavy quarks( see figure  \ref{fig:example1} ) .

We showed that the non-local shrinkable boundary condition in the bulk geometry and the associated quantum group edge mode symmetry are mapped to a local boundary condition in the gauge theory and CFT edge modes that transform under a large $N$ Kacs-Moody symmetry.   In the same spirit, the q-deformed entropy that arises from cutting the bulk string worldsheets is mapped to the un-deformed defect entropy of Wilson loops in the boundary gauge theory.  

We can summarize these results by saying that Gopakumar-Vafa duality provides a geometric interpretation of the \emph{measure} on the entanglement brane edge modes as defined by the Drinfeld element via equation \eqref{Du}.   This is captured most concisely by the toric diagrams in figure \ref{toricdual}, in which the quantum trace 
\begin{align}\label{ResTr} 
Z_{\text{res}} = \tr(D e^{-H}) = \mathtikz{\pairA{0}{0}\copairA{0}{0}\draw (0cm,0cm) node {\footnotesize $e$};\draw (0cm,1cm) node {\footnotesize $e$};}
 \end{align}
over entanglement branes on the resolved conifold is reproduced by introducing fluxes on the resolved geometry and turning them into branes on a deformed geometry.  The open strings instantons stretched between these branes determines a thermal partition function which agrees with the partition sum in \eqref{ResTr}.    This provides a relation between the categorical description of entanglement brane edge modes defined in \cite{2020arXiv201015737D} and the worldsheet description of topological D branes of the A model string theory. 

Our ultimate motivation for studying entanglement in topological string theory was to understand entanglement in bulk quantum gravity in AdS/CFT.  It is thus natural to ask what features of entanglement in topological string theory is expected to generalize to the physical string.  The general picture of string entanglement which we have developed in this two-part paper (as well as in \cite{Donnelly:2016jet,Donnelly:2018ppr}) suggest that winding modes play a crucial role in determining the entanglement entropy of closed strings.  Indeed it is the sum over the winding patterns of the string around the stretched entangling surface that allows it to be closed up.  

We have also provided further evidence that open-closed string duality plays an important role in characterising the entanglement structure of closed strings, as originally proposed by Susskind and Uglum.  As in \cite{Donnelly:2016jet,Donnelly:2018ppr}, we show that this involves the introduction of entanglement branes which provide the entanglement cut for the closed strings.  It should be emphasized that while these branes seem to be rigid in the perturbative worldsheet description, we expect them to fluctuate dynamically in the low energy effective gravitational theory just like ordinary D branes.   Indeed, it has been shown from the analysis of the symplectic structure of classical gravity that gravitational edge modes contains degrees of freedom associated with the fluctuations of a co-dimension two brane \cite{Donnelly:2016auv}.  It would be interesting to see if this can be related to the entanglement branes, similar to the way that string theory D branes are related to black branes in supergravity.
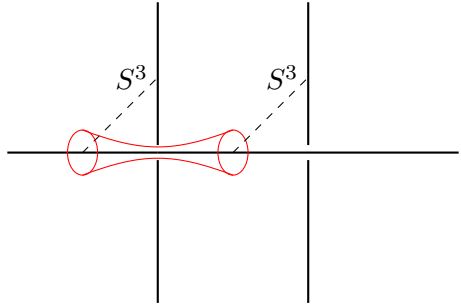
\begin{figure}[h]
\centering
\begin{tikzpicture}
\draw[thick](0,0.1)--(0,2) node [right]{};
\draw[thick](0,-0.1)--(0,-2) node [right]{};
\draw[thick](0,0)--(4,0) node [below]{};
\draw[thick](0,0)--(-2,0) node [below]{};
\draw[dashed](-1,0)--(0,1) node [left]{$S^3$};
\draw[red] (-1,0.3) .. controls (-0.2,0) and (0.2,0) .. (1,0.3);
\draw[red] (-1,-0.3) .. controls (-0.2,-0) and (0.2,-0) .. (1,-0.3);
\draw[red] (-1,0) ellipse (0.2 and 0.3);
\draw[red] (1,0) ellipse (0.2 and 0.3);
\draw[dashed](1,0)--(2,1)node[left]{$S^3$};
\draw[thick](2,0.1)--(2,2)node[right]{};
\draw[thick](2,-0.1)--(2,-2)node[right]{};
\end{tikzpicture}
\caption{Toric diagram for the conifold geometry}\label{er=epr coni}
\end{figure}

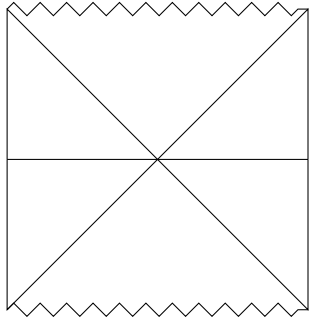
\begin{figure}[h]
\centering
\begin{tikzpicture}[node distance=2cm]
% coordinates for the nodes
\coordinate (A) at (0,0);
\coordinate[above = of A] (B);
\coordinate[right = 4cm of A] (G);
\coordinate[right = 4cm of B] (C);
\coordinate[below  = of C] (D);
\coordinate[below  = of A] (E);
\coordinate[right = 4cm of E] (F);
% some straight lines uning the coordinates and adding labels 
\draw (A) -- node[above]{} (B) -- node[below,sloped,pos=0.25]{} node[below,sloped,pos=0.75] {}
(F) -- (D);
\draw (A) -- (E) -- (C) -- (D);
% some decorated lines uning the coordinates and adding labels 
\draw[decorate,decoration=zigzag] (B) -- node[above] {} (C);
\draw[decorate,decoration=zigzag] (E) -- node[below] {} (F);
\draw (A) -- (G);
\end{tikzpicture}
\caption{Thermo-field double state and the Penrose diagram for the dual AdS-Schwarzschild geometry. The horizontal line in the middle represents the $t=0$ time slice. }\label{er=epr}
\end{figure}

\paragraph{Entanglement in Topoogical M-theory }
Just like in superstring theory where different consistent formulations are expected to be unified by M-theory \cite{1995NuPhB.443...85W, 1996NuPhB.460..506H}, it was proposed that topological string theories can be reduced from topological M-theory \cite{Dijkgraaf:2004te}. It would be interesting to see if our formalism can be generalized to calculate entanglement entropy in topological M-theory. Interestingly, the simplest state we can consider on a six-dimensional time slice would be the A model partition function deformed conifold geometry in figure  \ref{er=epr coni},  viewed as a wavefunction for topological M-theory in one higher dimension\footnote{Historically, the wavefunction behavior of the topological string partition function \cite{Bershadsky:1993cx, 1993hep.th....6122W, Ooguri:2005vr} was observed first and led to the conjecture of the existance of topological M-theory.}.  Figures  \ref{er=epr coni} and  \ref{er=epr}  suggest a parallel between the conifold geometry and the thermal field double state associated with the AdS-Schwarzschild geometry\footnote{We thank Tom Hartman for the discussion on this observation.}. In AdS/CFT, the entanglement between the two boundary CFT's in the TFD state is captured holographically by a spatial wormhole which connects the two boundaries through the  eternal black hole geometry  \cite{2003JHEP...04..021M}.   This is an example of the "ER=EPR" \cite{2013ForPh..61..781M} slogan, which relates  quantum entanglement and geometric connections.  For strings propagating on the conifold geometry in figure \ref{er=epr coni} ,   it would seem that "ER=EPR" manifests itself through worldsheets connecting entangled subsystems corresponding  to the two $S^3$'s.

\begin{figure}
\centering
\includegraphics[width=16cm,height=14cm,keepaspectratio]{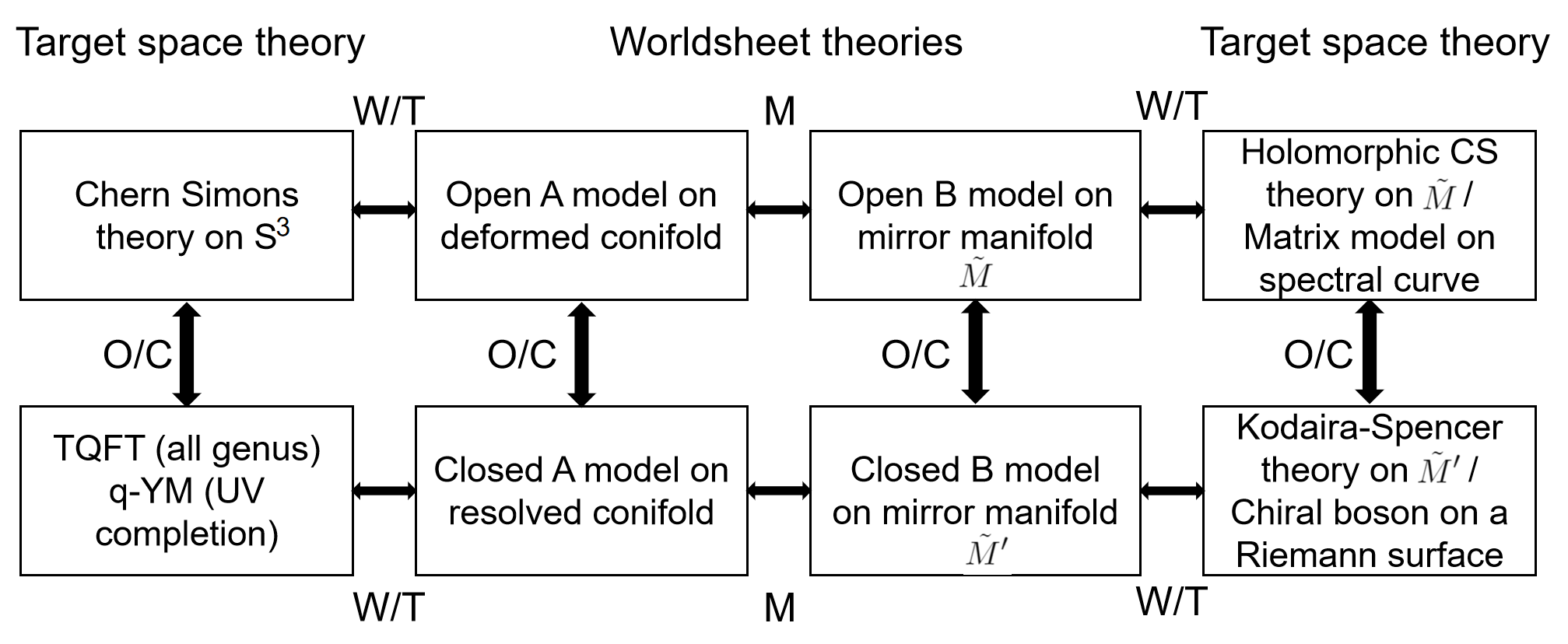}
\caption{Web of dualities for topological string theory. W/T: worldsheet/target space duality. O/C: open/closed duality (large N duality); M : mirror symmetry.}
\label{duality}
\end{figure}

\paragraph{Duality web and the B model}
Finally, we want to point out that almost all corners of the duality web for the topological string are well understood.  This is summarized in Fig. \ref{duality} and warrants further study.  One particularly interesting corner is the proposed UV completion of the A model via q-deformed 2D Yang Mills at large N.  This theory has non perturbative corrections (of order $e^{-N}$ ) which describes baby universes corresponding to  topology changing processes in string theory  \cite{2004hep.th....6058V, 2005NuPhB.715..304A, 2006PhRvD..73f6002D, 2007NuPhB.778...36A, 2020JHEP...08..044M, 2020arXiv200406738M}.  %to understand the ``missing corners'' of the explored regime of quantum gravity \cite{2017arXiv171100864B}. 
It would also be interesting to consider how entanglement entropy behaves under mirror symmetry, which maps the A model to the B model.    Although the $B$ model has the same closed string Hilbert space as the $A$ model, they have different local properties.  The chiral boson theory which describes the $B$ model has a ``pair of pants" amplitude which differs from the one which defines the $A$ model TQFT.   We thus expect that the associated cobordism theory would define a different notion of factorization. 

Another motivation for studying the $B$ model comes from an interesting connection with JT gravity. It was shown in the seminal paper that open topological B models are equivalent to matrix models on Calabi-Yau manifolds that can be written as a fibration over the spectral curve of the matrix model \cite{2002NuPhB.644....3D}. For example, the JT matrix model \cite{2019arXiv190311115S}, whose spectral curve is
\be
F(x,y)=y^2-\sin(\sqrt{x})^2
\ee
can be realized as topological B-model on Calabi-Yau\cite{2019arXiv190311115S, 2020arXiv200406738M}
\be
y^2-\sin(\sqrt{x})^2+u^2+v^2=0.
\ee
It would be interesting to compare how local properties of the topological string and JT gravity emerges from the matrix model and see if the $B$ model offers further insight into JT gravity.   In particular, the B model admits a Nekrasov deformation \cite{2002hep.th....6161N, 2009JHEP...10..069I, 2010maph.conf..265N} that is related to q-deformations.  Perhaps this can be related to q-deformations of JT gravity.  \footnote{We thank Cumrun Vafa for pointing this possible connection to us.}.

 %Given that the B-model is closely related to matrix models \cite{2002NuPhB.644....3D}, perhaps studying entanglement entropy in the  B model will teach us how notions like locality \cite{PhysRevLett.115.121602} and subregion duality \cite{2012CQGra..29o5009C, Bousso:2012mh} emerge in matrix models.    It would also be interesting to understand the connection between the Nekrasov deformation \cite{2002hep.th....6161N, 2009JHEP...10..069I, 2010maph.conf..265N} of the matrix models and the q-deformation\footnote{We thank Cumrun Vafa for pointing this possible connection to us.}. Finally, as JT gravity is dual to a matrix theory \cite{2019arXiv190311115S}, perhaps the B-model point of view new insights into JT gravity \cite{2019arXiv190311115S, 2020arXiv200406738M}.

%an interesting connection between topological string theory and JT gravity. It was shown in the seminal paper that open topological B models are equivalent to matrix models on Calabi-Yau manifolds that can be written as a fibration over the spectral curve of the matrix model \cite{2002NuPhB.644....3D}. For example, the JT matrix model \cite{2019arXiv190311115S}, whose spectral curve is
%\be
%F(x,y)=y^2-\sin(\sqrt{x})^2
%\ee
%can be realized as topological B-model on Calabi-Yau\cite{2019arXiv190311115S, 2020arXiv200406738M}
%\be
%y^2-\sin(\sqrt{x})^2+u^2+v^2=0.
%\ee

\section*{Acknowledgements}
We thank Thomas Hartman and William Donnelly for collaboration in the early stages of this work and providing countless suggestions and discussions. We thank Cumrun Vafa for the discussion on the B model. We thank Zhenhao Zhou for plotting the 3d diagrams of string worldsheets. The work of M.K. was supported in part by nsf grant PHY-1719877. The work of Y.J. is supported by the Simons Foundation through the Simons Collaboration on the Nonperturbative Bootstrap. G.W. is supported by Fudan University and the Thousands Young Talents Program. 
\appendix

\section{Topological twist and topological sigma model on the worldsheet}

Before we move on to the topological sigma model, let us briefly review $N=2$ supersymmetric non-linear sigma model defined on a Riemann surface $\Sigma$ with a Kahler manifold $X$ as a target space. This theory consists of the following data: holomorphic map/coordinate function $\Phi:\Sigma \rightarrow TX,$ superpartners of $\Phi.$ Because of the complex structure of $X,$ the complexified tangent bundle $TX$ decomposes into holomorhpic and anti-holomorphic tangent bundle
\begin{equation}
    TX=T^{1,0}X\oplus T^{0,1}X.
\end{equation}
Respective to the decomposition of the complexified tangent bundle, we denote the holomorphic components of $\Phi$ by $\phi^i\in T^{1,0}X$ and similarly for the anti-holomorphic components. With this holomorphic decomposition, we can think of $\phi^i$ as a holomorphic tangent vector, of the target space, valued scalar field on the worldsheet. A superpartner of such field then should live in holomorphic tangent vector valued $spin$ bundle, which reads
\begin{equation}
    \sqrt{K_\Sigma}\otimes(\mathcal{O}_\Sigma\oplus \Omega_\Sigma ^{0,1})\otimes\Phi^*(TX^{1,0}), 
\end{equation}
where $\sqrt{K_\Sigma}$ is an algebraic square root of canonical bundle of $\Sigma,$ $\mathcal{O}_\Sigma$ is structure sheaf of $\Sigma,$ and $\Omega^{0,1}_\Sigma\equiv\overline{K_\Sigma}$ is anti-holomorphic cotangent bundle of $\Sigma.$ As anti-holomorphic canonical bundle is dual of canonical bundle, the corresponding spinor bundle can be written as
\begin{equation}
    (K_\Sigma^{1/2}\oplus \overline{K_\Sigma}^{1/2})\otimes \Phi^*(TX^{1,0}).
\end{equation}
We will then denote the fermions living in $K_\Sigma^{1/2}\otimes \Phi^*(TX^{1,0})$ and $\overline{K}_\Sigma^{1/2}\otimes \Phi^*(TX^{1,0})$ by $\psi_+^i$ and $\psi_-^i,$ respectively. We will use the similar convention for $\psi_+^{\bar{i}}$ and $\psi_-^{\bar{i}}.$ Given the field contents, the worldsheet action is
\begin{equation}
    S=2t \int_\Sigma \left(\frac{1}{2} g_{IJ}\partial_z\phi^I\partial_{\bar{z}}\phi^J+ig_{i\bar{i}}\psi_-^{\bar{i}}D_z\psi_-^i +ig_{i\bar{i}}\psi_+^{\bar{i}}D_{\bar{z}}\psi_+^i+R_{i\bar{i}j\bar{j}}\psi_+^i\psi_+^{\bar{i}}\psi_-^{j}\psi_-^{\bar{j}}\right),
\end{equation}
where $g$ is the hermitian metric of the target space.

Topological string model is then obtained by a topological twist to the bundle \cite{Witten:1991zz}, in which fermionic fields live in, that preserves the form of kinetic terms of fermionic fields. 
The topological twist of A model can be understood as moving the non-trivial bundle $\sqrt{K_\Sigma}$ from $K_\Sigma^{1/2}\otimes\Phi^*(TX^{1,0})$ to $K_\Sigma^{1/2}\otimes \Phi^*(TX^{0,1})$ and similarly for $\overline{K}_{\Sigma}^{1/2}.$ 
As a result of this topological twist, $\psi_+^i$ and $ \psi_-^i$ becomes (anti)-holomorphic tangent vector valued scalar field on the worldsheet. 
Then we can focus on transformation that transforms $\phi^i$ into $\psi_+^i$ and $\phi^{\bar{i}}$ into $\psi_-^{\bar{i}},$ as those transformations can be represented by a globally well defined functions and others not in general\footnote{For high genus curves, there are still more non-trivial supersymmetry transformations. But, I have no idea what will happen if I take those non-trivial transformations into account. Perhaps, BRST operator will just go away}.

Given the topological twist, let us rename the fermionic fields as $\chi^i=\psi_+^i$ and $\chi^{\bar{i}}=\psi_-^{\bar{i}}.$ Supersymmetry transformation is concisely repackaged as
\begin{align}
    \{Q,\Phi\}=&\chi,\nonumber\\
    \{Q,\chi\}=&0,\nonumber\\
    \{Q,\psi_{-}^I\}=&i\partial_{\bar{z}}\Phi^I-\chi^J\Gamma^{I}_{JK}\psi_{-}^K,\nonumber\\
    \{Q,\psi_{+}^{\bar{I}}\}=&i\partial_{\bar{z}}\Phi^{\bar{I}}-\chi^{\bar{J}}\Gamma^{\bar{I}}_{\bar{J}\bar{K}}\psi_{+}^{\bar{K}},
\end{align}
where $Q^2=0$ on-shell thus supersymmetry becomes BRST symmetry. The action is 
\begin{equation}
    S=2t \int_\Sigma \left(\frac{1}{2} g_{IJ}\partial_z\phi^I\partial_{\bar{z}}\phi^J+ig_{i\bar{i}}\psi_-^{i}D_z\chi_-^{\bar{i}} +ig_{i\bar{i}}\psi_+^{\bar{i}}D_{\bar{z}}\chi^i-R_{i\bar{i}j\bar{j}}\psi_-^i\psi_+^{\bar{i}}\chi^j\chi{\bar{j}}\right).
\end{equation}
A very important observation is that this action is a sum of a Q-exact term and a topological term
\begin{equation}
    S=it\int_\Sigma d^2z\{Q,V\}+t\int_\Sigma \Phi^*(J),
\end{equation}
where $V=g_{i\bar{j}} (\psi_+^{\bar{i}}\partial_z\phi^j+\partial_z\phi^{\bar{i}}\psi_-^j)$ and $\Phi^*(J)$ is pullback of the K\"ahler form defined on $X.$ One can add pullback of two-form tensor $B$ to the action to complexfy the K\"ahler form.

We have not specified yet if $\Sigma$ has boundaries or not. If $\Sigma$ does not attain a boundary, then the worldsheet theory is a closed string theory. Similarly, if $\Sigma$ has boundaries, then the worldsheet theory is an open string theory.

Topological strings wrap ``volume minimizer," which is energetically stable, among homologous 2 cycles in $X.$ Which means that for closed string theory, worldsheet instanton is classfied by homology class 
\begin{equation}
    \Phi_*([\Sigma])\in H_2(X,\Bbb{Z}).
\end{equation}

This classification can be generalized to open string theory directly. Open string worlsheet can be regarded as a Riemann surface with $h$ holes due to the conformal invariance. As there are $h$ boundaries of the Riemann surface, one should impose boundary conditions. Let us denote $h$ boundaries of $\Sigma$ by $C_i,$ where $i=1,\dots, h.$ In \cite{Witten:1992fb}, Witten showed that the physical boundary condition is given by
\begin{equation}
    \Phi(C_i)\subset\mathcal{L}
\end{equation}
for some $\mathcal{L}$ which is a Lagrangian submanifold of $X.$ Note that a submanifold $\mathcal{L}$ is Lagrangian if $J|_\mathcal{L}=0.$ This condition implies that supersymmetric D-branes in topological A model wrap Lagrangian three-cycles in $X$\footnote{In this work, we do not focus on torsion one or five cycles.}. Therefore, open string worldsheet instanton is naturally classified by relative homology class
\begin{equation}
    \Phi_*(\Sigma)\in H_2(X,\mathcal{L}).
\end{equation}

One important class of observable in closed A model is a three points function which has various interpretations in physical string theory. Let us consider a non-trivial 2 form $[D_i]\in H^2(X).$ Then one can consider an operator
\begin{equation}
    \mathcal{O}_{D_i}=(D_i)_{i_1,i_2}\chi^{i_1}\chi^{i_2}.
\end{equation}
If we assume that $X$ is a Calabi-Yau threefolds, when computed on string worlsheet $\Bbb{P}^1,$ the three points function of $\mathcal{O}(D_i)$ is \cite{Candelas:1990rm}
\begin{equation}
    \langle\mathcal{O}_{D_1}\mathcal{O}_{D_2}\mathcal{O}_{D_3} \rangle = \mathcal{K}_{D_1D_2D_3}+\sum_\beta N_{0,\beta}(D_1,D_2,D_3)\prod_i \int_\beta [D_i]Q^{\beta},
\end{equation}
where $\mathcal{K}_{D_1D_2D_3}$ is an intersection number and $N_{0,\beta}(D_1,D_2,D_3)$ is a genus 0 Gromov-Witten invariant for an integral curve $\beta\in H_2(X),$ and $Q= e^{-\int_\beta J}.$ Note that this three points function can be obtained from the third derivative of the genus 0 prepotential, which is free energy of genus 0 worldsheet theory,
\begin{equation}
    \partial_{t_1}\partial_{t_2}\partial_{t_3}F_0(t)=\langle\mathcal{O}_{D_1}\mathcal{O}_{D_2}\mathcal{O}_{D_3} \rangle,\label{prepotential yukawa}
\end{equation}
where $t_i= \int_{D^i}J.$ Genus 0 prepotential receives classical and instanton contributions
\begin{equation}
    F_0=F_0^{cl}+F_0^{inst},
\end{equation}
where
(to add prepotential at LCS).
Coupling to gravity \cite{Witten:1992fb}, genus g free energy can be computed as well which reads
\begin{equation}
    F_g(t)=\sum_\beta N_{g,\beta}Q^\beta,
\end{equation}
where $N_{g,\beta}$ is a genus g Gromov-Witten invariant. Combining all genera prepotential, we get a generating functional the all genera free energy
\begin{equation}
    F(g_s,t)=\sum_g F_g(t)g_s^{2g-2},
\end{equation}
which will prove to be useful.

\section{Topological String on Conifolds and Geometric Transition}

Let us review briefly the geometric transition from open string to closed string theory of interest. Let us consider A-model open topological string theory on the deformed conifold $T^*S^3.$ We wrap N D-branes on $S^3,$ whose low energy effective theory is $SU(N)$ Chern-Simons theory \cite{Witten:1992fb}. Wilson lines can be introduced, if M D-branes wrap on a lagrangian submanifold\footnote{In topological string theory, Lagrangian is good enough to ensure supersymmetry whereas in physical string theory special Lagrangian is required. Note that in the conifold, Lagrangian submanifolds we consider are special Lagrangian actually.} $\mathcal{L}$ of $T^*S^3$ which intersects $S^3$ at $S^1.$ This corresponds to U(N) Chern-Simons theory on $S^3$ with M knots on $S^1.$ Under the geometric transition at large N, we obtain A-model topological string theory on the resolved conifold $\mathcal{O}(-1)\oplus\mathcal{O}(-1)\rightarrow \Bbb{P}^1,$ in which the N D-branes are desolved into B-flux and M D-branes are still wrapped on the same special lagrangian $\mathcal{L}$ and intersect $S^2$ at $S^1$ \cite{Ooguri:1999bv}.

Using the topological vertex formalism, one can obtain partition function on $\mathcal{O}(-1)\oplus\mathcal{O}(-1)\rightarrow D^2,$ where $\partial D^2=S^1$ which implies that the partition function can be understood as a wave function of topological string theory on $S^1$ with the fiber. Now, we need to subdivide $S^1$ into two line segments to compute the entanglement entropy\footnote{For replica trick, this subdivision is not needed. But still, understanding on the Wilson line dual to the cutting is absolutely necessary.}. In order to subdivide $S^1,$ one needs D-brane/anti-D-brane pair intersecting $S^1$ at two points. The D-brane/anti-D-brane pair cannot wrap any submanifold of the resolved conifold rather the D-brane/anti-D-brane pair should wrap a special lagrangian submanifold. Previously, we have said that there is a special lagrangian submanifold, on which a flavour D-brane can wrap to generate Wilson loop in the dual Chern-Simons theory. Hence, it is natural to conjecture that D-brane/anti-D-brane pair needed to cut $S^1$ to two line segments is dual to flavour D-brane/anti-D-brane pair in the dual open string theory.

The conjecture implies that the local degrees of freedom counted in closed string theory should result from open strings extended between the intersecting D-branes. Furthermore, cutting through $S^1,$ which is wrapped by the flavour D-brane/anti-D-brane pair, corresponds to cutting through the Wilson line/anti-Wilson-line pair, so we obtain nice interpretation in open string theory as well.

Let us first study the deformed conifold. Cotangent bundle of $S^3$ can be embedded into $\Bbb{C}^4$ by an equation
\begin{equation}
y_1^2+y_2^2+y_3^2+y_4^2=a^2,
\end{equation}
$y_i$'s$\in\Bbb{C}.$ We assume that $a$ is a real number. The bundle structure is more vivid when we write $y_i=x_i+i p_i,$ then the embedding equation is written as
\begin{equation}
\sum_i x_i^2=a^2+\sum_i p_i^2,~~~ \sum_i x_i p_i=0.
\end{equation}
It is then clear when $p_i=0,$ for all $i,$ then the equations are reduced to
\begin{equation}
\sum_i x_i^2=a^2.
\end{equation}
Thus $a$ describes radius of $S^3.$ When $a$ is sent to 0, the deformed conifold in the limit described by 
\begin{equation}
y_1^2+y_2^2+y_3^2+y_4^2=0.\label{eqn:sing con}
\end{equation}
As Jacobian of the defining equation vanishes at the origin $y_1=y_2=y_3=y_4=0,$ the conifold at the origin is singular. 

To fix the singularity at the origin, one can blow up/resolve the origin such that $y_1=y_2=y_3=y_4=0$ is replaced with a smooth manifold. If we reparametrize the coordinates by
\begin{equation}
    z_{ij}=\sum_n \sigma^n_{ij}y_n,
\end{equation}
then \eqref{eqn:sing con} is written as
\begin{equation}
    \det z_{ij}=0.\label{eqn:sing con2}
\end{equation}
In this presentation, the singularity occurs when the matrix coordinate $z_{ij}$ is trivial. It is important to note that we can view \eqref{eqn:sing con2} as a condition for the following equation to have a non-trivial solution
\begin{equation}
    \left(\begin{array}{cc} z_{11} & z_{12} \\ z_{21} & z_{22} \end{array}\right)\left(\begin{array}{c} \lambda_1 \\ \lambda_2 \end{array}\right)=0,\label{eqn:res con}
\end{equation}
for some complex variable $\lambda_1$ and $\lambda_2$ which cannot be simultaneously zero, because $\lambda_1=\lambda_2=0$ results in no constraints on $z_{ij}$ matrix. Furthermore, \eqref{eqn:res con} provides a resolution of the singularity because when $z_{ij}$ is non trivial $\lambda_1$ and $\lambda_2$ are fixed up to rescaling and $z_{ij}=0$ is replaced with coordinates $(\lambda_1,\lambda_2).$ This implies that equation \eqref{eqn:res con} is an embedding of the resolved conifold into $\Bbb{C}^4\times \Bbb{P}^1$ in which $z_{ij}$ is a coordinate of $\Bbb{C}^4$ and $[\lambda_1,\lambda_2]$ is a homogeneous coordinate of $\Bbb{P}^1.$ Note that, when $z_{ij}$ is generic, the non-homogeneous coordinate $z$ of $\Bbb{P}^1$ is related to the rest of the coordinates by
\begin{equation}
z:=\frac{\lambda_1}{\lambda_2}=\frac{y_1+iy_2}{y_3-iy_4}=\frac{y_3+iy_4}{iy_2-y_1}.
\end{equation}

Lagrangian submanifolds can easily be found by finding symmetric locus of an anti-holomorphic involution. We consider an anti-holomorphic involution
\begin{equation}
y_{1,2}=\overline{y}_{1,2},~~~y_{3,4}=-\overline{y}_{3,4}.\label{eqn:knot}
\end{equation}
There are more anti-holomorphic involutions, but it will be clear that one can choose \eqref{eqn:knot} without loss of generality to find a special lagrangian intersecting $S^3$ at $S^1.$ In the deformed conifolde, for example, the invariant locus of \eqref{eqn:knot}, a lagrangian submanifold $\mathcal{L},$ is
\begin{equation}
p_{1,2}=0,~~~x_{3,4}=0.
\end{equation}
At the symmetric locus of \eqref{eqn:knot}, the embedding equation becomes
\begin{equation}
x_1^2+x_2^2=a^2+p_3^2+p_4^2.
\end{equation} 
Hence $\mathcal{L}$ intersects $S^3$ at
\begin{equation}
x_1^2+x_2^2=a^2,
\end{equation}
which is a $S^1.$

The Wilson loop or knot on $S^3$ is $S^1$ which is described by
\begin{equation}
x_1^2+x_2^2=a^2, x_3=x_4=p_1=p_2=p_3=p_4=0.\label{eqn:knot 2}
\end{equation}
We now want to show that \eqref{eqn:knot 2} is homologous to $S^1$ on $S^2.$ We will use homotopy equivalence, with an understanding that two homotopically equivalent cycles are homologous. First let us consider a parametrization $C_a(t)$ for $t\in[0,1],$
\begin{equation}
(x_1(t),x_2(t))=a(\cos(2\pi t),\sin(2\pi t)).
\end{equation}
Let us then consider a homotopy $\gamma_a(t,l)$ for $t,l\in [0,1],$
\begin{equation}
(x_1(t),x_2(t),p_3(t),p_4(t))=( \sqrt{a^2+l^2}\cos(2\pi t),\sqrt{a^2+l^2}\sin (2\pi t), l\cos (-4\pi t),l\sin(-4\pi t)).
\end{equation}
From this homotopy equivalence, we have checked that the curve \eqref{eqn:knot 2} is homologous to a curve $\gamma_a(t,1)$
\begin{equation}
(x_1(t),x_2(t),p_3(t),p_4(t))=( \sqrt{a^2+1}\cos(2\pi t),\sqrt{a^2+1}\sin (2\pi t), \cos (-4\pi t),\sin(-4\pi t)).
\end{equation}
Let us take the singular limit $a\rightarrow 0,$ and blow up the singular point. Under this geometric transition, $\gamma_a(t,1)$ becomes $\gamma_0(t,1)$
\begin{equation}
(x_1(t),x_2(t),p_3(t),p_4(t))=( \cos(2\pi t),\sin (2\pi t), \cos (-4\pi t),\sin(-4\pi t)).
\end{equation}

What is left for us to check is if $\mathcal{L}$ intersects $S^2$ of the blow up, and if $\mathcal{L}$ and $S^2$ intersect whether or not the intersection is $\gamma_0(t,1).$ Let us first check if $\mathcal{L}$ intersects $S^2.$ In the resolved conifold, $\mathcal{L}$ is given by
\begin{equation}
x_1^2+x_2^2=p_3^2+p_4^2,~~~ p_{1,2}=x_{3,4}=0.
\end{equation}
As $(y_1+iy_2)/(y_3-iy_4)=(y_3+iy_4)/(iy_2-y_1)$ at $\mathcal{L}$ it is manifest that $\mathcal{L}$ intersects $S^2.$ To study if $\gamma_a(t,1)$ intersects $S^2,$ let us parametrize $z=(y_1+iy_2)/(y_3-iy_4)$
\begin{equation}
z=-i(\cos(-2\pi t)+i \sin(-2\pi t)).
\end{equation}
Hence, we have shown that $\gamma_0(t,1)$ is an equator of $S^2$ thus proved the claim\footnote{It still remains ambiguous what other choices of homotopy equivalences mean. For example, if I choose a homotopy equivalence $(x_1(t),x_2(t),p_3(t),p_4(t))=( \cos(2\pi t),\sin (2\pi t), \cos (-2\pi t),\sin(-2\pi t))$ then I could send $S^1$ on $S^3$ in the deformed conifold to a point on $S^2$ in the resolved conifold. This shows that there could be many distinct choices, which partially results from the fact that $S^1$ in $S^2$ is trivial in homology, or more precisely $H_1(S^2,\Bbb{Z})=0$.}.
\section{$\Bbb{C}^3$ as a toric variety}
$\Bbb{C}^3$ can be regarded as a Calabi-Yau manifold in a sense that the first chern class of $\Bbb{C}^3$ is trivial. We want to find a $T^2\times \Bbb{R}$ fiber hiding in $\Bbb{C}^3.$ Let $z_i$ be complex coordinates of $\Bbb{C}^3,$ for $i=1,2,3.$ We first introduce three Hamiltonians, which encode the fibration structure,
\begin{align}
r_\alpha(z)&=|z_1|^2-|z_3|^2,\\
r_\beta(z)&=|z_2|^2-|z_3|^2,\\
r_\gamma(z)&=\text{Im}(z_1z_2z_3).
\end{align}
Given the symplectic form $\omega$
\begin{equation}
\omega=dz_i\wedge d\bar{z}_i,
\end{equation}
the Hamiltonians, which can be understood as base coordinates, uniquely determine the fiber coordinates through Possion bracket
\begin{equation}
\partial_v z_i=\{r_v,z_i\}_{\omega}.
\end{equation}
More explicitly, let $\alpha$ be a coordinate of an $\alpha$-cycle, which is generated by $r_\alpha,$ and $\beta$ be a coordinate of a $\beta$-cycle of $T^2,$ both of which parametrize phases of the complex coordinates of $\Bbb{C}^3$ by following group action
\begin{equation}
r_\alpha\otimes r_\beta:(z_1,z_2,z_3)\mapsto (e^{i\alpha}z_1,e^{i\beta}z_2,e^{-i(\alpha+\beta)}z_3).
\end{equation}
In a similar way, translation along the remaining fiber direction $\Bbb{R}$ induces 
\begin{equation}
(z_1,z_2,z_3)\mapsto (z_1+\gamma \bar{z}_2\bar{z}_3,z_2+\gamma \bar{z}_1\bar{z}_3,z_3+\gamma\bar{z}_1\bar{z}_2).
\end{equation}

As we have identified the fiber $T^2\times\Bbb{R},$ let us now study at which places in the base manifold the fiber degenerates, meaning that some cycles in $T^2$ shirinks to a point. An easy way to see if some cycle shrinks to a point is to check if the flow generated by a Hamiltonian is trivial. For example, $r_\alpha$ generates trivial flow if $z_1=z_3=0.$ In a similar way, $r_\beta$ generates trivial flow if $z_2=z_3=0.$ Now consider a cycle generated by $r_\alpha-r_\beta.$ The flow generated by $r_\alpha-r_\beta$ is
\begin{equation}
(z_1,z_2,z_3)\mapsto (e^{i\theta}z_1,e^{-i\theta}z_2,z_3).
\end{equation}
Which shows that the cycle generated by $r_\alpha-r_\beta$ degenerates at $z_1=z_2=0.$ More generally, one can consider a cycle generated by $p r_\alpha+q r_\beta,$ which degenerates $z_1=z_2=z_3=0.$ 

Now let us analyze the degeneration loci in the base manifold $\Bbb{R}^3.$ $\alpha$ cycle degenerates at $z_1=z_3=0,$ which then implies that the degeneration locus in the base manifold for $\alpha$ cycle is $r_\alpha=r_\gamma=0.$ One can show that $\beta$ cycle degenerates at $r_\beta=r_\gamma=0$ in the base manifold. For a generic cycle generated by $pr_\alpha+qr_\beta,$ the degeneration locus in the base manifold is simply $pr_\alpha+qr_\beta=r_\gamma=0.$ Thus it is manifest that the degeneration loci in the base manifold can be compactly encoded in the two dimensional space, at $r_\gamma=0,$ which is parametrized by $(r_\alpha,r_\beta).$ We denote a cycle generated by $(p r_\alpha+qr_\beta)$ as $(-q,p)$ cycle. With the understanding on the degeneration loci of $T^2$ cycles, we represent $\Bbb{C}^3$ by 
\begin{center}
\begin{tikzpicture}
\draw[thick,->](0,0)--(1,0)node[anchor=north west]{(1,0)};
\draw[thick,->](0,0)--(0,1)node[anchor=south east]{(0,1)} ;
\draw[thick,->](0,0)--(-1,-1)node[anchor=south, below]{(-1,-1)};
\end{tikzpicture}
\end{center}
Note that equivalent graph can be obtained by applying a $SL(2,Z)$ transformation on all the vectors. For example, upon acting
\begin{equation}
\left(\begin{array}{cc} 1&0\\-1&1 \end{array}\right),
\end{equation}
we obtain equivalent toric diagrams
\begin{center}
\begin{tikzpicture}
\draw[thick,->](0,0)--(0,1)node[right]{(0,1)};
\draw[thick,->](0,0)--(1,-1)node[below]{(1,-1)};
\draw[thick,->](0,0)--(-1,0)node[below]{(-1,0)};
\end{tikzpicture}
\end{center}
\begin{center}
\begin{tikzpicture}
\draw[thick,->](0,0)--(0,-1)node[right]{(0,-1)};
\draw[thick,->](0,0)--(1,1)node[above]{(1,1)};
\draw[thick,->](0,0)--(-1,0)node[below]{(-1,0)};
\end{tikzpicture}
\end{center}

\section{Replica trick in Chern-Simons theory} 

We first introduce an identity that provides the basis for the surgery method. For a manifold $M$ which is the connected sum of two manifolds $M_1$ and $M_2$ glued along a boundary $S^2,$ Chern-Simons theory partition function obeys an identity \cite{Dong:2008ft, Witten:1988hf}.
\begin{equation}\label{eqn:surgery}
    Z(M) =\frac{Z(M_1)Z(M_2)}{Z(S^3)}.
\end{equation}
\eqref{eqn:surgery} can be generalized to a manifold $M$ which is glued along n $S^2$'s
\begin{equation}\label{eqn:surgery2}
    Z(M)=\frac{Z(M_1)Z(M_2)}{Z(S^3)^n}.
\end{equation}
A straightforward generalization of the surgery formula \eqref{eqn:surgery} for $M_1$ and $M_2$ with unknots of representations $R_1\in M_1$ and $R_2 \in M_2$ is \cite{Dong:2008ft}
\begin{equation}
    Z(M;R_1,R_2)=\frac{Z(M_1;R_1)Z(M_2;R_2)}{Z(S^3,R_{S^3})},
\end{equation}
where $R_{S^3}$ is the representation of the Wilson line going through the gluing boundary $S^2.$ An important identity we will use at various steps in the large N-limit is
\begin{equation}
\lim_{N \to \infty}    Z(S^3;R)Z(S^3;\overline{R})=S_{00}(t')^2d_q(R)^2.
\end{equation}

\begin{figure}[h]
\centering
\begin{tikzpicture}
  \draw (-1,0) to[bend left] (1,0);
  \draw (-1.2,.1) to[bend right] (1.2,.1);
  \draw[rotate=0] (0,0) ellipse (100pt and 50pt);
  \draw[blue,thick] (0,0) ellipse (70pt and 30pt) ;
  \node[blue] at (0,-1.3) {R};
\end{tikzpicture}
\caption{Solid torus with a Wilson loop operator inserted.}
\label{fig:torus with wilson}
\end{figure}
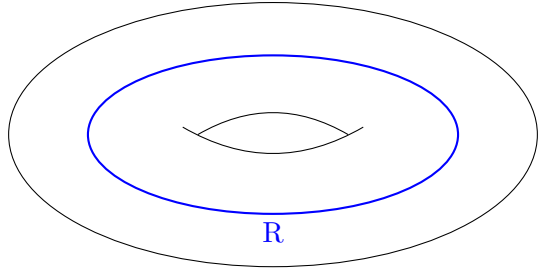
\begin{figure}[h]
\centering
\begin{tikzpicture}
  \draw (-1,0) to[bend left] (1,0);
  \draw (-1.2,.1) to[bend right] (1.2,.1);
  \draw[rotate=0] (0,0) ellipse (100pt and 50pt);
  \draw[blue,thick] (0,0) ellipse (70pt and 30pt) ;
  \node[blue] at (0,-1.3) {R};
  \node at (-2,0){$A$};
  \node at (2,0){$\bar{A}$};
  \draw[dashed] (0,-50pt) to[bend left] (0,-0.26);
  \draw (0,-50pt) to[bend right] (0,-0.26);
  \draw (0,50pt) to[bend left] (0,0.27);
  \draw[dashed] (0,50pt) to[bend right] (0,0.27);
\end{tikzpicture}
\caption{Separated solid torus with a Wilson loop operator inserted.}
\label{fig:torus with wilson2}
\end{figure}
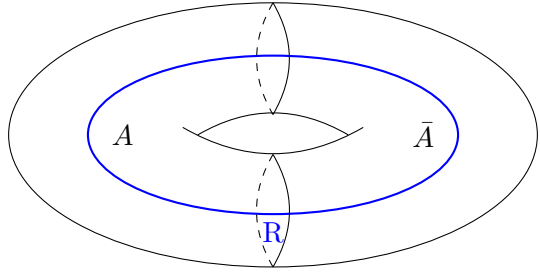

Let us consider a single state without superposition $|R\rangle_{CS} \in \mathcal{H}(T^2),$ see Fig. \ref{fig:torus with wilson}. We separate the solid torus into two regions $A$ and $\bar{A},$ see Fig. \ref{fig:torus with wilson2}. An overlap $\langle R_i |R_j\rangle$ is equivalent to a path integral on $S^2\times S^1$ with insertions of Wilson loops with representations $R_j$ and $\overline{R}_i$ along $S^1.$ 

\begin{figure}
    \centering
    \includegraphics[width=0.7\textwidth]{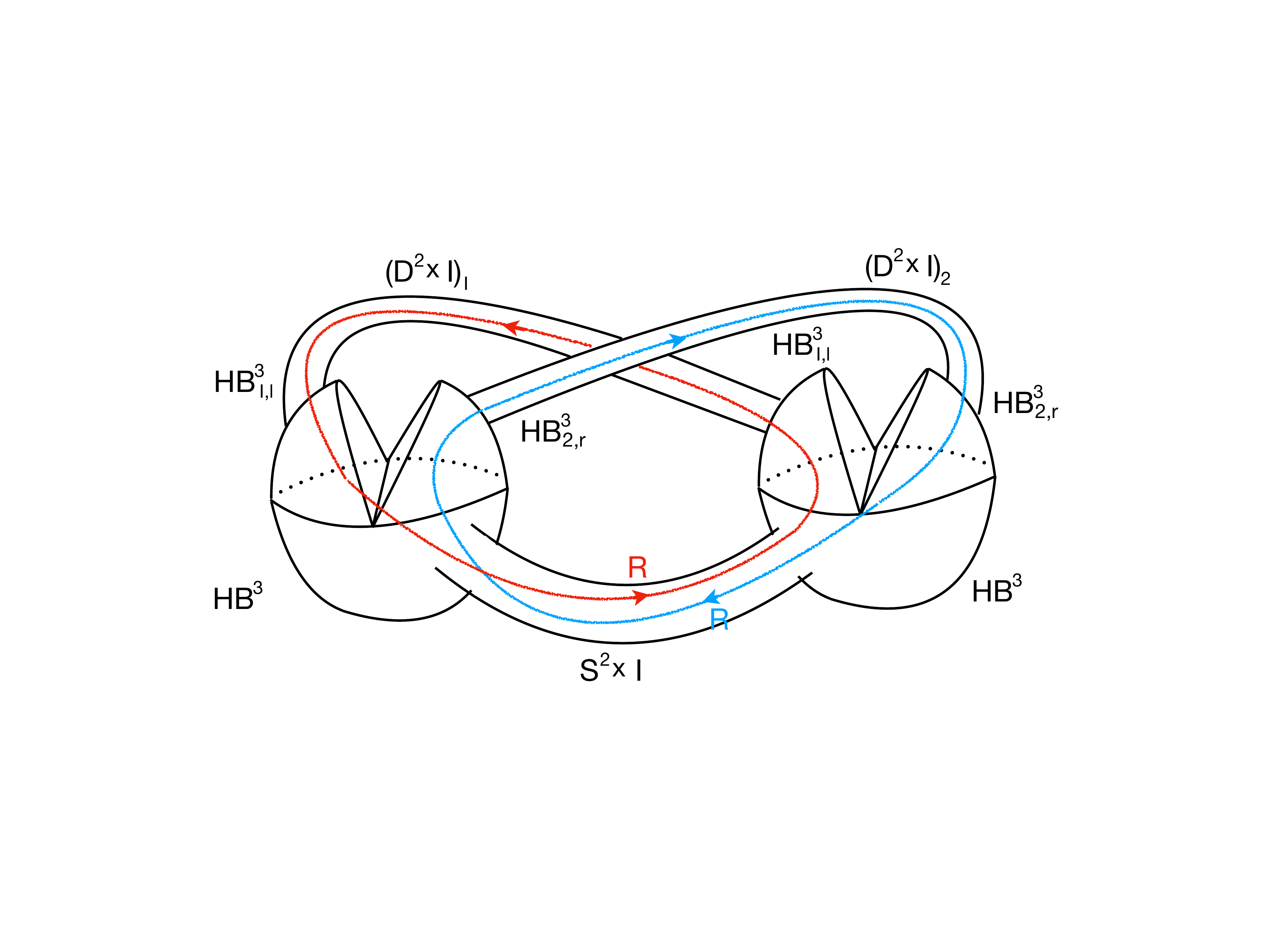}
    \caption{Geometric representation of the reduced density matrix}
    \label{fig:reduced123}
\end{figure}

\begin{figure}
    \centering
    \includegraphics[width=0.7\textwidth]{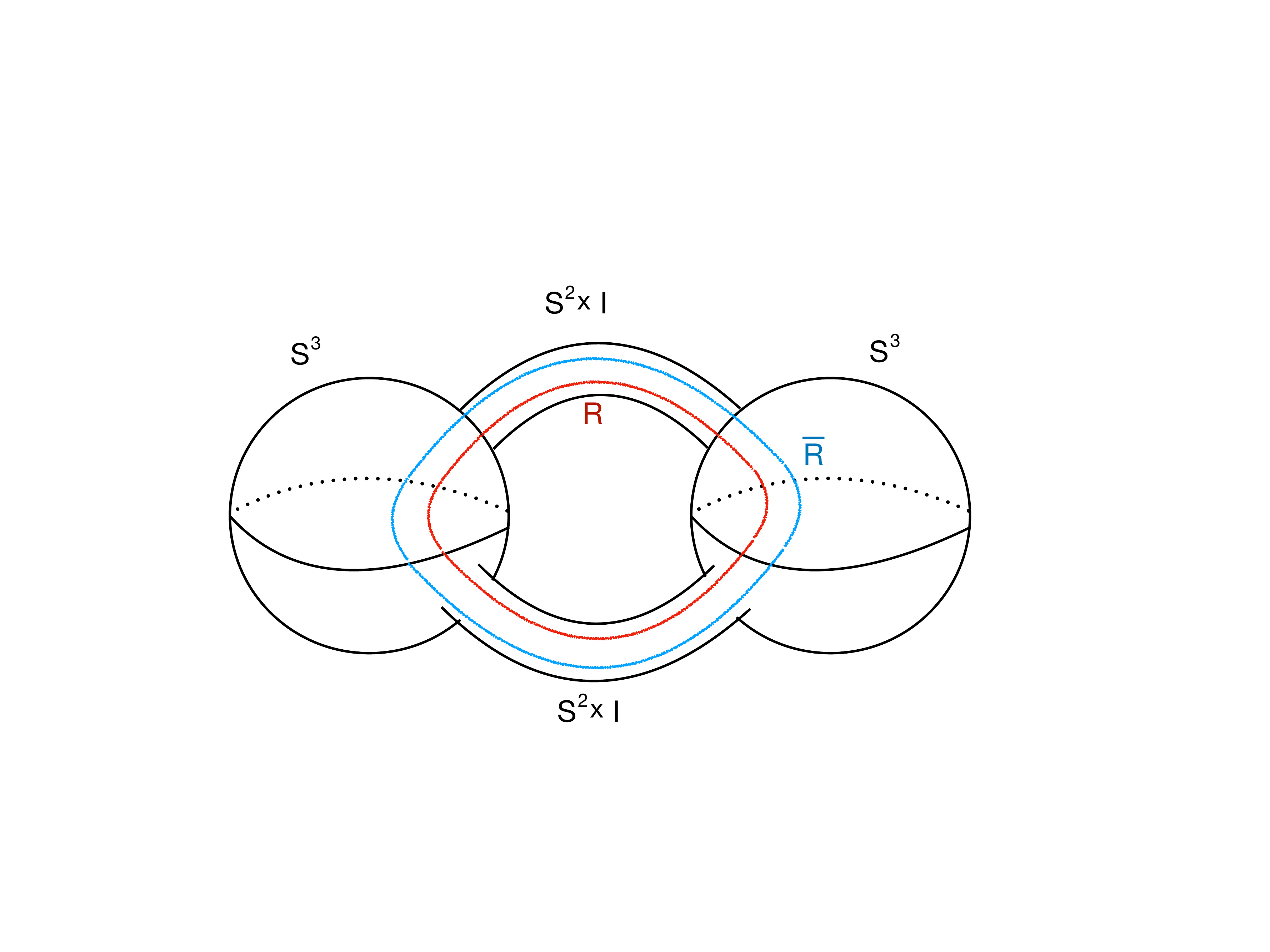}
    \caption{Geometric representation of $Z_1.$}
    \label{fig:z1}
\end{figure}

To gain intuition on the replicated geometry, let us deform Fig. \ref{fig:torus with wilson2} and compute $\langle R_i|R_j \rangle$ performing the surgery \cite{Witten:1988hf}. One can understand $A$ or $\overline{A}$ as two three-dimensional half solid-balls, $\Bbb{H}B^3_l$ and $\Bbb{H}B^3_r$, that are connected by a three-dimensional solid cylinder $D^2\times I.$ It is useful to note that the boundary of $\Bbb{H}B^3$ consists of two two-dimensional disks glued along $S^1,$ $\partial(\Bbb{H}B^3)=D^2_u\cup D_d^2$. To prepare a reduced density matrix, we prepare two copies of Fig. \ref{fig:torus with wilson2} and glue $\overline{A}_1$ and $\overline{A}_2$ as follows. First, we glue $\Bbb{H}B^3_{\overline{1},l}\in \overline{A}_1$ and $\Bbb{H}B^3_{\overline{2},l}\in \overline{A}_2$ along $D^2_{\overline{1},d} \in \Bbb{H}B^3_{\overline{1},l}$ and $D^2_{\overline{2},d} \in \Bbb{H}B^3_{\overline{2},l}.$ As a result of this gluing, we again obtain a three dimensional half solid-ball whose boundary is $S^2=D^2_{1,u}\cup D^2_{2,u}.$ Similarly, we can glue $\Bbb{H}B^3_{\overline{1},r} \in\overline{A}_1$ and $\Bbb{H}B^3_{\overline{2},r}$ following the same procedure to obtain one more three dimensional half solid ball. Finally, we glue two solid cylinders along $S^1\times I$ to produce $S^2\times I.$ As a result, we obtain the reduced density matrix $\rho_A,$ c.f. Fig. \ref{fig:reduced123}.
By repeating the surgery for the region $A,$ $\tr_A(\rho_A)$ can be obtained similarly, c.f. Fig. \ref{fig:z1}.

The geometric representation of $\langle R_i |R_i\rangle$ constitutes of two large $S^3$'s, two thin $S^3$'s hosting the Wilson loops \cite{Dong:2008ft}. Note that one can pinch the end points of $S^2\times I$ to deform the tube into $S^3$ glued to the large $S^3$'s.

Because each one of the small large $S^3$'s is glued along two $S^2,$ from \eqref{eqn:surgery2} we obtain
\begin{equation}\label{eqn:surgery3}
   \tr{\rho_A(R)}= \langle R_i| R_i \rangle = \frac{Z(S^3;R_i)^2 Z(S^3; \overline{R}_i)^2}{Z(S^3;R_i)^2 Z(S^3;\overline{R}_i)^2}=1.
\end{equation}

An n-sheeted copy of \ref{fig:z1} can be similarly obtained by the surgery operation. Each replication adds two Wilson loops, $R_i$ and $\overline{R}_i,$ each of which is going through two thin $S^3$'s successively. As a result, the n-sheeted copy contains $2+2n$ $S^3$'s and the large $S^3$'s are connected to the thin $S^3$'s via gluing along $4n$ $S^2$'s. Hence, we obtain
\begin{equation}\label{eqn:surgery4}
    \tr{\rho_A(R)^n}= Z(S;R_i)^{1-n} Z(S;\overline{R}_i)^{1-n}.
\end{equation}

%For n copies of Fig.\ref{fig:torus with wilson2}, we successively glue the solid tori with the gluing map . \gabriel{should draw/describe the density matrix and then explain the usual cyclic gluing}
\section{Wilson loops in AdS/CFT} 
In this section, we want to further explain that the open/closed duality between the Hartle-Hawking states in each duality frame exhibited in the previous sections is not special to topological string theory. In fact, it has striking similarity to the AdS/CFT correspondence \cite{1998PhRvL..80.4859M, 2001EPJC...22..379R}. We first review the holographic dictionary between Wilson loops in four-dimensional $\mathcal{N}=4$ $SU(N)$ super Yang-Mills theory (SYM$_4$) and worldsheets in $AdS_5$. Then we argue that the duality between the Wilson loops and the worldsheets is the AdS/CFT analogue of the duality between the HH states in closed/open topological string theories.

Let's begin with the Wilson lines in $\mathcal{N}=4$ $SU(N)$ SYM$_4,$ in particular in the fundamental representation. The Wilson lines in the fundamental representation are interpreted as world lines of heavy quarks in SYM$_4$ as follows. The Wilon lines can be introduced to SYM$_4,$ in  string theory, by first placing $N+1$ D3-branes to get a $SU(N+1)$ SYM and then displacing one of the D3-branes. The separated D3-brane at distance $d$ probes the background generated by the $N$ D3-branes. The open string connecting N D3-brane stack and the probe brane breaks the gauge symmetry from $SU(N+1)$ to $SU(N) \times U(1)$. The boundary of the worldsheet in the N D-branes stack can be naturally understood as a Wilson loop in the fundamental representation. On a similar note, the end point of the open string in the N D-brane stack can be understood as a heavy quark in the $SU(N)$ SYM$_4$. Thus, the worldline of the heavy quark in the SYM$_4$ corresponds to a Wilson loop in the fundamental representation.

We can ask for the closed string dual by invoking the geometric transition to replace the $N$ D-branes with the non-trivial background $AdS_5 \times S^5$. The string worldsheet connecting the stack N D-branes and the probe D-brane, after the geometric transition, remains to be a string worldsheet ending on the probe D-brane. Once we identify the boundary of $AdS_5$ with where the SYM$_4$ is, these two different point of views provide us the Wilson loop/worldsheet duality in AdS/CFT. Later, the duality was generalized by \cite{2006JHEP...08..074G} beyond the fundamental representation to all
representations.

It was suggested in \cite{Gopakumar:1998ki} and later shown by \cite{Ooguri:1999bv, Gomis:2006mv, 2002math......1219T} that the duality between the worldsheet and the Wilson loops continues to hold in topological string theory under geometric transitions. The similarity and differences of the dualities in these two cases can be seen directly by comparing Fig. \ref{fig:example1} and Fig. \ref{fig:example2}, and summarized in the table below. 

\begin{center}
    \begin{tabular}{c|c|c}
 &  Type IIB Superstring& Topological string\\ \hline
Open string geometry & $\Bbb{R}^{1,9}$ & Deformed conifold \\ \hline
Dynamical brane topology & $\Bbb{R}^{1,3}$ & $S^3$ \\ \hline
Target space field theory & SYM$_4$(worldvolume EFT) & Chern-Simons theory(exact) \\ \hline
Closed string geometry & $AdS_5 \times S^5$(decoupling limit) & Resolved conifold \\ \hline
Probe brane topology & $\Bbb{R}^{1,3}$ & $\Bbb{C} \times S^1$  \\ \hline External heavy particles & Heavy quarks & Anyons  \\ \hline
Where probe brane ends & $\Bbb{R}^{1,3}$ at infinity of $AdS_5$& $S^3$ at infinity of resolved conifold\\ \hline
    \end{tabular}
\end{center}

So the correspondence between the entanglement entropy from the Wilson loop/worldsheet duality is in the same spirit as \cite{Lewkowycz:2013laa}. However, we are not looking at the extra entanglement entropy from a single probe Wilson loop, we are instead calculating the entanglement entropy from a superposition of Wilson loops that build up a geometrical dual spacetime!

\newpage
%\printbibliography
%\end{spacing}

\bibliographystyle{utphys}
\bibliography{draftZEEtopostrings.bib}

\providecommand{\href}[2]{#2}\begingroup\raggedright\begin{thebibliography}{10}

\bibitem{1999IJTP...38.1113M}
J.~{Maldacena}, ``{The Large-N Limit of Superconformal Field Theories and
  Supergravity},'' \href{http://dx.doi.org/10.1023/A:1026654312961}{{\em
  International Journal of Theoretical Physics} {\bfseries 38} (Jan., 1999)
  1113--1133}, \href{http://arxiv.org/abs/hep-th/9711200}{{\ttfamily
  arXiv:hep-th/9711200 [hep-th]}}.

\bibitem{1998PhLB..428..105G}
S.~S. {Gubser}, I.~R. {Klebanov}, and A.~M. {Polyakov}, ``{Gauge theory
  correlators from non-critical string theory},''
  \href{http://dx.doi.org/10.1016/S0370-2693(98)00377-3}{{\em Physics Letters
  B} {\bfseries 428} no.~1-2, (May, 1998) 105--114},
  \href{http://arxiv.org/abs/hep-th/9802109}{{\ttfamily arXiv:hep-th/9802109
  [hep-th]}}.

\bibitem{1998AdTMP...2..253W}
E.~{Witten}, ``{Anti-de Sitter space and holography},'' {\em Advances in
  Theoretical and Mathematical Physics} {\bfseries 2} (Jan., 1998) 253--291,
  \href{http://arxiv.org/abs/hep-th/9802150}{{\ttfamily arXiv:hep-th/9802150
  [hep-th]}}.

\bibitem{2006PhRvL..96r1602R}
S.~{Ryu} and T.~{Takayanagi}, ``{Holographic Derivation of Entanglement Entropy
  from the anti de Sitter Space/Conformal Field Theory Correspondence},''
  \href{http://dx.doi.org/10.1103/PhysRevLett.96.181602}{{\em Phys. Rev. Lett.}
  {\bfseries 96} no.~18, (May, 2006) 181602},
  \href{http://arxiv.org/abs/hep-th/0603001}{{\ttfamily arXiv:hep-th/0603001
  [hep-th]}}.

\bibitem{2007JHEP...07..062H}
V.~E. {Hubeny}, M.~{Rangamani}, and T.~{Takayanagi}, ``{A covariant holographic
  entanglement entropy proposal},''
  \href{http://dx.doi.org/10.1088/1126-6708/2007/07/062}{{\em Journal of High
  Energy Physics} {\bfseries 2007} no.~7, (July, 2007) 062},
  \href{http://arxiv.org/abs/0705.0016}{{\ttfamily arXiv:0705.0016 [hep-th]}}.

\bibitem{2013JHEP...11..074F}
T.~{Faulkner}, A.~{Lewkowycz}, and J.~{Maldacena}, ``{Quantum corrections to
  holographic entanglement entropy},''
  \href{http://dx.doi.org/10.1007/JHEP11(2013)074}{{\em Journal of High Energy
  Physics} {\bfseries 2013} (Nov., 2013) 74},
  \href{http://arxiv.org/abs/1307.2892}{{\ttfamily arXiv:1307.2892 [hep-th]}}.

\bibitem{2015JHEP...01..073E}
N.~{Engelhardt} and A.~C. {Wall}, ``{Quantum extremal surfaces: holographic
  entanglement entropy beyond the classical regime},''
  \href{http://dx.doi.org/10.1007/JHEP01(2015)073}{{\em Journal of High Energy
  Physics} {\bfseries 2015} (Jan., 2015) 73},
  \href{http://arxiv.org/abs/1408.3203}{{\ttfamily arXiv:1408.3203 [hep-th]}}.

\bibitem{2013JHEP...08..090L}
A.~{Lewkowycz} and J.~{Maldacena}, ``{Generalized gravitational entropy},''
  \href{http://dx.doi.org/10.1007/JHEP08(2013)090}{{\em Journal of High Energy
  Physics} {\bfseries 2013} (Aug., 2013) 90},
  \href{http://arxiv.org/abs/1304.4926}{{\ttfamily arXiv:1304.4926 [hep-th]}}.

\bibitem{PhysRevD.15.2752}
G.~W. Gibbons and S.~W. Hawking, ``Action integrals and partition functions in
  quantum gravity,'' \href{http://dx.doi.org/10.1103/PhysRevD.15.2752}{{\em
  Phys. Rev. D} {\bfseries 15} (May, 1977) 2752--2756}.
  \url{https://link.aps.org/doi/10.1103/PhysRevD.15.2752}.

\bibitem{2020arXiv201010539H}
D.~{Harlow} and E.~{Shaghoulian}, ``{Global symmetry, Euclidean gravity, and
  the black hole information problem},'' {\em arXiv e-prints} (Oct., 2020)
  arXiv:2010.10539, \href{http://arxiv.org/abs/2010.10539}{{\ttfamily
  arXiv:2010.10539 [hep-th]}}.

\bibitem{2010GReGr..42.2323V}
M.~{van Raamsdonk}, ``{Building up spacetime with quantum entanglement},''
  \href{http://dx.doi.org/10.1007/s10714-010-1034-0}{{\em General Relativity
  and Gravitation} {\bfseries 42} no.~10, (Oct., 2010) 2323--2329},
  \href{http://arxiv.org/abs/1005.3035}{{\ttfamily arXiv:1005.3035 [hep-th]}}.

\bibitem{Susskind:1994sm}
L.~Susskind and J.~Uglum, ``{Black hole entropy in canonical quantum gravity
  and superstring theory},''
  \href{http://dx.doi.org/10.1103/PhysRevD.50.2700}{{\em Phys. Rev.} {\bfseries
  D50} (1994) 2700--2711},
\href{http://arxiv.org/abs/hep-th/9401070}{{\ttfamily arXiv:hep-th/9401070
  [hep-th]}}.
%%CITATION = HEP-TH/9401070;%%.

\bibitem{Donnelly:2018ppr}
W.~Donnelly and G.~Wong, ``{Entanglement branes, modular flow, and extended
  topological quantum field theory},''
  \href{http://dx.doi.org/10.1007/JHEP10(2019)016}{{\em JHEP} {\bfseries 10}
  (2019) 016}, \href{http://arxiv.org/abs/1811.10785}{{\ttfamily
  arXiv:1811.10785 [hep-th]}}.

\bibitem{Donnelly:2016jet}
W.~Donnelly and G.~Wong, ``{Entanglement branes in a two-dimensional string
  theory},'' \href{http://dx.doi.org/10.1007/JHEP09(2017)097}{{\em JHEP}
  {\bfseries 09} (2017) 097}, \href{http://arxiv.org/abs/1610.01719}{{\ttfamily
  arXiv:1610.01719 [hep-th]}}.

\bibitem{Jafferis:2019wkd}
D.~L. Jafferis and D.~K. Kolchmeyer, ``{Entanglement Entropy in
  Jackiw-Teitelboim Gravity},''
\href{http://arxiv.org/abs/1911.10663}{{\ttfamily arXiv:1911.10663 [hep-th]}}.
%%CITATION = ARXIV:1911.10663;%%.

\bibitem{2020arXiv201015737D}
W.~{Donnelly}, Y.~{Jiang}, M.~{Kim}, and G.~{Wong}, ``{Entanglement entropy and
  edge modes in topological string theory: I},'' {\em arXiv e-prints} (Oct.,
  2020) arXiv:2010.15737, \href{http://arxiv.org/abs/2010.15737}{{\ttfamily
  arXiv:2010.15737 [hep-th]}}.

\bibitem{Bryan:2004iq}
J.~Bryan and R.~Pandharipande, ``{The Local Gromov-Witten theory of curves},''
  \href{http://dx.doi.org/10.1090/S0894-0347-06-00545-5}{{\em J. Am. Math.
  Soc.} {\bfseries 21} (2008) 101--136},
  \href{http://arxiv.org/abs/math/0411037}{{\ttfamily arXiv:math/0411037}}.

\bibitem{Aganagic:2004js}
M.~Aganagic, H.~Ooguri, N.~Saulina, and C.~Vafa, ``{Black holes, q-deformed 2d
  Yang-Mills, and non-perturbative topological strings},''
  \href{http://dx.doi.org/10.1016/j.nuclphysb.2005.02.035}{{\em Nucl. Phys.}
  {\bfseries B715} (2005) 304--348},
\href{http://arxiv.org/abs/hep-th/0411280}{{\ttfamily arXiv:hep-th/0411280
  [hep-th]}}.
%%CITATION = HEP-TH/0411280;%%.

\bibitem{Gopakumar:1998ki}
R.~Gopakumar and C.~Vafa, ``{On the gauge theory / geometry correspondence},''
  \href{http://dx.doi.org/10.4310/ATMP.1999.v3.n5.a5}{{\em AMS/IP Stud. Adv.
  Math.} {\bfseries 23} (2001) 45--63},
  \href{http://arxiv.org/abs/hep-th/9811131}{{\ttfamily arXiv:hep-th/9811131}}.

\bibitem{Hubeny:2019bje}
V.~E. Hubeny, R.~Pius, and M.~Rangamani, ``{Topological string entanglement},''
  \href{http://dx.doi.org/10.1007/JHEP10(2019)239}{{\em JHEP} {\bfseries 10}
  (2019) 239}, \href{http://arxiv.org/abs/1905.09890}{{\ttfamily
  arXiv:1905.09890 [hep-th]}}.

\bibitem{Witten:1992fb}
E.~Witten, ``{Chern-Simons gauge theory as a string theory},'' {\em Prog.
  Math.} {\bfseries 133} (1995) 637--678,
\href{http://arxiv.org/abs/hep-th/9207094}{{\ttfamily arXiv:hep-th/9207094
  [hep-th]}}.
%%CITATION = HEP-TH/9207094;%%.

\bibitem{1998PhRvL..80.4859M}
J.~{Maldacena}, ``{Wilson Loops in Large N Field Theories},''
  \href{http://dx.doi.org/10.1103/PhysRevLett.80.4859}{{\em prl} {\bfseries 80}
  no.~22, (Jun, 1998) 4859--4862},
  \href{http://arxiv.org/abs/hep-th/9803002}{{\ttfamily arXiv:hep-th/9803002
  [hep-th]}}.

\bibitem{2001EPJC...22..379R}
S.~J. {Rey} and J.~T. {Yee}, ``{Macroscopic strings as heavy quarks: Large- N
  gauge theory and anti-de Sitter supergravity},''
  \href{http://dx.doi.org/10.1007/s100520100799}{{\em European Physical Journal
  C} {\bfseries 22} no.~2, (Nov, 2001) 379--394},
  \href{http://arxiv.org/abs/hep-th/9803001}{{\ttfamily arXiv:hep-th/9803001
  [hep-th]}}.

\bibitem{Ooguri:1999bv}
H.~Ooguri and C.~Vafa, ``{Knot invariants and topological strings},''
  \href{http://dx.doi.org/10.1016/S0550-3213(00)00118-8}{{\em Nucl. Phys.}
  {\bfseries B577} (2000) 419--438},
\href{http://arxiv.org/abs/hep-th/9912123}{{\ttfamily arXiv:hep-th/9912123
  [hep-th]}}.
%%CITATION = HEP-TH/9912123;%%.

\bibitem{Gomis:2006mv}
J.~Gomis and T.~Okuda, ``{Wilson loops, geometric transitions and bubbling
  Calabi-Yau's},'' \href{http://dx.doi.org/10.1088/1126-6708/2007/02/083}{{\em
  JHEP} {\bfseries 02} (2007) 083},
\href{http://arxiv.org/abs/hep-th/0612190}{{\ttfamily arXiv:hep-th/0612190
  [hep-th]}}.
%%CITATION = HEP-TH/0612190;%%.

\bibitem{2013arXiv1307.1132J}
K.~{Jensen} and A.~{Karch}, ``{The holographic dual of an EPR pair has a
  wormhole},'' {\em arXiv e-prints} (July, 2013) arXiv:1307.1132,
  \href{http://arxiv.org/abs/1307.1132}{{\ttfamily arXiv:1307.1132 [hep-th]}}.

\bibitem{2013PhRvD..88j6006J}
K.~{Jensen} and A.~{O'Bannon}, ``{Holography, entanglement entropy, and
  conformal field theories with boundaries or defects},''
  \href{http://dx.doi.org/10.1103/PhysRevD.88.106006}{{\em Phys. Rev. D}
  {\bfseries 88} no.~10, (Nov., 2013) 106006},
  \href{http://arxiv.org/abs/1309.4523}{{\ttfamily arXiv:1309.4523 [hep-th]}}.

\bibitem{Lewkowycz:2013laa}
A.~Lewkowycz and J.~Maldacena, ``{Exact results for the entanglement entropy
  and the energy radiated by a quark},''
  \href{http://dx.doi.org/10.1007/JHEP05(2014)025}{{\em JHEP} {\bfseries 05}
  (2014) 025},
\href{http://arxiv.org/abs/1312.5682}{{\ttfamily arXiv:1312.5682 [hep-th]}}.
%%CITATION = ARXIV:1312.5682;%%.

\bibitem{Witten:1988hf}
E.~Witten, ``{Quantum Field Theory and the Jones Polynomial},''
  \href{http://dx.doi.org/10.1007/BF01217730}{{\em Commun. Math. Phys.}
  {\bfseries 121} (1989) 351--399}.

\bibitem{Guadagnini:1989tj}
E.~Guadagnini, M.~Martellini, and M.~Mintchev, ``{Braids and Quantum Group
  Symmetry in {Chern-Simons} Theory},''
  \href{http://dx.doi.org/10.1016/0550-3213(90)90443-H}{{\em Nucl. Phys. B}
  {\bfseries 336} (1990) 581--609}.

\bibitem{Slingerland:2001aa}
J.~K. Slingerland and F.~A. Bais, ``Quantum groups and nonabelian braiding in
  quantum hall systems,''
  \href{http://dx.doi.org/10.1016/S0550-3213(01)00308-X}{{\em Nucl.Phys.}
  {\bfseries B612} (2001) 229--290},
  \href{http://arxiv.org/abs/cond-mat/0104035}{{\ttfamily cond-mat/0104035}}.
  \url{https://arxiv.org/pdf/cond-mat/0104035.pdf}.

\bibitem{Ooguri:2002gx}
H.~Ooguri and C.~Vafa, ``{World sheet derivation of a large N duality},''
  \href{http://dx.doi.org/10.1016/S0550-3213(02)00620-X}{{\em Nucl. Phys.}
  {\bfseries B641} (2002) 3--34},
\href{http://arxiv.org/abs/hep-th/0205297}{{\ttfamily arXiv:hep-th/0205297
  [hep-th]}}.
%%CITATION = HEP-TH/0205297;%%.

\bibitem{2005CMaPh.254..425A}
M.~{Aganagic}, A.~{Klemm}, M.~{Mari{\~n}o}, and C.~{Vafa}, ``{The Topological
  Vertex},'' \href{http://dx.doi.org/10.1007/s00220-004-1162-z}{{\em
  Communications in Mathematical Physics} {\bfseries 254} no.~2, (Mar., 2005)
  425--478}, \href{http://arxiv.org/abs/hep-th/0305132}{{\ttfamily
  arXiv:hep-th/0305132 [hep-th]}}.

\bibitem{Witten:1988xj}
E.~Witten, ``{Topological Sigma Models},''
  \href{http://dx.doi.org/10.1007/BF01466725}{{\em Commun. Math. Phys.}
  {\bfseries 118} (1988) 411}.

\bibitem{Aganagic:2002qg}
M.~Aganagic, M.~Marino, and C.~Vafa, ``{All loop topological string amplitudes
  from Chern-Simons theory},''
  \href{http://dx.doi.org/10.1007/s00220-004-1067-x}{{\em Commun. Math. Phys.}
  {\bfseries 247} (2004) 467--512},
  \href{http://arxiv.org/abs/hep-th/0206164}{{\ttfamily arXiv:hep-th/0206164}}.

\bibitem{Aganagic:2003db}
M.~Aganagic, A.~Klemm, M.~Marino, and C.~Vafa, ``{The Topological vertex},''
  \href{http://dx.doi.org/10.1007/s00220-004-1162-z}{{\em Commun. Math. Phys.}
  {\bfseries 254} (2005) 425--478},
  \href{http://arxiv.org/abs/hep-th/0305132}{{\ttfamily arXiv:hep-th/0305132}}.

\bibitem{Klemm:1999gm}
A.~Klemm and E.~Zaslow, ``{Local mirror symmetry at higher genus},'' {\em
  AMS/IP Stud. Adv. Math.} {\bfseries 23} (2001) 183--207,
  \href{http://arxiv.org/abs/hep-th/9906046}{{\ttfamily arXiv:hep-th/9906046}}.

\bibitem{Dijkgraaf:2002fc}
R.~Dijkgraaf and C.~Vafa, ``{Matrix models, topological strings, and
  supersymmetric gauge theories},''
  \href{http://dx.doi.org/10.1016/S0550-3213(02)00766-6}{{\em Nucl. Phys. B}
  {\bfseries 644} (2002) 3--20},
  \href{http://arxiv.org/abs/hep-th/0206255}{{\ttfamily arXiv:hep-th/0206255}}.

\bibitem{Aganagic:2003qj}
M.~Aganagic, R.~Dijkgraaf, A.~Klemm, M.~Marino, and C.~Vafa, ``{Topological
  strings and integrable hierarchies},''
  \href{http://dx.doi.org/10.1007/s00220-005-1448-9}{{\em Commun. Math. Phys.}
  {\bfseries 261} (2006) 451--516},
  \href{http://arxiv.org/abs/hep-th/0312085}{{\ttfamily arXiv:hep-th/0312085}}.

\bibitem{Cota:2019cjx}
C.~F. Cota, A.~Klemm, and T.~Schimannek, ``{Topological strings on genus one
  fibered Calabi-Yau 3-folds and string dualities},''
  \href{http://dx.doi.org/10.1007/JHEP11(2019)170}{{\em JHEP} {\bfseries 11}
  (2019) 170}, \href{http://arxiv.org/abs/1910.01988}{{\ttfamily
  arXiv:1910.01988 [hep-th]}}.

\bibitem{Huang:2015sta}
M.-x. Huang, S.~Katz, and A.~Klemm, ``{Topological String on elliptic CY
  3-folds and the ring of Jacobi forms},''
  \href{http://dx.doi.org/10.1007/JHEP10(2015)125}{{\em JHEP} {\bfseries 10}
  (2015) 125}, \href{http://arxiv.org/abs/1501.04891}{{\ttfamily
  arXiv:1501.04891 [hep-th]}}.

\bibitem{Vafa:2004qa}
C.~Vafa, ``{Two dimensional Yang-Mills, black holes and topological strings},''
  \href{http://arxiv.org/abs/hep-th/0406058}{{\ttfamily arXiv:hep-th/0406058}}.

\bibitem{2008arXiv0809.3976M}
D.~{Maulik}, A.~{Oblomkov}, A.~{Okounkov}, and R.~{Pandharipand e},
  ``{Gromov-Witten/Donaldson-Thomas correspondence for toric 3-folds},'' {\em
  arXiv e-prints} (Sept., 2008) arXiv:0809.3976,
  \href{http://arxiv.org/abs/0809.3976}{{\ttfamily arXiv:0809.3976 [math.AG]}}.

\bibitem{hori2003mirror}
K.~Hori, R.~Thomas, S.~Katz, C.~Vafa, R.~Pandharipande, A.~Klemm, R.~Vakil, and
  E.~Zaslow, {\em Mirror symmetry}, vol.~1.
\newblock American Mathematical Soc., 2003.

\bibitem{Witten:1991zz}
E.~Witten, ``{Mirror manifolds and topological field theory},'' {\em AMS/IP
  Stud. Adv. Math.} {\bfseries 9} (1998) 121--160,
  \href{http://arxiv.org/abs/hep-th/9112056}{{\ttfamily arXiv:hep-th/9112056}}.

\bibitem{2001hep.th....1218V}
C.~{Vafa}, ``{Brane/anti-Brane Systems and U(N|M) Supergroup},'' {\em arXiv
  e-prints} (Jan., 2001) hep--th/0101218,
  \href{http://arxiv.org/abs/hep-th/0101218}{{\ttfamily arXiv:hep-th/0101218
  [hep-th]}}.

\bibitem{de_Haro_2007}
S.~de~Haro, S.~Ramgoolam, and A.~Torrielli, ``Large n expansion of q-deformed
  two-dimensional yang-mills theory and hecke algebras,''
  \href{http://dx.doi.org/10.1007/s00220-007-0232-4}{{\em Communications in
  Mathematical Physics} {\bfseries 273} no.~2, (May, 2007) 317–355}.
  \url{http://dx.doi.org/10.1007/s00220-007-0232-4}.

\bibitem{Couvreur_2017}
R.~Couvreur, J.~L. Jacobsen, and H.~Saleur, ``Entanglement in nonunitary
  quantum critical spin chains,''
  \href{http://dx.doi.org/10.1103/physrevlett.119.040601}{{\em Physical Review
  Letters} {\bfseries 119} no.~4, (Jul, 2017) }.
  \url{http://dx.doi.org/10.1103/PhysRevLett.119.040601}.

\bibitem{Quella:2020aa}
T.~Quella, ``Symmetry protected topological phases beyond groups: The
  q-deformed aklt model,'' \href{http://arxiv.org/abs/2005.09072}{{\ttfamily
  2005.09072}}. \url{https://arxiv.org/pdf/2005.09072.pdf}.

\bibitem{2002math......1219T}
C.~H. {Taubes}, ``{Lagrangians for the Gopakumar-Vafa conjecture},'' {\em arXiv
  Mathematics e-prints} (Jan., 2002) math/0201219,
  \href{http://arxiv.org/abs/math/0201219}{{\ttfamily arXiv:math/0201219
  [math.DG]}}.

\bibitem{Wen:2016snr}
X.~Wen, S.~Matsuura, and S.~Ryu, ``{Edge theory approach to topological
  entanglement entropy, mutual information and entanglement negativity in
  Chern-Simons theories},''
  \href{http://dx.doi.org/10.1103/PhysRevB.93.245140}{{\em Phys. Rev.}
  {\bfseries B93} no.~24, (2016) 245140},
\href{http://arxiv.org/abs/1603.08534}{{\ttfamily arXiv:1603.08534
  [cond-mat.mes-hall]}}.
%%CITATION = ARXIV:1603.08534;%%.

\bibitem{Wong:2017pdm}
G.~Wong, ``{A note on entanglement edge modes in Chern Simons theory},''
  \href{http://dx.doi.org/10.1007/JHEP08(2018)020}{{\em JHEP} {\bfseries 08}
  (2018) 020},
\href{http://arxiv.org/abs/1706.04666}{{\ttfamily arXiv:1706.04666 [hep-th]}}.
%%CITATION = ARXIV:1706.04666;%%.

\bibitem{Das:2015oha}
D.~Das and S.~Datta, ``{Universal features of left-right entanglement
  entropy},'' \href{http://dx.doi.org/10.1103/PhysRevLett.115.131602}{{\em
  Phys. Rev. Lett.} {\bfseries 115} no.~13, (2015) 131602},
\href{http://arxiv.org/abs/1504.02475}{{\ttfamily arXiv:1504.02475 [hep-th]}}.
%%CITATION = ARXIV:1504.02475;%%.

\bibitem{Dong:2008ft}
S.~Dong, E.~Fradkin, R.~G. Leigh, and S.~Nowling, ``{Topological Entanglement
  Entropy in Chern-Simons Theories and Quantum Hall Fluids},''
  \href{http://dx.doi.org/10.1088/1126-6708/2008/05/016}{{\em JHEP} {\bfseries
  05} (2008) 016}, \href{http://arxiv.org/abs/0802.3231}{{\ttfamily
  arXiv:0802.3231 [hep-th]}}.

\bibitem{Fliss:2020cos}
J.~R. Fliss and R.~G. Leigh, ``{Interfaces and the extended Hilbert space of
  Chern-Simons theory},'' \href{http://arxiv.org/abs/2004.05123}{{\ttfamily
  arXiv:2004.05123 [hep-th]}}.

\bibitem{Elitzur:1989nr}
S.~Elitzur, G.~W. Moore, A.~Schwimmer, and N.~Seiberg, ``{Remarks on the
  Canonical Quantization of the Chern-Simons-Witten Theory},''
  \href{http://dx.doi.org/10.1016/0550-3213(89)90436-7}{{\em Nucl. Phys. B}
  {\bfseries 326} (1989) 108--134}.

\bibitem{Marino:2004uf}
M.~Marino, ``{Chern-Simons theory and topological strings},''
  \href{http://dx.doi.org/10.1103/RevModPhys.77.675}{{\em Rev. Mod. Phys.}
  {\bfseries 77} (2005) 675--720},
  \href{http://arxiv.org/abs/hep-th/0406005}{{\ttfamily arXiv:hep-th/0406005}}.

\bibitem{2004CMaPh.247..467A}
M.~{Aganagic}, M.~{Mari{\~n}o}, and C.~{Vafa}, ``{All Loop Topological String
  Amplitudes from Chern-Simons Theory},''
  \href{http://dx.doi.org/10.1007/s00220-004-1067-x}{{\em Communications in
  Mathematical Physics} {\bfseries 247} no.~2, (Jan., 2004) 467--512},
  \href{http://arxiv.org/abs/hep-th/0206164}{{\ttfamily arXiv:hep-th/0206164
  [hep-th]}}.

\bibitem{Diaconescu:2002sf}
D.-E. Diaconescu, B.~Florea, and A.~Grassi, ``{Geometric transitions and open
  string instantons},''
  \href{http://dx.doi.org/10.4310/ATMP.2002.v6.n4.a2}{{\em Adv. Theor. Math.
  Phys.} {\bfseries 6} (2003) 619--642},
  \href{http://arxiv.org/abs/hep-th/0205234}{{\ttfamily arXiv:hep-th/0205234}}.

\bibitem{Dupic_2018}
T.~Dupic, B.~Estienne, and Y.~Ikhlef, ``Entanglement entropies of minimal
  models from null-vectors,''
  \href{http://dx.doi.org/10.21468/scipostphys.4.6.031}{{\em SciPost Physics}
  {\bfseries 4} no.~6, (Jun, 2018) }.
  \url{http://dx.doi.org/10.21468/SciPostPhys.4.6.031}.

\bibitem{2018arXiv180706575L}
J.~{Lin}, ``{Entanglement entropy in Jackiw-Teitelboim Gravity},'' {\em arXiv
  e-prints} (July, 2018) arXiv:1807.06575,
  \href{http://arxiv.org/abs/1807.06575}{{\ttfamily arXiv:1807.06575
  [hep-th]}}.

\bibitem{2020arXiv200313117V}
H.~{Verlinde}, ``{ER = EPR revisited: On the Entropy of an Einstein-Rosen
  Bridge},'' {\em arXiv e-prints} (Mar., 2020) arXiv:2003.13117,
  \href{http://arxiv.org/abs/2003.13117}{{\ttfamily arXiv:2003.13117
  [hep-th]}}.

\bibitem{2011JHEP...04..113K}
E.~{Kiritsis} and V.~{Niarchos}, ``{Large-N limits of 2d CFTs, quivers and
  AdS$_{3}$ duals},'' \href{http://dx.doi.org/10.1007/JHEP04(2011)113}{{\em
  Journal of High Energy Physics} {\bfseries 2011} (Apr., 2011) 113},
  \href{http://arxiv.org/abs/1011.5900}{{\ttfamily arXiv:1011.5900 [hep-th]}}.

\bibitem{2005NuPhB.715..304A}
M.~{Aganagic}, H.~{Ooguri}, N.~{Saulina}, and C.~{Vafa}, ``{Black holes,
  q-deformed 2d Yang Mills, and non-perturbative topological strings},''
  \href{http://dx.doi.org/10.1016/j.nuclphysb.2005.02.035}{{\em Nuclear Physics
  B} {\bfseries 715} no.~1, (May, 2005) 304--348},
  \href{http://arxiv.org/abs/hep-th/0411280}{{\ttfamily arXiv:hep-th/0411280
  [hep-th]}}.

\bibitem{Donnelly:2016auv}
W.~Donnelly and L.~Freidel, ``{Local subsystems in gauge theory and gravity},''
  \href{http://dx.doi.org/10.1007/JHEP09(2016)102}{{\em JHEP} {\bfseries 09}
  (2016) 102}, \href{http://arxiv.org/abs/1601.04744}{{\ttfamily
  arXiv:1601.04744 [hep-th]}}.

\bibitem{1995NuPhB.443...85W}
E.~{Witten}, ``{String theory dynamics in various dimensions},''
  \href{http://dx.doi.org/10.1016/0550-3213(95)00158-O}{{\em Nuclear Physics B}
  {\bfseries 443} no.~1, (Feb., 1995) 85--126},
  \href{http://arxiv.org/abs/hep-th/9503124}{{\ttfamily arXiv:hep-th/9503124
  [hep-th]}}.

\bibitem{1996NuPhB.460..506H}
P.~{Hor{\v{r}}ava} and E.~{Witten}, ``{Heterotic and Type I string dynamics
  from eleven dimensions},''
  \href{http://dx.doi.org/10.1016/0550-3213(95)00621-4}{{\em Nuclear Physics B}
  {\bfseries 460} no.~3, (Feb., 1996) 506--524},
  \href{http://arxiv.org/abs/hep-th/9510209}{{\ttfamily arXiv:hep-th/9510209
  [hep-th]}}.

\bibitem{Dijkgraaf:2004te}
R.~Dijkgraaf, S.~Gukov, A.~Neitzke, and C.~Vafa, ``{Topological M-theory as
  unification of form theories of gravity},''
  \href{http://dx.doi.org/10.4310/ATMP.2005.v9.n4.a5}{{\em Adv. Theor. Math.
  Phys.} {\bfseries 9} no.~4, (2005) 603--665},
  \href{http://arxiv.org/abs/hep-th/0411073}{{\ttfamily arXiv:hep-th/0411073}}.

\bibitem{Bershadsky:1993cx}
M.~Bershadsky, S.~Cecotti, H.~Ooguri, and C.~Vafa, ``{Kodaira-Spencer theory of
  gravity and exact results for quantum string amplitudes},''
  \href{http://dx.doi.org/10.1007/BF02099774}{{\em Commun. Math. Phys.}
  {\bfseries 165} (1994) 311--428},
  \href{http://arxiv.org/abs/hep-th/9309140}{{\ttfamily arXiv:hep-th/9309140}}.

\bibitem{1993hep.th....6122W}
E.~{Witten}, ``{Quantum Background Independence In String Theory},'' {\em arXiv
  e-prints} (June, 1993) hep--th/9306122,
  \href{http://arxiv.org/abs/hep-th/9306122}{{\ttfamily arXiv:hep-th/9306122
  [hep-th]}}.

\bibitem{Ooguri:2005vr}
H.~Ooguri, C.~Vafa, and E.~P. Verlinde, ``{Hartle-Hawking wave-function for
  flux compactifications},''
  \href{http://dx.doi.org/10.1007/s11005-005-0022-x}{{\em Lett. Math. Phys.}
  {\bfseries 74} (2005) 311--342},
  \href{http://arxiv.org/abs/hep-th/0502211}{{\ttfamily arXiv:hep-th/0502211}}.

\bibitem{2003JHEP...04..021M}
J.~{Maldacena}, ``{Eternal black holes in anti-de Sitter},''
  \href{http://dx.doi.org/10.1088/1126-6708/2003/04/021}{{\em Journal of High
  Energy Physics} {\bfseries 2003} no.~4, (Apr., 2003) 021},
  \href{http://arxiv.org/abs/hep-th/0106112}{{\ttfamily arXiv:hep-th/0106112
  [hep-th]}}.

\bibitem{2013ForPh..61..781M}
J.~{Maldacena} and L.~{Susskind}, ``{Cool horizons for entangled black
  holes},'' \href{http://dx.doi.org/10.1002/prop.201300020}{{\em Fortschritte
  der Physik} {\bfseries 61} no.~9, (Sept., 2013) 781--811},
  \href{http://arxiv.org/abs/1306.0533}{{\ttfamily arXiv:1306.0533 [hep-th]}}.

\bibitem{2004hep.th....6058V}
C.~{Vafa}, ``{Two Dimensional Yang-Mills, Black Holes and Topological
  Strings},'' {\em arXiv e-prints} (June, 2004) hep--th/0406058,
  \href{http://arxiv.org/abs/hep-th/0406058}{{\ttfamily arXiv:hep-th/0406058
  [hep-th]}}.

\bibitem{2006PhRvD..73f6002D}
R.~{Dijkgraaf}, R.~{Gopakumar}, H.~{Ooguri}, and C.~{Vafa}, ``{Baby universes
  in string theory},'' \href{http://dx.doi.org/10.1103/PhysRevD.73.066002}{{\em
  Phys. Rev. D} {\bfseries 73} no.~6, (Mar., 2006) 066002},
  \href{http://arxiv.org/abs/hep-th/0504221}{{\ttfamily arXiv:hep-th/0504221
  [gr-qc]}}.

\bibitem{2007NuPhB.778...36A}
M.~{Aganagic}, T.~{Okuda}, and H.~{Ooguri}, ``{Quantum entanglement of baby
  universes},'' \href{http://dx.doi.org/10.1016/j.nuclphysb.2007.04.006}{{\em
  Nuclear Physics B} {\bfseries 778} no.~1-2, (Aug., 2007) 36--68},
  \href{http://arxiv.org/abs/hep-th/0612067}{{\ttfamily arXiv:hep-th/0612067
  [hep-th]}}.

\bibitem{2020JHEP...08..044M}
D.~{Marolf} and H.~{Maxfield}, ``{Transcending the ensemble: baby universes,
  spacetime wormholes, and the order and disorder of black hole information},''
  \href{http://dx.doi.org/10.1007/JHEP08(2020)044}{{\em Journal of High Energy
  Physics} {\bfseries 2020} no.~8, (Aug., 2020) 44},
  \href{http://arxiv.org/abs/2002.08950}{{\ttfamily arXiv:2002.08950
  [hep-th]}}.

\bibitem{2020arXiv200406738M}
J.~{McNamara} and C.~{Vafa}, ``{Baby Universes, Holography, and the
  Swampland},'' {\em arXiv e-prints} (Apr., 2020) arXiv:2004.06738,
  \href{http://arxiv.org/abs/2004.06738}{{\ttfamily arXiv:2004.06738
  [hep-th]}}.

\bibitem{2002NuPhB.644....3D}
R.~{Dijkgraaf} and C.~{Vafa}, ``{Matrix models, topological strings, and
  supersymmetric gauge theories},''
  \href{http://dx.doi.org/10.1016/S0550-3213(02)00766-6}{{\em Nuclear Physics
  B} {\bfseries 644} no.~1, (Nov., 2002) 3--20},
  \href{http://arxiv.org/abs/hep-th/0206255}{{\ttfamily arXiv:hep-th/0206255
  [hep-th]}}.

\bibitem{PhysRevLett.115.121602}
S.~A. Hartnoll and E.~A. Mazenc, ``Entanglement entropy in two-dimensional
  string theory,'' \href{http://dx.doi.org/10.1103/PhysRevLett.115.121602}{{\em
  Phys. Rev. Lett.} {\bfseries 115} (Sep, 2015) 121602}.
  \url{https://link.aps.org/doi/10.1103/PhysRevLett.115.121602}.

\bibitem{2012CQGra..29o5009C}
B.~{Czech}, J.~L. {Karczmarek}, F.~{Nogueira}, and M.~{Van Raamsdonk}, ``{The
  gravity dual of a density matrix},''
  \href{http://dx.doi.org/10.1088/0264-9381/29/15/155009}{{\em Classical and
  Quantum Gravity} {\bfseries 29} no.~15, (Aug., 2012) 155009},
  \href{http://arxiv.org/abs/1204.1330}{{\ttfamily arXiv:1204.1330 [hep-th]}}.

\bibitem{Bousso:2012mh}
R.~Bousso, B.~Freivogel, S.~Leichenauer, V.~Rosenhaus, and C.~Zukowski, ``{Null
  Geodesics, Local CFT Operators and AdS/CFT for Subregions},''
  \href{http://dx.doi.org/10.1103/PhysRevD.88.064057}{{\em Phys. Rev. D}
  {\bfseries 88} (2013) 064057},
  \href{http://arxiv.org/abs/1209.4641}{{\ttfamily arXiv:1209.4641 [hep-th]}}.

\bibitem{2002hep.th....6161N}
N.~A. {Nekrasov}, ``{Seiberg-Witten Prepotential From Instanton Counting},''
  {\em arXiv e-prints} (June, 2002) hep--th/0206161,
  \href{http://arxiv.org/abs/hep-th/0206161}{{\ttfamily arXiv:hep-th/0206161
  [hep-th]}}.

\bibitem{2009JHEP...10..069I}
A.~{Iqbal}, C.~{Koz{\c{c}}az}, and C.~{Vafa}, ``{The refined topological
  vertex},'' \href{http://dx.doi.org/10.1088/1126-6708/2009/10/069}{{\em
  Journal of High Energy Physics} {\bfseries 2009} no.~10, (Oct., 2009) 069},
  \href{http://arxiv.org/abs/hep-th/0701156}{{\ttfamily arXiv:hep-th/0701156
  [hep-th]}}.

\bibitem{2010maph.conf..265N}
N.~A. {Nekrasov} and S.~L. {Shatashvili},
  \href{http://dx.doi.org/10.1142/9789814304634_0015}{``{Quantization of
  Integrable Systems and Four Dimensional Gauge Theories},''} in {\em XVITH
  INTERNATIONAL CONGRESS ON MATHEMATICAL PHYSICS. Held 3-8 August 2009 in
  Prague}, pp.~265--289.
\newblock Mar., 2010.
\newblock \href{http://arxiv.org/abs/0908.4052}{{\ttfamily arXiv:0908.4052
  [hep-th]}}.

\bibitem{2019arXiv190311115S}
P.~{Saad}, S.~H. {Shenker}, and D.~{Stanford}, ``{JT gravity as a matrix
  integral},'' {\em arXiv e-prints} (Mar., 2019) arXiv:1903.11115,
  \href{http://arxiv.org/abs/1903.11115}{{\ttfamily arXiv:1903.11115
  [hep-th]}}.

\bibitem{Candelas:1990rm}
P.~Candelas, X.~C. De~La~Ossa, P.~S. Green, and L.~Parkes, ``{A Pair of
  Calabi-Yau manifolds as an exactly soluble superconformal theory},''
  \href{http://dx.doi.org/10.1016/0550-3213(91)90292-6}{{\em AMS/IP Stud.\
  Adv.\ Math.} {\bfseries 9} (1998) 31--95}.

\bibitem{2006JHEP...08..074G}
J.~{Gomis} and F.~{Passerini}, ``{Holographic Wilson loops},''
  \href{http://dx.doi.org/10.1088/1126-6708/2006/08/074}{{\em Journal of High
  Energy Physics} {\bfseries 2006} no.~8, (Aug., 2006) 074},
  \href{http://arxiv.org/abs/hep-th/0604007}{{\ttfamily arXiv:hep-th/0604007
  [hep-th]}}.

\end{thebibliography}\endgroup

\end{document}